\documentclass[useAMS,usenatbib]{mn2e}
\usepackage{savesym}
\usepackage{txfonts}
\savesymbol{iint}
\savesymbol{iiint}
\savesymbol{iiiint}
\savesymbol{idotsint}
\usepackage{graphicx}
\usepackage{amssymb}
\usepackage[below]{placeins}
\usepackage{txfonts}
\usepackage{natbib}
\usepackage{amsmath}
\usepackage{float}
\restoresymbol{TXF}{iint}
\restoresymbol{TXF}{iiint}
\restoresymbol{TXF}{iiiint}
\restoresymbol{TXF}{idotsint}
\usepackage{multirow}
\usepackage{array}
\usepackage{xcolor}
\usepackage[percent]{overpic}
\bibpunct{(}{)}{;}{a}{}{,}

\def \aap{A\&A}
\def \apjl{ApJ}

\def \apjs{ApJS}

\def\spose#1{\hbox to 0pt{#1\hss}}
\def\approxlt{\mathrel{\spose{\lower 3pt\hbox{$\sim$}}
        \raise 2.0pt\hbox{$<$}}}
\def\approxgt{\mathrel{\spose{\lower 3pt\hbox{$\sim$}}
        \raise 2.0pt\hbox{$>$}}}
\def\approxpropto{\mathrel{\spose{\lower 3pt\hbox{$\sim$}}
        \raise 2.0pt\hbox{$\propto$}}}
\mathchardef\twiddle="2218

\def\multleft#1{\hbox to size{\vbox {\halign {\lft{##}\cr #1}}\hfill}\par}
\def\multright#1{\hbox to size{\vbox {\halign {\rt{##}\cr #1}}\hfill}\par}

\def\today{\ifcase\month\or January\or February\or March\or April\or May\or
      June\or July\or August\or September\or October\or November\or December\fi
      \space\number\day, \number\year}
\def\<{\thinspace}

\def\arcsec{{\rm\thinspace arcsec}}

\def\ga{{\rm\thinspace gauss}}

\newcommand\field{\Pi_{\rm F}}
\newcommand\viscint{\Pi_{\rm v}}
\newcommand\nelec{n_{\rm e}}
\newcommand\nh{n_{\rm H}}

\newcommand\tcool{t_{\rm c}}

\def\arcsec {\hbox{$^{\prime\prime}$}}

\makeatletter
\newcommand{\thickhline}{%
    \noalign {\ifnum 0=`}\fi \hrule height 1.2pt
    \futurelet \reserved@a \@xhline
}
\newcolumntype{"}{@{\hskip\tabcolsep\vrule width 1pt\hskip\tabcolsep}}
\makeatother

\newcommand{\ion}[2]{#1\,{\sc{#2}}}

\bibpunct{(}{)}{;}{a}{}{,}

\title[Cold gas in giant ellipticals]{The origin of cold gas in giant elliptical galaxies \\and its role in fueling radio-mode AGN feedback}

\author[Werner et al.]{N. Werner$^{1,2}$, J. B. R. Oonk$^{3}$, M. Sun$^{4}$, P.~E.~J.~Nulsen$^{5}$, S. W. Allen$^{1,2,6}$, R. E. A. Canning$^{1,2}$,\newauthor A. 
Simionescu$^{7}$, A. Hoffer$^{8}$, T. Connor$^{8}$, M.~Donahue$^{8}$, A.~C.~Edge$^{9}$, A.~C.~Fabian$^{10}$, \newauthor A.~von~der~Linden$^{1,2,11}$, 
C.~S.~Reynolds$^{12}$, M.~Ruszkowski$^{13,14}$ \\
$^1$Kavli Institute for Particle Astrophysics and Cosmology, Stanford University, 452 Lomita Mall, Stanford, CA 94305-4085, USA \\
$^2$Department of Physics, Stanford University, 382 Via Pueblo Mall, Stanford, CA 94305-4060, USA \\
$^3$ASTRON, Netherlands Institute for Radio Astronomy, P.O. Box 2, 7990 AA Dwingeloo, The Netherlands \\
$^4$Department of Physics, University of Alabama in Huntsville, Huntsville, AL 35899, USA \\
$^{5}$Harvard-Smithsonian Center for Astrophysics, 60 Garden Street, Cambridge, MA 02138, USA \\
$^6$SLAC National Accelerator Laboratory, 2575 Sand Hill Road, Menlo Park, CA 94025, USA \\
$^7$Institute of Space and Astronautical Science (ISAS), JAXA, 3-1-1 Yoshinodai, Chuo-ku, Sagamihara, Kanagawa, 252-5210 Japan \\
$^8$Physics \& Astronomy Department, Michigan State University, East Lansing, MI 48824-2320, USA \\
$^9$Institute for Computational Cosmology, Department of Physics, Durham University, Durham, DH1 3LE, UK \\
$^{10}$Institute of Astronomy, Madingley Road, Cambridge CB3 0HA, UK \\
$^{11}$Dark Cosmology Centre, Niels Bohr Institute, University of Copenhagen, Juliane Maries Vej 30, DK-2100 Copenhagen, Denmark \\
$^{12}$Department of Astronomy and the Maryland Astronomy Center for Theory and Computation, University of Maryland, College Park, MD 20742, USA \\
$^{13}$Department of Astronomy, University of Michigan, 500 Church Street, Ann Arbor, MI 48109, USA \\
$^{14}$Michigan Center for Theoretical Physics, 3444 Randall Lab, 450 Church St, Ann Arbor, MI 48109, USA \\
}

\begin{document}
\maketitle

\begin{abstract}
The nature and origin of the cold interstellar medium (ISM) in early type galaxies are still a matter of debate, and understanding the role of this component in galaxy evolution 
and in fueling the central supermassive black holes requires more observational constraints. Here, we present a multi-wavelength study of the ISM in eight nearby, X-ray and 
optically bright, giant elliptical galaxies, all central dominant members of relatively low mass groups. 
Using far-infrared spectral imaging with the {\it Herschel} Photodetector Array Camera \& Spectrometer (PACS), we map the emission of cold gas in the cooling lines of [\ion{C}
{ii}]$\lambda157\mu$m, [\ion{O}{i}]$\lambda$63$\mu$m, and [\ion{O}{i}b]$\lambda$145$\mu$m. Additionally, we present H$\alpha$+[\ion{N}{ii}] imaging of warm ionized gas 
with the Southern Astrophysical Research (SOAR) telescope, and a study of the thermodynamic structure of the hot X-ray emitting plasma with {\it Chandra}. 
All systems with extended H$\alpha$ emission in our sample (6/8 galaxies) display significant [\ion{C}{ii}] line emission indicating the presence of reservoirs of cold gas. This 
emission is co-spatial with the optical H$\alpha$+[\ion{N}{ii}] emitting nebulae and the lowest entropy soft X-ray emitting plasma. 
The entropy profiles of the hot galactic atmospheres show a clear dichotomy, with the systems displaying extended emission line nebulae having lower entropies beyond $r
\gtrsim1$~kpc than the cold-gas-poor systems. We show that while the hot atmospheres of the cold-gas-poor galaxies are thermally stable outside of their innermost cores, the 
atmospheres of the cold-gas-rich systems are prone to cooling instabilities. This provides considerable weight to the argument that cold gas in giant ellipticals is produced 
chiefly by cooling from the hot phase. 
We show that cooling instabilities may develop more easily in rotating systems and discuss an alternative condition for thermal instability for this case.
The hot atmospheres of cold-gas-rich galaxies display disturbed morphologies indicating that the accretion of clumpy multiphase gas in these systems may result in variable 
power output of the AGN jets, potentially triggering sporadic, larger outbursts. 
In the two cold-gas-poor, X-ray morphologically relaxed galaxies of our sample, NGC 1399 and NGC 4472, powerful AGN outbursts may have destroyed or removed most of 
the cold gas from the cores, allowing the jets to propagate and deposit most of their energy further out, increasing the entropy of the hot galactic atmospheres and leaving their 
cores relatively undisturbed.
\end{abstract}

\begin{keywords}
galaxies: active -- galaxies: elliptical and lenticular, cD -- galaxies: ISM -- accretion -- infrared: galaxies -- X-rays: galaxies
\end{keywords}

\section{Introduction}

Recent observations and simulations strongly suggest that the growth and evolution of giant early type galaxies is closely tied to that of their central supermassive black holes 
through a well regulated feedback cycle \citep[e.g.][]{silk1998,magorrian1998,croton2006,sijacki2007}. Many fundamental aspects of this feedback process, such as the nature 
of the material feeding the black holes in active galactic nuclei (AGN), jet formation, and the heating, cooling and detailed physics of the interstellar medium (ISM) are not well 
understood. Among the best laboratories to test theories for the formation and growth of massive galaxies are nearby giant ellipticals, groups and clusters of galaxies: systems 
where hot X-ray emitting plasma may be cooling and accreting onto the galaxies and interacting with their AGN.

\begin{table*}
\caption{Summary of the sample of galaxies. The distances in column 2 are from $^\dagger$\citet{tonry2001} and $^\ddagger$\citet{blakeslee2009}. The K-band luminosities 
in column 5 were determined using the 2MASS survey \citep{jarrett2003}. The H$\alpha$+[\ion{N}{ii}] luminosities in column 6 are from \citet{macchetto1996}. The luminosities 
marked with `$\star $' have been revised in Sect.~\ref{results} and Table~\ref{SOARfluxes}. The 1.4~GHz radio luminosities in column 7 are from \citet{condon1998} and 
\citet{condon2002}. Bolometric X-ray luminosities in column 8 are from \citet{osullivan2001}. The jet powers for NGC~4636, NGC~4472, and NGC~5846 are from 
\citet{allen2006}, for NGC~5044 from \citet{david2009}, for NGC~1399 from \citet{shurkin2008}, and for NGC~5813 from \citet{randall2011}. For the jet-powers in NGC~5044 
and NGC~5813 no errorbars were published \citep{randall2011,david2009}. Star-formation rates were estimated using a combination of FUV data from {\it Galex} and MIR 
data from {\it Wise}, employing the technique developed by \citet{hao2011}.}
\begin{center}
\begin{tabular}{lccccccccc}

\hline\hline
Galaxy  			& 	$d$		& Scale				&       $z$	& $L_{\rm K}$						& $L_{\rm{H}\alpha+[\ion{N}{II}]}$ 		& $L_{\rm radio}$ 
				& $L_{\rm X}$  					&  $P_{\rm jet}$  & SFR \\ %& $N_{\mathrm{H}}$
				&  (Mpc)		& (arcsec~kpc$^{-1}$)	& 		&	($\times10^{40}$~erg~s$^{-1}$)	&	($\times10^{39}$~erg~s$^{-1}$) 	& ($
\times10^{38}$~erg~s$^{-1}$) 	& ($\times10^{41}$~erg~s$^{-1}$)  	& ($\times10^{41}$~erg~s$^{-1}$)   & $M_\odot$~yr$^{-1}$\\  %& (10$^{20}$~cm$^{-2}$)
\hline 
NGC~1399		&	20.9$^\ddagger $		& 9.9		&	0.004753	& 2.69  & 	9.3$^\star $	&	1.52 & 5.68  &	$21.9\pm7$		& 0.171	\\ % & 1.50
NGC~4472		&	16.7$^\ddagger $		& 12.4	&	0.003326	&  3.97 & 	5.8	&	1.20 & 2.96  &	$80.7\pm23.5$		& 0.099	\\  %& 1.53
NGC~4636 		&	14.7$^\dagger$	& 14.0	&	0.003129	& 1.20  & 	5.5$^\star $	&	0.28 & 3.32  &	$3.0\pm0.8$		& 0.027	\\ %& 1.90
NGC~5044 		&	31.2$^\dagger$	& 6.6   	&  	0.009280	&  1.65 &  45.6$^\star $	&	0.59 & 5.87  & 	$6.0$			& 0.073	\\ %& 4.87
NGC~5813		&	32.2$^\dagger$	& 6.4		& 	0.006578	&  2.30 & 	14.0$^\star $	&	0.26 &10.63 &	$1.2$			& 0.063	\\ %& 4.37
NGC~5846 		&	27.1$^\dagger$	& 7.6		&	0.005717	&  2.11 & 	24.6	&	0.26 	& 6.25 &	$7.4\pm2.6$		& 0.074	\\ %& 4.29
NGC~6868		&       26.8$^\dagger$	& 5.6		&       0.009520 &  1.74 &	23.8	&	1.00	& 1.01 &	--				& 0.078	\\ %& 3.91
NGC~7049		&	29.9$^\dagger$	& 7.1		&       0.007622 &  2.32 & 	31.8	&	0.84	& 1.23  &	--				& 0.117	\\ %& 2.70
\hline
\label{galaxies}
\end{tabular}
\end{center}
\end{table*}

In massive galaxy clusters, with central cooling times shorter than the Hubble time, the central brightest cluster galaxies (BCG) are often surrounded by spectacular, 
filamentary optical H$\alpha$+[\ion{N}{ii}] emission-line nebulae extending up to 70~kpc from the core of the BCG \citep[e.g.][]
{cowie1983,johnstone1987,heckman1989,donahue1992,crawford1999,mcdonald2010}. These nebulae of ionized gas seem to be co-spatial with warm (1000-2000~K) 
molecular hydrogen seen in the near-infrared \citep[e.g.][]{jaffe1997,falcke1998,donahue2000,edge2002,hatch2005,jaffe2005,johnstone2007,oonk2010,lim2012} and with 
large quantities ($10^{8}-10^{11.5}$~M$_{\odot}$) of cold ($<50$~K) molecular gas traced by CO \citep[e.g.][]{edge2001,edge2003,salome2003,mcdonald2012}. Recently, 
observations with the {\it Herschel Space Observatory} \citep{pilbratt2010} also revealed the presence of far-infrared (FIR) cooling lines of [\ion{C}{ii}], [\ion{O}{i}], and [\ion{N}
{ii}] in the X-ray bright cores of Abell~1068, Abell~2597 \citep{edge2010}, and the Centaurus, Perseus, and Virgo clusters \citep{mittal2011,mittal2012,werner2013}. In nearby 
systems, such as the Centaurus, Perseus, and Virgo clusters, where the spatial distribution of the [\ion{C}{ii}] line could be mapped, it was found to be extended and co-spatial 
with optical H$\alpha$+[\ion{N}{ii}], far-ultraviolet \ion{C}{iv} \citep[in M~87,][]{sparks2012}, and soft X-ray emission. 
The filamentary nebulae in the centers of massive cool-core galaxy clusters thus contain multi-phase material spanning a temperature range of over 5 orders of magnitude, 
from $<50$~K to $\sim10^{7}$~K. 
This gas often appears to be interacting with the jets and the buoyant relativistic plasma from the central AGN \citep[][]{fabian2003b,hatch2006,canning2013}. Despite the large 
quantities of cold gas, many extended nebulae show no evidence for recent star-formation.

The majority of AGN in these systems are in a quiescent, so called `radio'-mode, where they are accreting at a modest rate. The accretion flows in this mode, though optically 
faint, often drive powerful relativistic jets, extending to large distances, which can have a profound impact on their surroundings. Using {\it Chandra} X-ray observations of 
nine nearby, X-ray luminous elliptical galaxies, \citet{allen2006} found a tight correlation between the Bondi accretion rates \citep[see][]{bondi1952} from the hot X-ray emitting 
ISM and the power in the relativistic jets \citep[though see][]{russell2013}. The jet powers were determined from the work required to inflate bubbles of relativistic plasma 
associated with the most recent cavities in the hot X-ray emitting atmospheres of the galaxies \citep{churazov2002}. The relationship between jet power and accretion rate of 
the hot ISM, however, raises the question: what is the role of the cold gas in the AGN feedback cycle? 

\begin{table*}
\caption{Summary of observations.}
\centering
\begin{tabular}{lcccccccc}
\hline\hline
Galaxy  			&  {\it Herschel} & {\it Herschel} & {\it Chandra} &  Detector	& {\it Chandra} & SOAR/SOI	&	SOAR/SOI	&	SOAR/SOI\\
				& Obs. ID		&  Exp. (s)		&	Obs. ID	&			& Exp. (ks)	& Obs. date	&  filter			&	Exp (s)	 \\
\hline 
NGC1399			& 1342239492 & 8161 	&	319		& ACIS-S  &	49.9 & 2012 Oct. 9 & CTIO 656375-4/6916-78 & $3\times600$ / $3\times600$	\\
				& 			&	     	&	4172		& ACIS-I	&	36.7 & 2013 Aug. 4 & Goodman 1.68$''$ long slit, KOSI600 grating & $3\times480$ \\ 
				&			&	     	&	9530		&  ACIS-S &	59.3 & 2013 Sep. 8 & Goodman 3.0$''$ long slit, KOSI600 grating & $4\times480$ \\
\hline
NGC4472			& 1342234992 & 8161 	&	321		& ACIS-S	&	19.1 \\
\hline
NGC4636 		& 1342236884 & 5442 	&	323		& ACIS-S	&	35.8 & 2009 June 1	& CTIO 6600/75 & $3\times780$	\\
				&			&		&			&		&		&			& CTIO 6120/140& $3\times500$   \\
\hline
NGC5044 		& 1342238376 & 24723	&	9399		& ACIS-S  &	53.4	& 2010 April 10& CTIO 6649/76 & $3\times1200$	\\
				& 			& 	      	&	3926		& ACIS-I	&	54.5 &			& CTIO 6520/76 & $3\times720$ 	 \\
				& 			&	      	&	4415		& ACIS-I	&	73.3 &			&			&	\\
\hline
NGC5813			& 1342238158 &10880 	&	5907		& ACIS-S  &	47.4		& 2008 July 6	&CTIO 6600/75 & $3\times900$\\
				& 			& 	      	&	9517  	& ACIS-S  &	98.8 &			&CTIO 6120/140 & $3\times720$\\
\hline
NGC5846 		& 1342238157 & 5442  	&	788	& ACIS-S  &	17.3 & 2009 June 1	& CTIO 6600/75  & $3\times900$\\
				& 			&	      	&	7923   	& ACIS-I	&	75.9 &			& CTIO 6120/140 & $3\times540$ \\
\hline
NGC6868			& 1342215929 & 5442 	&	3191	& ACIS-I   &        18.6		& 2012 Oct. 9 & CTIO 6600/75& $2\times900$, $1\times600$  \\
				&			&		&	11753	& ACIS-I	&	56.8  	& 2012 Oct. 9 & CTIO 6563/75& $2\times900$, $1\times600$ \\
\hline
NGC7049			& 1342216658 & 5442 	&	5895		& ACIS-I	&	2.2 &  2012 Oct. 9 & CTIO 6600/75 & $2\times900$, $1\times600$  \\
				&			&		&			&		&	       &  2012 Oct. 9 & CTIO 6563/75 & $1\times900$, $2\times600$	\\
\hline
\label{obs}
\end{tabular}\\
\end{table*}

Here we present a multi-wavelength study of 8 nearby, X-ray bright, giant elliptical galaxies, all central dominant members of relatively low mass groups. We use FIR data obtained with the {\it Herschel} Photodetector Array Camera \& Spectrometer (PACS) in the lines of [\ion{C}{ii}]$\lambda157$~$\mu$m, [\ion{O}{i}]$\lambda$63$\mu$m, and [\ion{O}{i}b]$\lambda$145$\mu$m that are excellent probes of cold $\sim100$~K gas; optical data from the Southern Astrophysical Research (SOAR) telescope in the lines of H$\alpha$+[\ion{N}{ii}] probing warm, ionized $\sim$10,000~K gas; and X-ray data from the {\it Chandra X-ray Observatory}, tracing the hot 5-20 million K plasma permeating these systems.

\subsection{The galaxy sample}
\label{sample}

Our target list is drawn from the parent sample of \citet{dunn2010}, who identified the optically and X-ray brightest giant elliptical/S0 galaxies within a distance $d\leq100$~Mpc 
and with declination $\rm{Dec.} \geq -45$. Motivated by the goal of studying the properties of cold gas in these types of systems in detail, we selected only those galaxies with 
distance $d<35$~Mpc and with relatively bright H$\alpha$+[\ion{N}{ii}] emission as reported by \citet{macchetto1996}. All of the systems in our sample are the central 
dominant galaxies of their respective groups. The properties of the galaxies are summarized in Table~\ref{galaxies}. The table lists the distances of the galaxies 
\citep{tonry2001,blakeslee2009}, their K-band near-infrared luminosities determined using the magnitudes measured by the 2MASS survey \citep{jarrett2003},  H$\alpha$+
[\ion{N}{ii}] luminosities \citep{macchetto1996}, radio luminosities at 1.4 GHz \citep{condon1998,condon2002}, bolometric X-ray luminosities \citep{osullivan2001}, jet-powers 
\citep{allen2006,david2009,shurkin2008,randall2011},  and star-formation rates (SFR) estimated using a combination of far-ultraviolet (FUV) data from {\it Galex} and mid-
infrared (MIR) data from {\it Wise}, employing the technique developed by \citet{hao2011}.

The near-infrared K-band luminosities, and by implication the stellar masses, span a range of approximately a factor of three, with NGC~4636 being the least and NGC~4472 
the most luminous. The H$\alpha$+[\ion{N}{ii}] luminosities span a range of a factor of eight, with NGC~5044 being the most luminous, and with NGC~4636 and NGC~4472 on 
the low luminosity end. NGC~5044 boasts a particularly spectacular network of radial filaments, extending out to a radius of at least $r\sim$10~kpc 
\citep{gastaldello2009,david2011}. The SFRs are small, typically of the order of 0.1~$M_\odot$~yr$^{-1}$, and they do not correlate with the H$\alpha$ luminosities.  All eight 
galaxies have central, active radio jets, also spanning a range of radio luminosities and jet-powers.
The X-ray luminosities of the galaxies span a range of an order of magnitude. 

\section{Observations and data analysis}
\label{analysis}

\subsection{Far-infrared spectroscopy with {\it Herschel} PACS}

We observed the FIR cooling lines of [\ion{C}{ii}]$\lambda157\mu$m, [\ion{O}{i}]$\lambda63\mu$m, and [\ion{O}{i}b]$\lambda145\mu$m in our sample of 8 galaxies with the 
PACS integral-field spectrometer \citep{poglitsch2010} on the {\it Herschel Space Observatory}. 
Table~\ref{obs} gives a summary of the observations. The observations were taken in line spectroscopy mode with chopping-nodding to remove the telescope background, 
sky background and dark current. A chopper throw of 6 arcmin was used. For 7 systems, with a relatively compact H$\alpha$ emission region, the observations were taken in 
pointed mode targeting the centers of the galaxies. For NGC~5044, we used raster mapping in $3\times3$ steps of 23.5 arcsec to match the extent of the H$\alpha$ nebula.
The observations were reduced using the {\tt HIPE} software version 8.2.0, using the PACS {\tt ChopNodLineScan} pipeline script. This script processes the data from level 0 
(raw channel data) to level 2 (flux calibrated spectral cubes). 

During the final stage of the reduction the data were spectrally and spatially rebinned into $5\times5\times\lambda$ cubes. In the following we will refer to these cubes as the 
rebinned cubes. Each spatial pixel, termed spaxel, in these cubes has a size of $9.4\times9.4$~arcsec$^{2}$. The cubes thus provide us with a field of view (FoV) of 
$47\times47$~arcsec$^{2}$. For the wavelength regridding the parameters {\tt oversample} and {\tt upsample} were set to 
2 and 1, respectively. This means that one spectral bin corresponds to the native resolution of the PACS instrument.

The integrated [\ion{C}{ii}] line fluxes were in all cases, except for NGC~5044, obtained by spatially integrating the $5\times5$ spaxels from the rebinned cubes. For 
NGC~5044 the integrated line flux is obtained after projecting the individual rebinned cubes onto the sky. No point-spread function (PSF) correction is applied as the [\ion{C}
{ii}] emission, in all cases where it is detected, is found to be extended.

To visualize the extent of the [\ion{C}{ii}] emission in our galaxies, we have created sky maps of the [\ion{C}{ii}] emission by using the {\tt specProject} task in {\tt HIPE} and the 
{\tt hrebin} task in {\tt IDL}. In the following we will refer to these data cubes as the projected cubes. A pixel size of 6~arcsec was chosen in order to Nyquist sample the beam, 
the full-width-at-half-maximum (FWHM) of which is 12~arcsec at the observed wavelength of the [\ion{C}{ii}] line. We only consider spatial bins where the signal-to-noise ratio 
of the integrated line flux is greater than 2. We have compared the integrated line fluxes obtained from the level~2 rebinned cubes with those from the projected cubes and find 
that they are consistent. Velocity and velocity width maps were constructed by fitting a single Gaussian to the projected data.

For the [\ion{O}{i}] and [\ion{O}{i}b] lines, we report line fluxes obtained from either the central spaxel or the central $3\times3$ spaxels. The PSF corrected fluxes should be 
viewed as lower limits to the true [\ion{O}{i}] and [\ion{O}{i}b] line emission in these galaxies. For NGC~5044 the integrated [\ion{O}{i}] and [\ion{O}{i}b] fluxes are obtained after 
projecting the individual rebinned cubes onto the sky.

\begin{table*}
\begin{center}
\caption{Summary of observations with {\it Herschel} PACS and the properties of the FIR lines. }
\begin{tabular}{llccccc}
\hline\hline
galaxy & Line &  Observation duration & Line Flux in central spaxel &  Observed FWHM  & Line shift & Integrated Line Flux  \\
&	       &  (s) &  ($10^{-14}$~erg s$^{-1}$ cm$^{-2}$) & (km~s$^{-1}$) & (km~s$^{-1}$) & ($10^{-14}$~erg s$^{-1}$ cm$^{-2}$) \\
\hline
NGC~1399 &	\ion{C}{ii}$\lambda157.7\mu$m & 2250   &  $<0.2$ 			&  300$^{\rm f}$  & -	& - \\
		  &	\ion{O}{i}$\lambda63.2\mu$m &   2484  &  $<2.8$ 	 		&  300$^{\rm f}$  & -	 & - \\
		  &	\ion{O}{i}b$\lambda$145.5 &	     3360   &  $<0.2$ 			&  300$^{\rm f}$  & -	& - \\
NGC~4472&  \ion{C}{ii}$\lambda157.7\mu$m &   2250  &  $0.2\pm0.04$ 	&  $255\pm36$	   & $99\pm15$ & $1.04\pm0.16$	\\
		  &  \ion{O}{i}$\lambda63.2\mu$m &	2484	&  $<2.5$			&  255$^{\rm f}$ & -	& - \\
		  &  \ion{O}{i}b$\lambda145.5\mu$m & 3360	&  $<0.2$			&  255$^{\rm f}$ & -	& - \\
NGC~4636&  \ion{C}{ii}$\lambda157.7\mu$m &  1500	&  $2.6\pm0.06$ 	&  $361\pm6$     & $22\pm3$ & $10.52\pm0.37$\\
		  &  \ion{O}{i}$\lambda63.2\mu$m &	     1656 &  $1.3\pm0.2$ (1.9) 	&  $233\pm29$   & $75\pm12$ & - \\
		  &  \ion{O}{i}b $\lambda145.5\mu$m & 2240	&  $<0.5$			&   233$^{\rm f}$ & -	& - \\
NGC~5044&  \ion{C}{ii}$\lambda157.7\mu$m & 6750	&  	-			&	-		&- &	$31.73\pm4.76$ \\
		  &  \ion{O}{i}$\lambda63.2\mu$m &	   7452	&  	-			&	-		& -& $7.1\pm1.8$	\\
		  &  \ion{O}{i}b$\lambda145.5\mu$m &10080	&  	-			&	-		& -& $1.0\pm0.5$	\\
NGC~5813& \ion{C}{ii}$\lambda157.7\mu$m &  3000	&  $1.4\pm0.04$	&	$419\pm10$ & $96\pm4$ & $8.55\pm0.26$ \\
		  & \ion{O}{i}$\lambda63.2\mu$m &	   3312 	&  $1.0\pm0.2$ (1.5)	&	$273\pm36$  & $30\pm15$ & - \\
		  &  \ion{O}{i}b$\lambda145.5\mu$m &4480	&$0.2\pm0.04$ (0.4)&	$608\pm88$  & $100\pm37$ & $1.4\pm0.2$ \\ % (1.6)
NGC~5846& \ion{C}{ii}$\lambda157.7\mu$m & 1500	&  $2.2\pm0.06$ 	&	$477\pm10$ & $-25\pm4$ & $13.61\pm0.36$ \\
		  & \ion{O}{i}$\lambda63.2\mu$m &	   1656	&  $<2.5$		   	&	477$^{\rm f}$ & - & - \\
		  & \ion{O}{i}b$\lambda145.5\mu$m & 2240	&  $<0.3$		   	&	477$^{\rm f}$ & - & - \\
NGC~6868& \ion{C}{ii}$\lambda157.7\mu$ &	1500	&  $5.7\pm0.08$ 	&	$510\pm6$  & $125\pm3$ & $21.61\pm0.44$ \\
		  & \ion{O}{i}$\lambda63.2\mu$m &	1656	&  $3.0\pm0.3$ 	&	$506\pm31$ & $137\pm13$ & $5.9\pm0.8$ \\ % (6.3)
		  & \ion{O}{i}b$\lambda145.5\mu$m & 2240	& $0.5\pm0.06$ 	&	$579\pm50$ &  $125\pm21$ & $2.0\pm0.3$  \\ %(2.1)
NGC~7049& \ion{C}{ii}$\lambda157.7\mu$m & 1500	&  $2.5\pm0.06$ 	& $395\pm7$ & $78\pm3$ & $22.49\pm0.45$	\\
		  & \ion{O}{i}$\lambda63.2\mu$m &	1656	&  $<3.3$ 			& 395$^{\rm f}$	 & - & - \\
		  & \ion{O}{i}b$\lambda145.5\mu$m & 2240	&  $<0.2$ 			& 395$^{\rm f}$	 & - & - \\
\hline
\label{FIRlines}
\end{tabular}
\end{center}
Notes: In brackets, we indicate the PSF corrected fluxes determined assuming the emission comes from a point-source, correcting the measured spaxel flux upward to account 
for the beam size. The PSF corrected fluxes can therefore be considered a lower limit to the flux integrated over the whole PACS area. 
\end{table*}

\begin{table}
\caption{H$\alpha$+[\ion{N}{ii}] fluxes from the SOAR telescope.}
\centering
\begin{tabular}{lc}
\hline\hline
Galaxy  			&	$f_{\rm{H}\alpha+\rm{[NII]}}$	\\
				&  	($10^{-13}$~erg~s$^{-1}$~cm$^{-2}$)	\\
\hline 
NGC~1399		& $<0.34$ \\ 
NGC~4636 		& $2.7\pm0.4$	\\
NGC~5044 		& $7.6\pm0.9$	\\
NGC~5813		& $2.2\pm0.3$	\\
\hline
\label{SOARfluxes}
\end{tabular}
\end{table}

\begin{table*}
\caption{Deprojected thermodynamic properties at $r\sim0.5$~kpc.}
\centering
\begin{tabular}{lccccc}
\hline\hline
Galaxy  			&	$n_{\rm e}$	&	$kT$			& 	$P_{\rm e}$		& $K$			&       $t_{\rm cool}$ \\
				&  	(cm$^{-3}$)	&	(keV)		&	(keV~cm$^{-3}$)	& (keV~cm$^2$)	& 	($10^7$~yr)	\\
\hline 
NGC~1399		& $0.167\pm0.005$ &  $0.950\pm0.009$	&	$0.158\pm0.005$	& $3.14\pm0.07$	&	4.6	\\ 
NGC~4472		& $0.154\pm0.008$	& $0.860\pm0.012$	&	$0.132\pm0.007$	& $2.99\pm0.12$	&	4.3	\\
NGC~4636 		& $0.072\pm0.001$	& $0.535\pm0.020$	&	$0.039\pm0.001$	& $3.11\pm0.07$	&	5.2	\\
NGC~5044 		& $0.068\pm0.005$	& $0.582\pm0.037$	&	$0.039\pm0.004$	& $3.51\pm0.27$	&  	5.9	\\
NGC~5813		& $0.074\pm0.003$	& $0.616\pm0.041$	&	$0.046\pm0.005$	& $3.50\pm0.31$	& 	5.7	\\
NGC~5846 		& $0.076\pm0.006$	& $0.677\pm0.033$	&	$0.052\pm0.005$	& $3.77\pm0.26$	&	6.2	\\
NGC~6868		&  $0.072\pm0.002$	& $0.768\pm0.036$	&	$0.055\pm0.003$	& $4.45\pm0.22$	&       7.8	\\
\hline
\label{deprojected}
\end{tabular}
\end{table*}

\subsection{H$\alpha$+[\ion{N}{ii}] imaging and spectroscopy}
\label{SOARanalysis}

We performed narrow-band imaging observations for all systems in our sample, except for NGC~4472, using the SOAR Optical Imager (SOI) on the 4.1~m SOAR telescope. 
See Table~\ref{obs} for details on the observations. For each galaxy, we obtained two narrow-band images, one centered on the H$\alpha$ emission and the other in an 
adjacent emission-line free band. We reduced the images using standard procedures in the {\tt IRAF MSCRED} package. The pixels were binned by a factor of two, for a scale 
of 0.154 arcsec per pixel. The typical seeing was $\sim$1$''$. Spectrophotometric standard stars were observed for each exposure. More detail on the SOI data reduction 
can be found in \citet{sun2007}. For continuum subtraction, we follow the isophote fitting method described in \citet{goudfrooij1994}. We assumed line ratios of [\ion{N}{ii}]$
\lambda6583$/H$\alpha = 1.5$ and [\ion{N}{ii}]$\lambda6548$/[\ion{N}{ii}]$\lambda6583=1/3$, which are consistent with the typical values in \citet{goudfrooij1994}.

On August~4 and September~8 2013, we also took spectra of NGC~1399 using the Goodman spectrograph on the SOAR telescope. We used a 600 lines per mm grating for 
the wavelength range of 4350--6950 \AA. The first run was taken with a 1.68$''$ slit, along the NS direction. The second run was taken with a 3.0$''$ slit, at a position angle of 
302 degrees (almost along the major axis of the galaxy).  We took three 480 second exposures on the first night and four 480 second exposures on the second night. Before 
and after each exposure we took a quartz flat and a comparison spectrum of an Fe lamp. Bias correction, flat-fielding, and wavelength calibration were all performed using 
standard {\tt IRAF} procedures. For flux calibration, we took spectra of the spectrophotometric standard stars LTT7379 and LTT1020.

\subsection{{\it Chandra} X-ray data}

\begin{figure*}
\begin{minipage}{0.32\textwidth}
\includegraphics[width=1\textwidth,clip=t,angle=0.,bb=36 187 577 605]{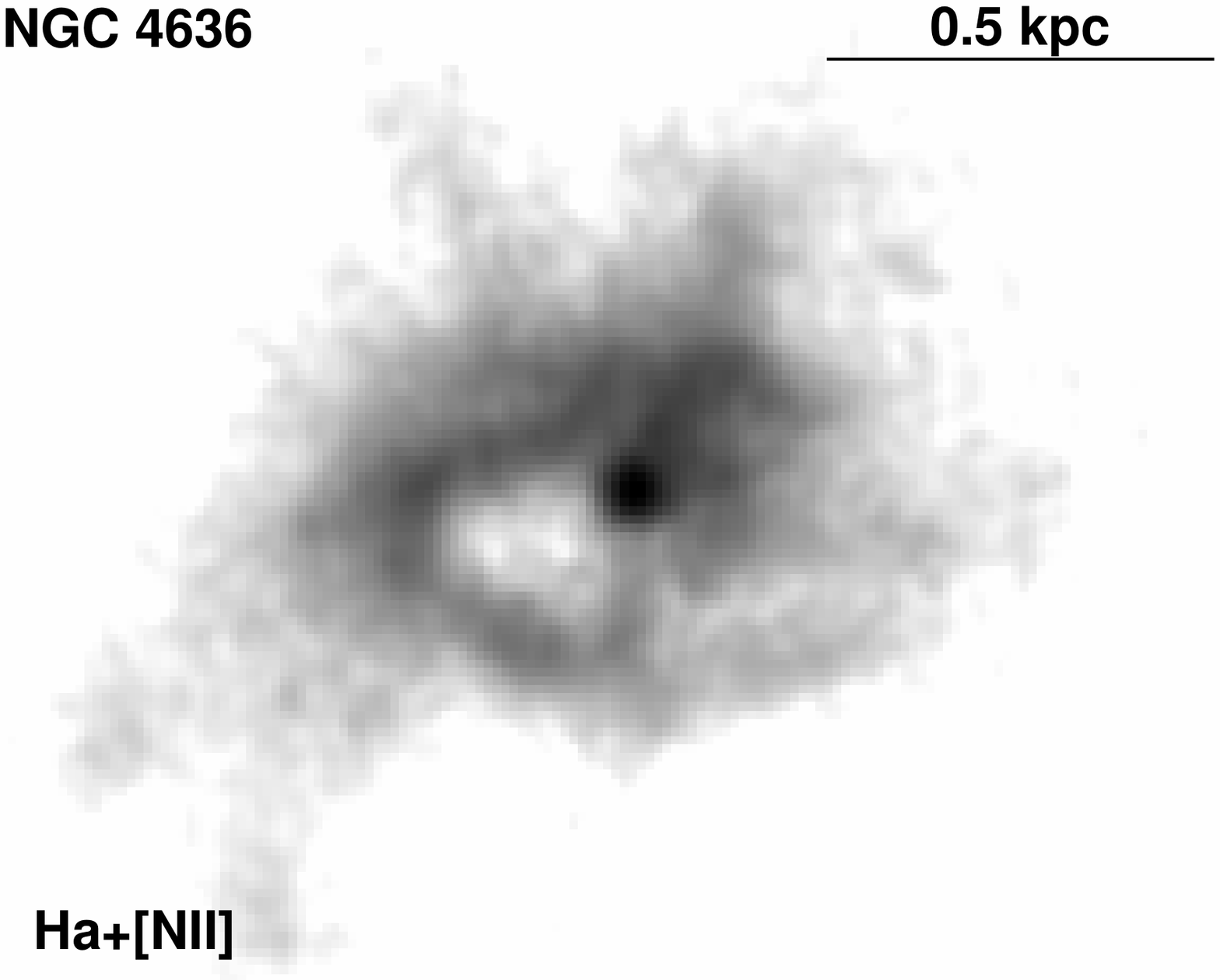}
\end{minipage}
\begin{minipage}{0.32\textwidth}
\includegraphics[width=1\textwidth,clip=t,angle=0.,bb=36 187 577 605]{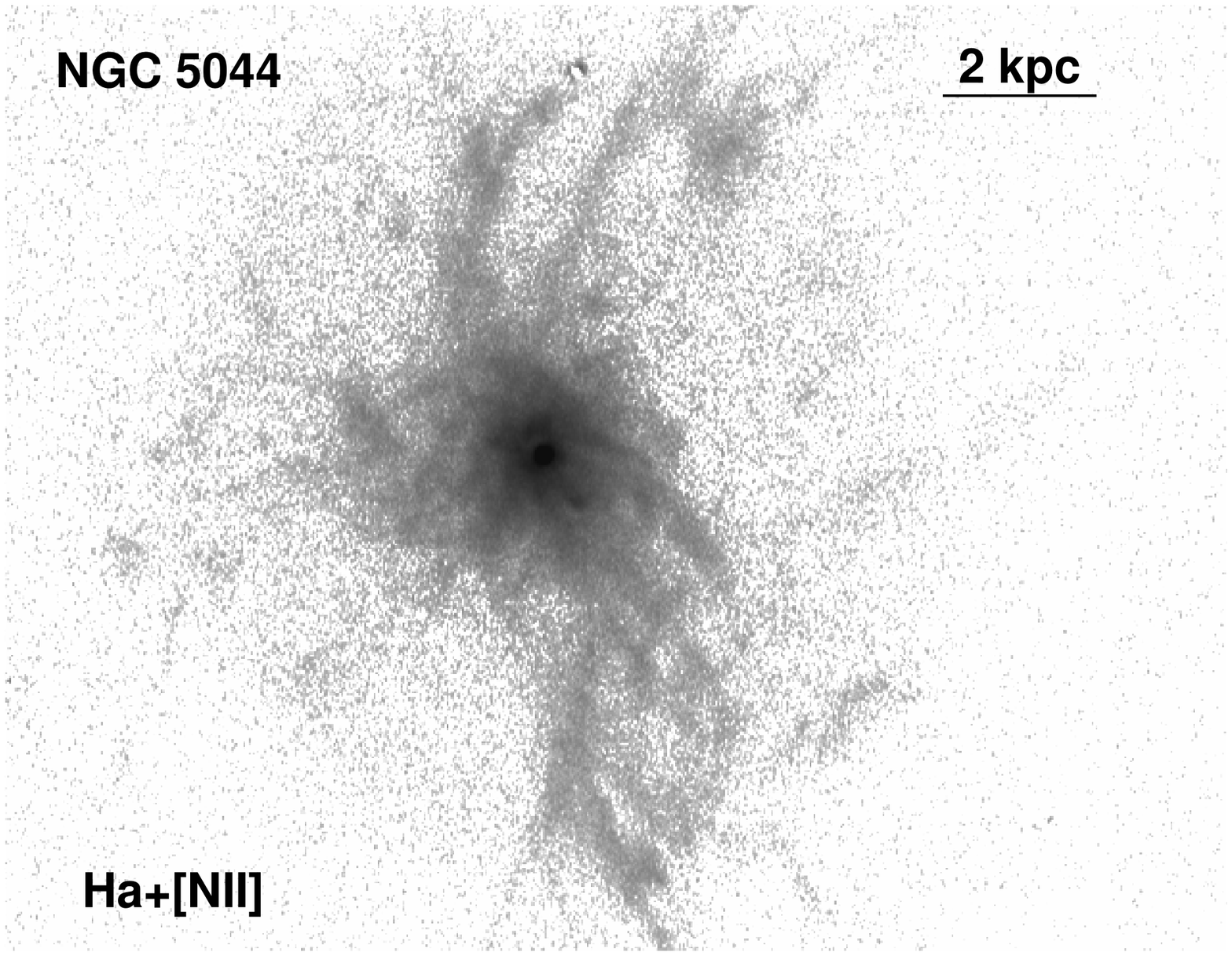}
\end{minipage}
\begin{minipage}{0.32\textwidth}
\includegraphics[width=1\textwidth,clip=t,angle=0.,bb=36 187 577 605]{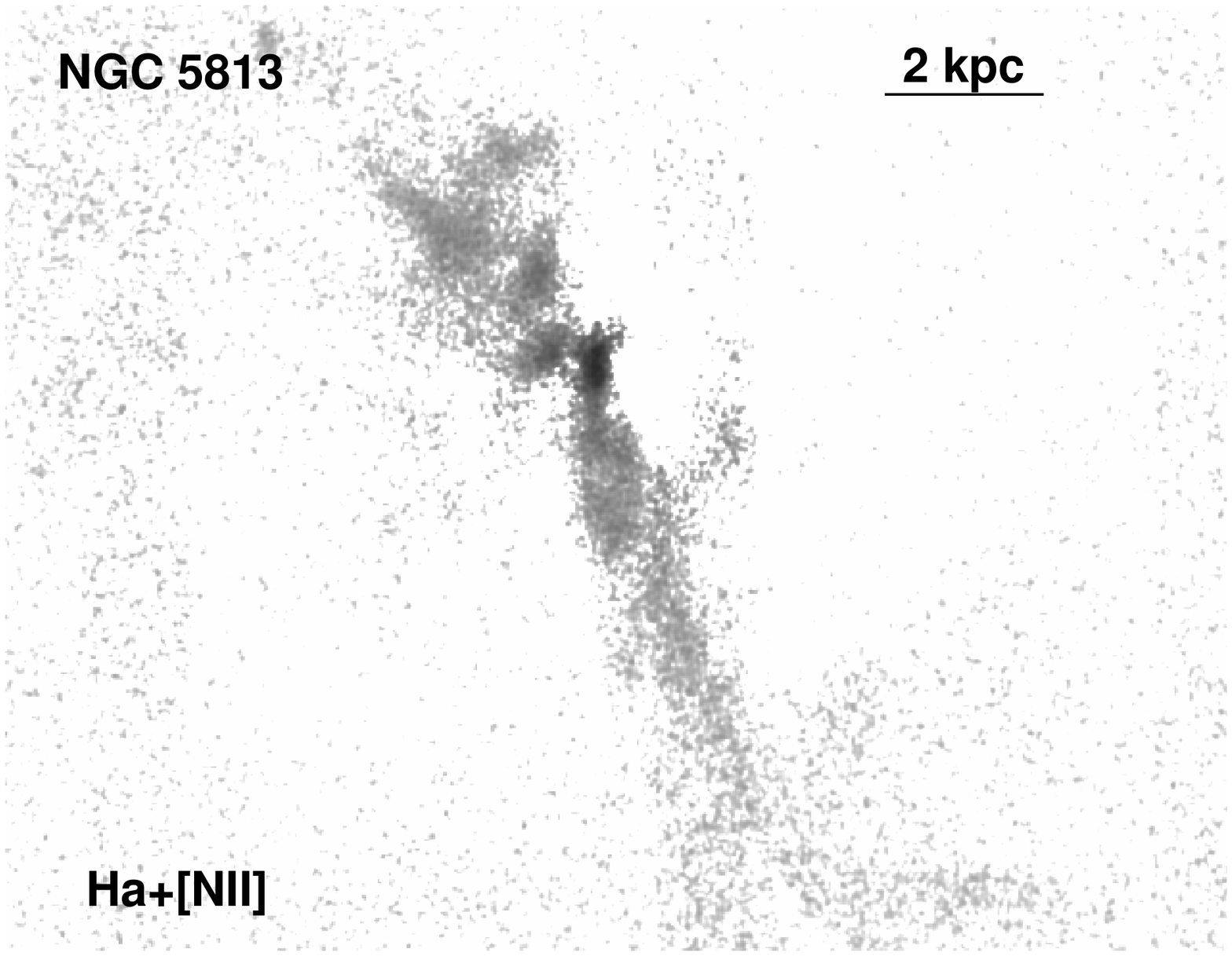}
\end{minipage}
\begin{minipage}{0.32\textwidth}
\includegraphics[width=1\textwidth,clip=t,angle=0.,bb=36 187 577 605]{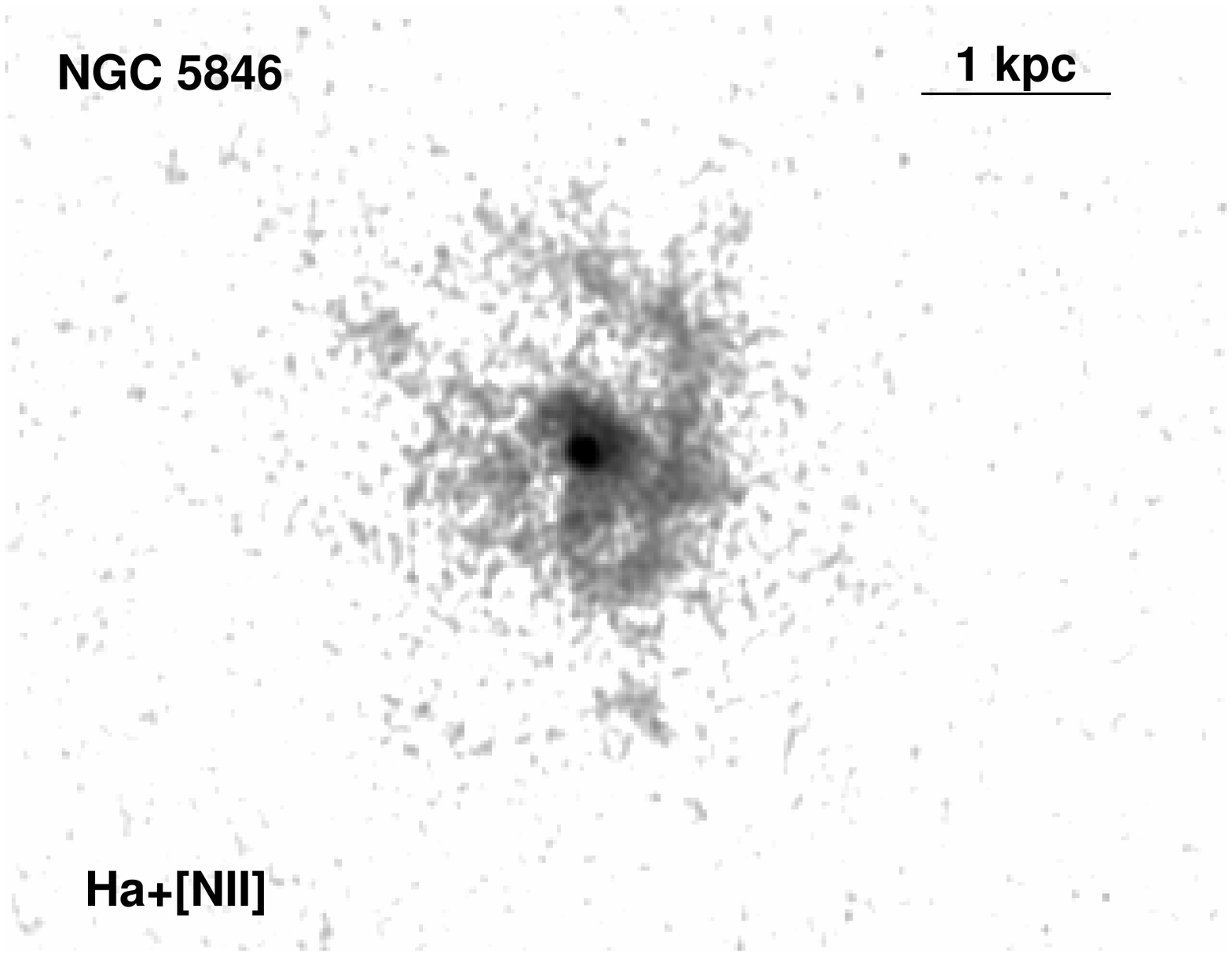}
\end{minipage}
\begin{minipage}{0.32\textwidth}
\includegraphics[width=1\textwidth,clip=t,angle=0.,bb=36 187 577 605]{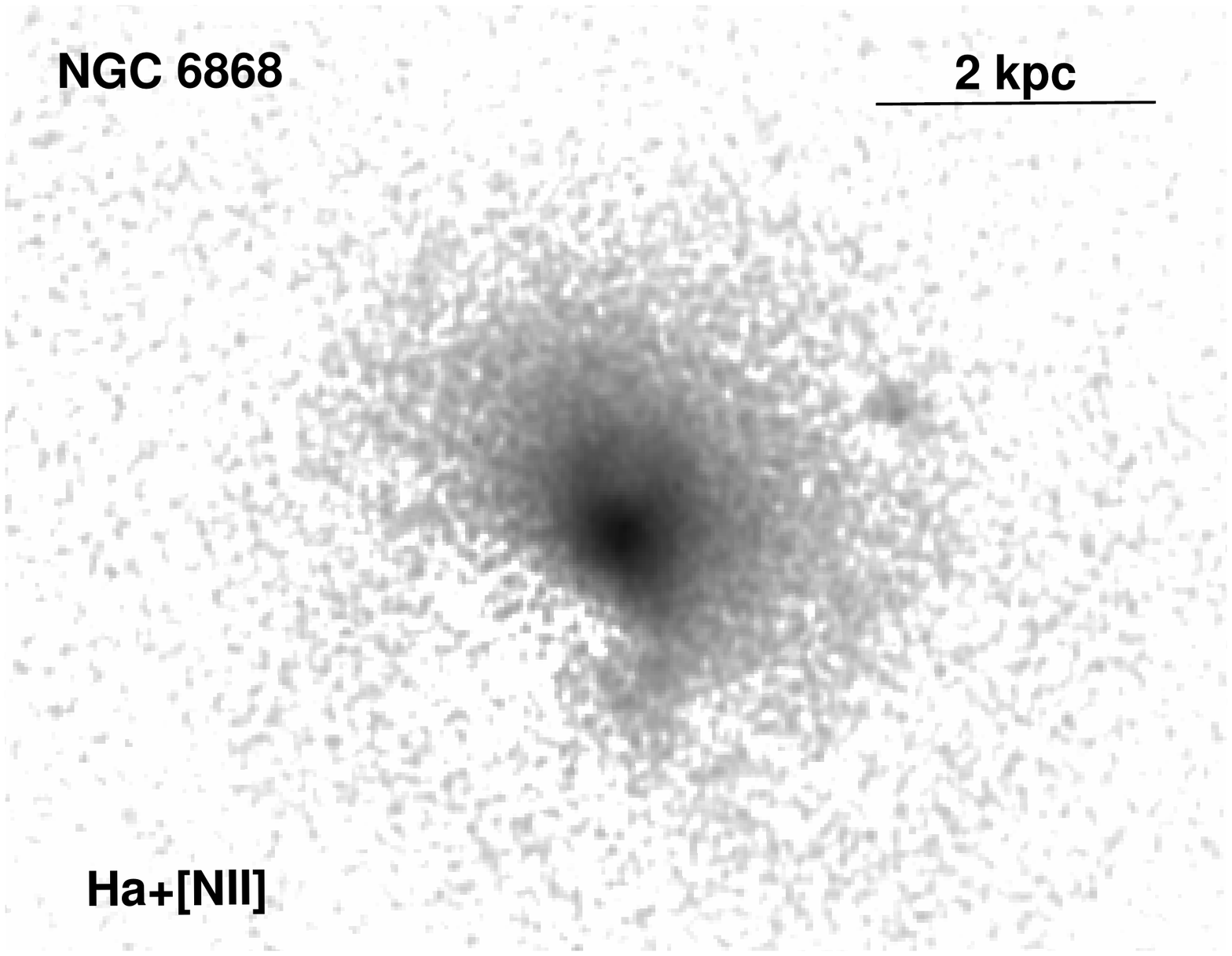}
\end{minipage}
\begin{minipage}{0.32\textwidth}
\includegraphics[width=1\textwidth,clip=t,angle=0.,bb=36 187 577 605]{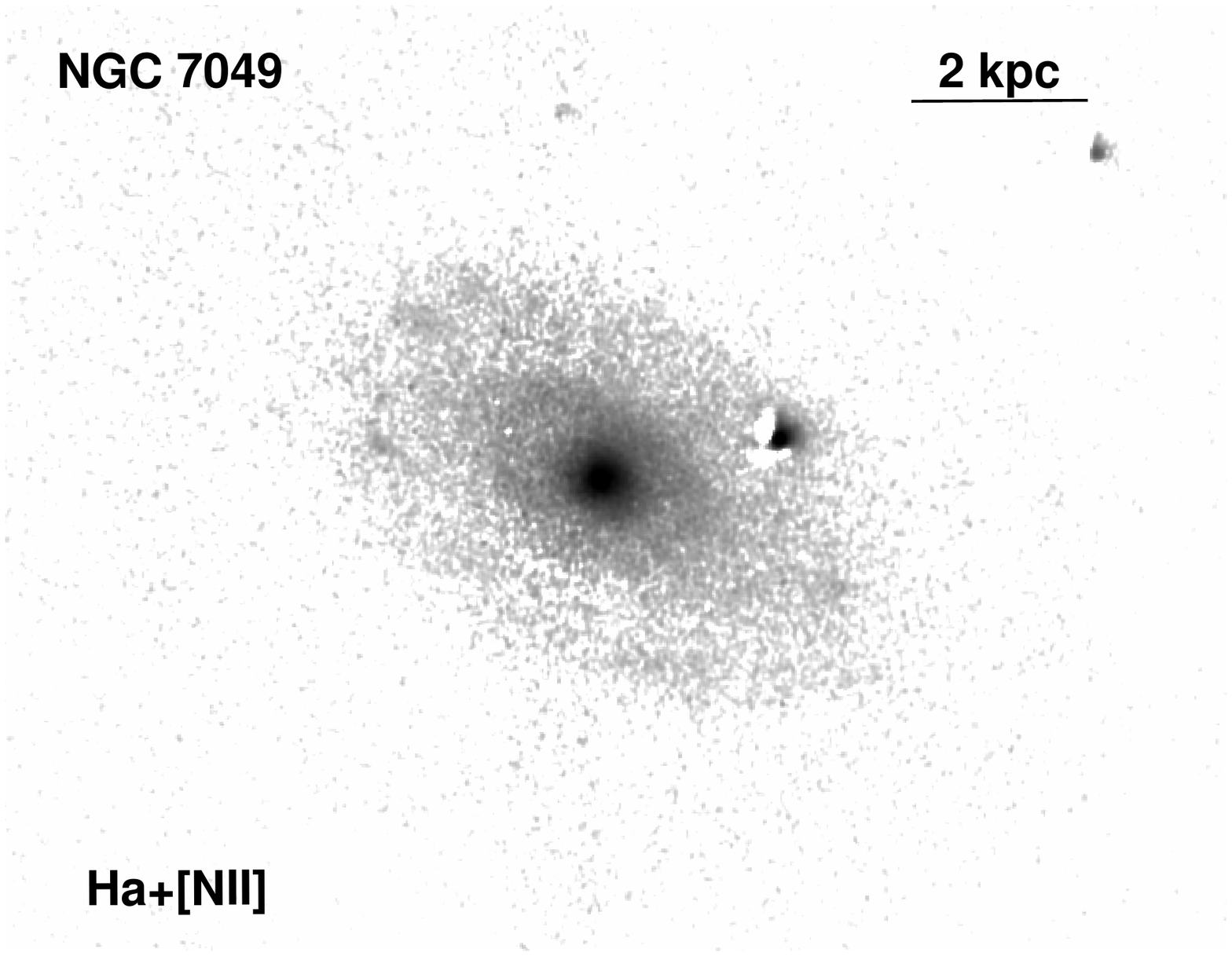}
\end{minipage}
\caption{H$\alpha$+[\ion{N}{ii}] images of the galaxies with extended emission line nebulae in our sample obtained with the 4.1 m SOAR telescope. } 
\label{fig:Ha}
\end{figure*}

The archival {\it Chandra} data were reprocessed using the CIAO (version 4.3) software package. We cleaned the data to remove periods of anomalously high background. 
Table~\ref{obs} summarizes the net exposure times after cleaning. Our data reduction procedures are described in detail in \citet{million2010a,million2010b}. 

To produce 2D maps of projected thermodynamic quantities, we identified spatial regions using the contour binning algorithm \citep{sanders2006b}, which groups 
neighbouring pixels of similar surface brightness until a desired signal-to-noise ratio threshold is met. We adopted a signal-to-noise ratio of 18--25 ($\sim$320--630 counts per 
region) in the 0.6--2.0 keV band, which gives us small enough regions to resolve substructure, yet provides enough counts to achieve better than 5 per cent precision on the 
temperature. Point sources were identified using the CIAO task WAVDETECT and excluded from all regions used for spectral analysis. Separate photon-weighted response 
matrices and effective area files were constructed for each spectral region. 

We modelled the spectra extracted from each spatial region with the SPEX package \citep{kaastra1996}. The spectrum for each region was fitted with a model consisting of an 
absorbed single-phase plasma in collisional ionization equilibrium, with the temperature and spectral normalization (emission measure) as free parameters. The line-of-sight 
absorption column densities, $N_{\rm H}$, were fixed to the values determined by the Leiden/Argentine/Bonn radio survey of \ion{H}{i} \citep[][]{kalberla2005}.

From the best fit emission measures, $Y=\int n_{\rm H} n_{\rm e} {\rm d}V$, where $n_{\rm H} = n_{\rm e} /1.2$ is the hydrogen number density and $V$ is the emitting plasma 
volume, we determined the projected electron densities, $n_{\rm e}$, for each region assuming a fixed column depth of $l=20$~kpc. Using these electron densities and 
the best fit plasma temperatures ($kT$), we determined the pressures ($P_{\rm e}=n_{\rm e}kT$) and entropies ($K=kT/n_{\rm e}^{2/3}$). We note that the choice of the 
assumed column depth for these projected maps is arbitrary and it does not affect the magnitude of the observed azimuthal variations in the derived thermodynamic 
properties.

Because the statistical quality of the data for the X-ray faintest galaxies in our sample, NGC~6868 and NGC~7049, does not allow us to produce 2D maps of thermodynamic 
properties, we only present X-ray images for these systems. Background-subtracted images were created in six narrow energy bands, spanning 0.5--2.0 keV. These were flat 
fielded with respect to the median energy for each image and then co-added to create the broad-band X-ray images. 

To determine the deprojected thermodynamic properties of the hot gas, we extracted, for each galaxy (except NGC~7049), a set of azimuthally averaged spectra from 
concentric annuli. We modeled these spectra simultaneously in the 0.6--2.0 keV band, with the {\small XSPEC} \citep[version 12.5][]{arnaud1996} spectral fitting package, 
using the {\small projct} model \citep[see][for details]{werner2012}. 
We have determined the azimuthally-averaged deprojected radial profiles of electron density and temperature, from which we derived the deprojected radial profiles for 
entropy, electron pressure, and cooling time. We define the cooling time as the gas enthalpy divided by the energy radiated per unit volume of the plasma:
\begin{equation}
t_{\mathrm{cool}}=\frac{\frac{5}{2}(n_{\mathrm{e}}+n_{\mathrm{i}})kT}{n_{\mathrm{e}}n_{\mathrm{i}}\Lambda(T)},
\end{equation}
where the ion number density $n_{\mathrm{i}}=0.92n_{\mathrm{e}}$, and $\Lambda(T)$ is the cooling function for Solar metallicity tabulated by \citet{schure2009}. 
We caution that, because the deprojected thermodynamic properties are determined assuming spherical symmetry, for the 5/7 analyzed galaxies with disturbed morphologies, 
the derived values have significant systematic uncertainties.

\section{Results}
\label{results}

Using {\it Herschel} PACS, we detect FIR [\ion{C}{ii}]$\lambda157\mu$m line emission in 7/8 systems (see Fig.~\ref{fig:C2sample} and Table~\ref{FIRlines}). The detections 
of [\ion{O}{i}]$\lambda63.2\mu$m and [\ion{O}{i}b]$\lambda145.5\mu$m lines are relatively weak, and are only seen in 4/8 and 3/8 systems, respectively (see Fig.~
\ref{fig:O1sample}, \ref{fig:O1bsample} and Table~\ref{FIRlines}). No cold neutral gas is detected in NGC~1399 and the [\ion{C}{ii}] detection in NGC~4472 is relatively 
weak. Table~\ref{FIRlines} lists the observed lines for all the target galaxies, their rest frame wavelengths and observation durations. For the central spaxel, we present the 
observed line fluxes and the $2\sigma$ upper limits where appropriate, the observed FWHM, and the observed line shifts with respect to the systemic velocity determined 
based on optical data. In the brackets, we give the PSF corrected fluxes determined assuming that the emission comes from a point-source, correcting the measured spaxel 
flux upward to account for the beam size. The PSF corrected fluxes can therefore be considered a lower limit to the flux integrated over the whole PACS area.  For sources that 
allow us to confirm that the emission is extended we report non-PSF corrected fluxes. The last column gives the line fluxes spatially integrated over the $5\times5$~spaxel 
($47''\times47''$) {\it Herschel} PACS field of view.

The [\ion{O}{i}]/[\ion{C}{ii}] line ratios determined from the central spaxel in systems with significant [\ion{O}{i}] detections are in the range of 0.4--0.8. 
The $2\sigma$ upper limits on the fluxes of the [\ion{O}{i}]$\lambda63\mu$m and [\ion{O}{i}b]$\lambda145\mu$m lines given in Table~\ref{FIRlines} were determined 
assuming that their velocity widths are equal to the FWHM of the [\ion{C}{ii}] line.

\begin{figure}
\includegraphics[width=1\columnwidth,clip=t,angle=0.,bb=18 144 592 718]{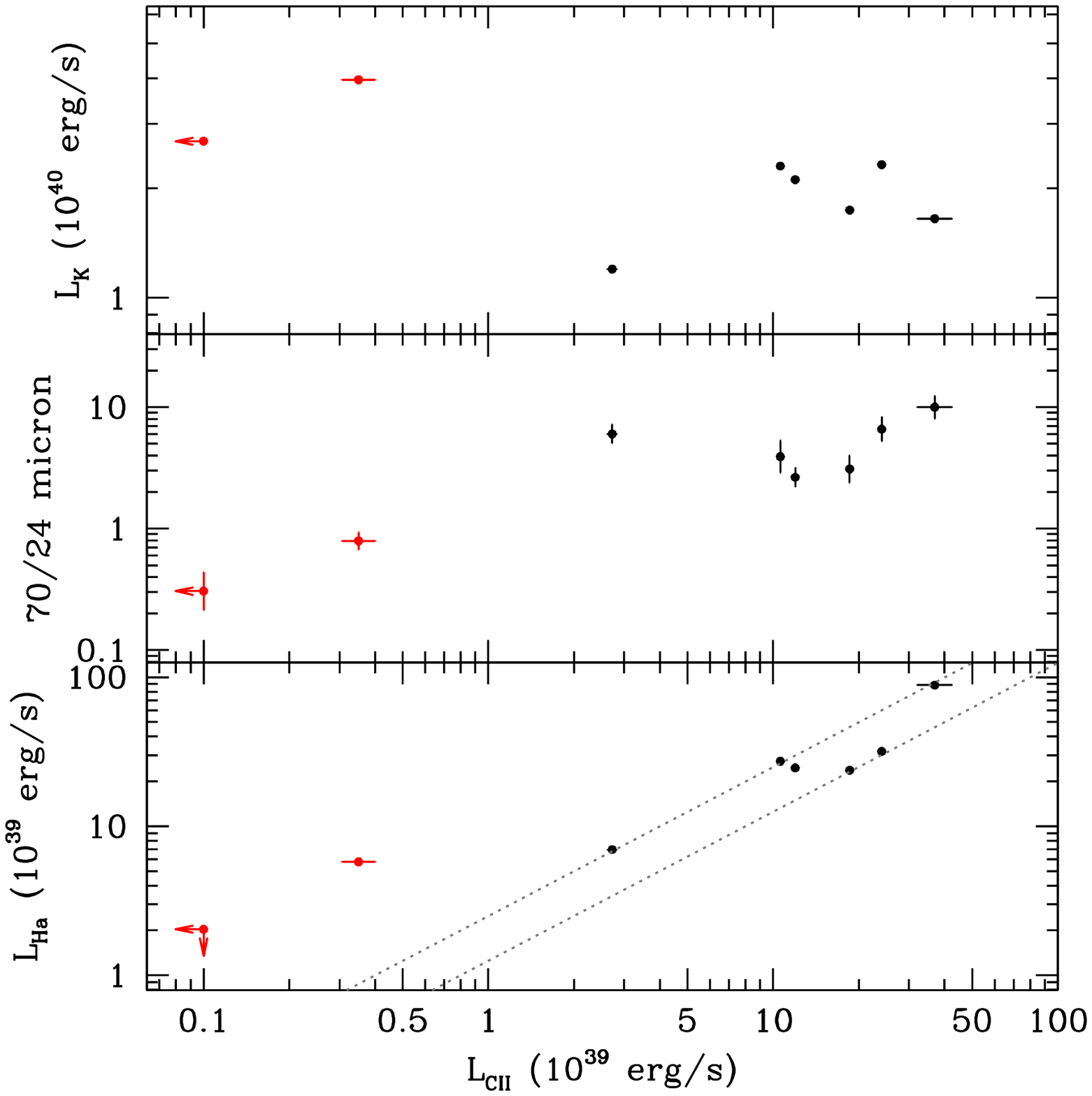}
\caption{The H$\alpha$+[\ion{N}{ii}] luminosity (bottom panel) from SOAR for NGC~1399, NGC~5044, NGC~5813, and NGC~4636 (see Table~\ref{SOARfluxes}) and from 
\citet{macchetto1996} for NGC~4472, NGC~5846, NGC~6868, and NGC~7049 (see Table~\ref{galaxies}), 70/24 $\mu$m infrared continuum ratio \citep[middle panel,][]
{temi2007}, and K-band luminosity \citep[top panel, based on the 2MASS survey,][]{jarrett2003} plotted against the spatially integrated  [\ion{C}{ii}] luminosity measured by {\it 
Herschel} PACS. The grey doted lines on the bottom panel indicate constant [\ion{C}{ii}] over H$\alpha$+[\ion{N}{ii}] luminosity ratios of 0.4 and 0.8. These ratios are 
remarkably similar for 6/8 galaxies in our sample. The outliers, NGC~4472 and NGC~1399, have little or no cold gas. They are indicated in red. } 
\label{Ha_CII}
\end{figure}

\begin{figure}
\includegraphics[width=1\columnwidth,clip=t,angle=0.,bb=18 144 592 718]{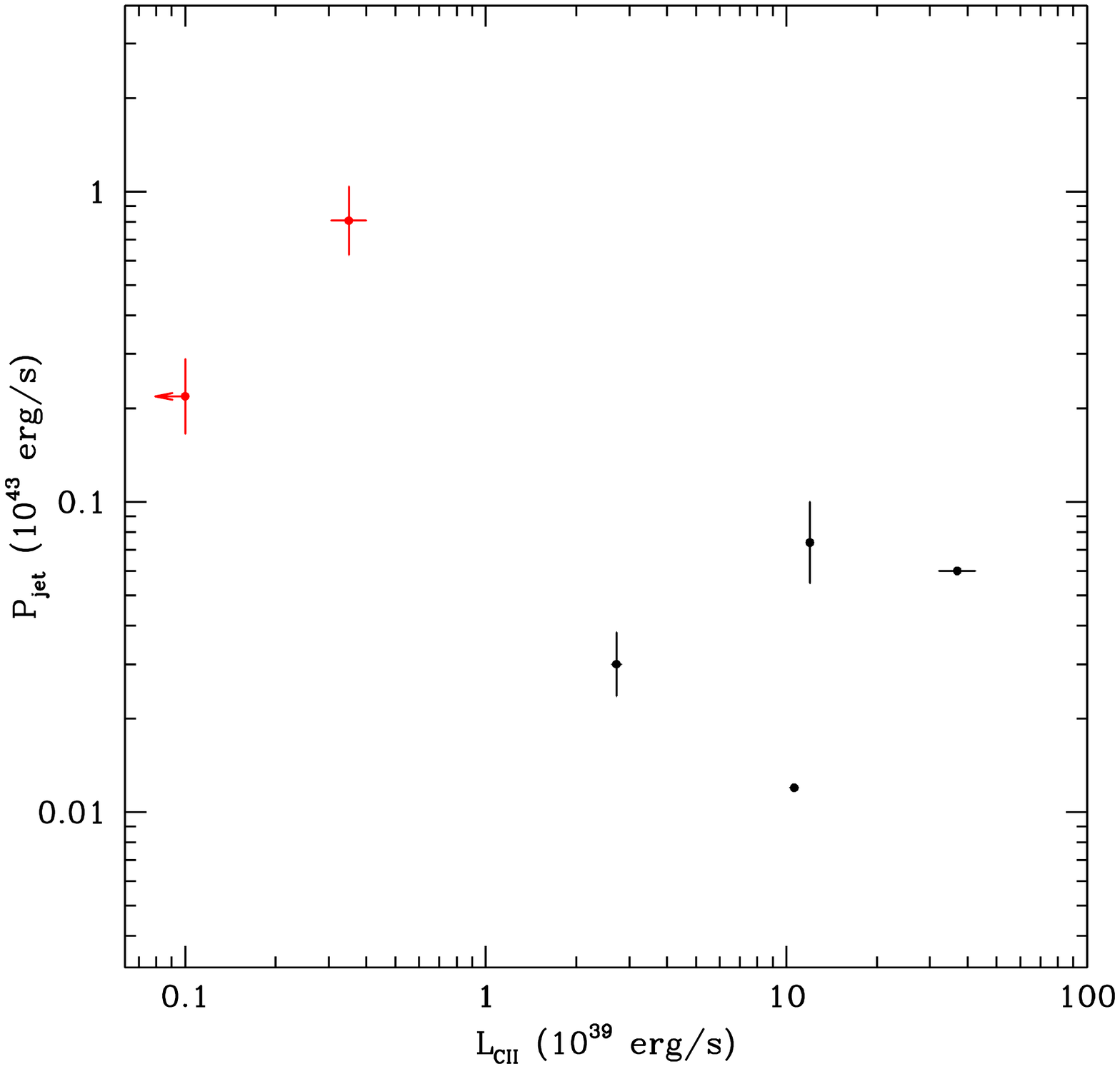}
\caption{Jet powers \citep[][]{allen2006,shurkin2008,david2009,randall2011} plotted against the spatially integrated  [\ion{C}{ii}] luminosity measured by {\it Herschel} PACS. 
Galaxies with little or no cold gas are indicated in red. For the jet-powers in NGC~5044 and NGC~5813 no errorbars were published \citep{randall2011,david2009}.} 
\label{Pj_CII}
\end{figure}

As indicated in Section~\ref{sample} and in Table~\ref{galaxies}, according to \citet{macchetto1996} all our targets are H$\alpha$ luminous. Fig.~\ref{fig:Ha} shows narrow 
band H$\alpha$+[\ion{N}{ii}] images obtained with the SOAR telescope of our six targets with extended line-emitting nebulae. The nebulae in NGC~6868 and NGC~7049 
show regular extended disk-like morphologies reaching radii $r\sim2-3$~kpc \citep[see also][]{macchetto1996}. On the other hand, NGC~5044, NGC~5813, NGC~5846, and 
NGC~4636 show filamentary emission. NGC~5044 has the largest optical emission line luminosity, with a rich, dense network of filaments extending from the core out to a 
radius of $\sim10$~kpc \citep[see also][]{david2011,gastaldello2009}. The nebulae in NGC~5813 show a radial bipolar distribution in the direction of buoyant bubbles rising in 
the hot X-ray emitting atmosphere of the galaxy \citep[see also][]{randall2011}. The bright H$\alpha$+[\ion{N}{ii}] emission in NGC~4636 and NGC~5846 appears relatively 
compact, concentrated in the innermost $r\lesssim1.5$~kpc of the galaxies, but the morphology of the nebulae is clearly filamentary exhibiting signs of interaction with the 
central AGN. 

Our new SOAR data allow robust constraints on the H$\alpha$+[\ion{N}{ii}] flux in four galaxies, NGC~5044, NGC~4636, NGC~5813, and NGC~1399 (see Table~
\ref{SOARfluxes}). The NGC~5846 data were not taken in photometric conditions, and for NGC~6868 and NGC~7049, the filters available at the time of the observations made 
the continuum subtraction less reliable than what is achieved for other galaxies. Our H$\alpha$+[\ion{N}{ii}] flux for NGC~4636 is consistent with the values published by 
\citet{macchetto1996} and \citet{buson1993}. For NGC~5813, our result is consistent with \citet{goudfrooij1994} but is almost a factor of two higher than the flux measured by 
\citet{macchetto1996}. For NGC~5044, our flux is significantly higher than the values measured by \citet{goudfrooij1994} and \citet{macchetto1996}, but is consistent with the 
result of \citet{rickes2004}. The large spatial extent of the nebula surrounding NGC~5044 combined with the small field-of-view of the instrument (1.4$'\times2.2'$) may have 
prevented \citet{macchetto1996} from a robust determination of the local background.

\begin{figure*}
\begin{minipage}{0.32\textwidth}
\vspace{-1.3cm}
\includegraphics[width=0.95\textwidth,clip=t,angle=0.]{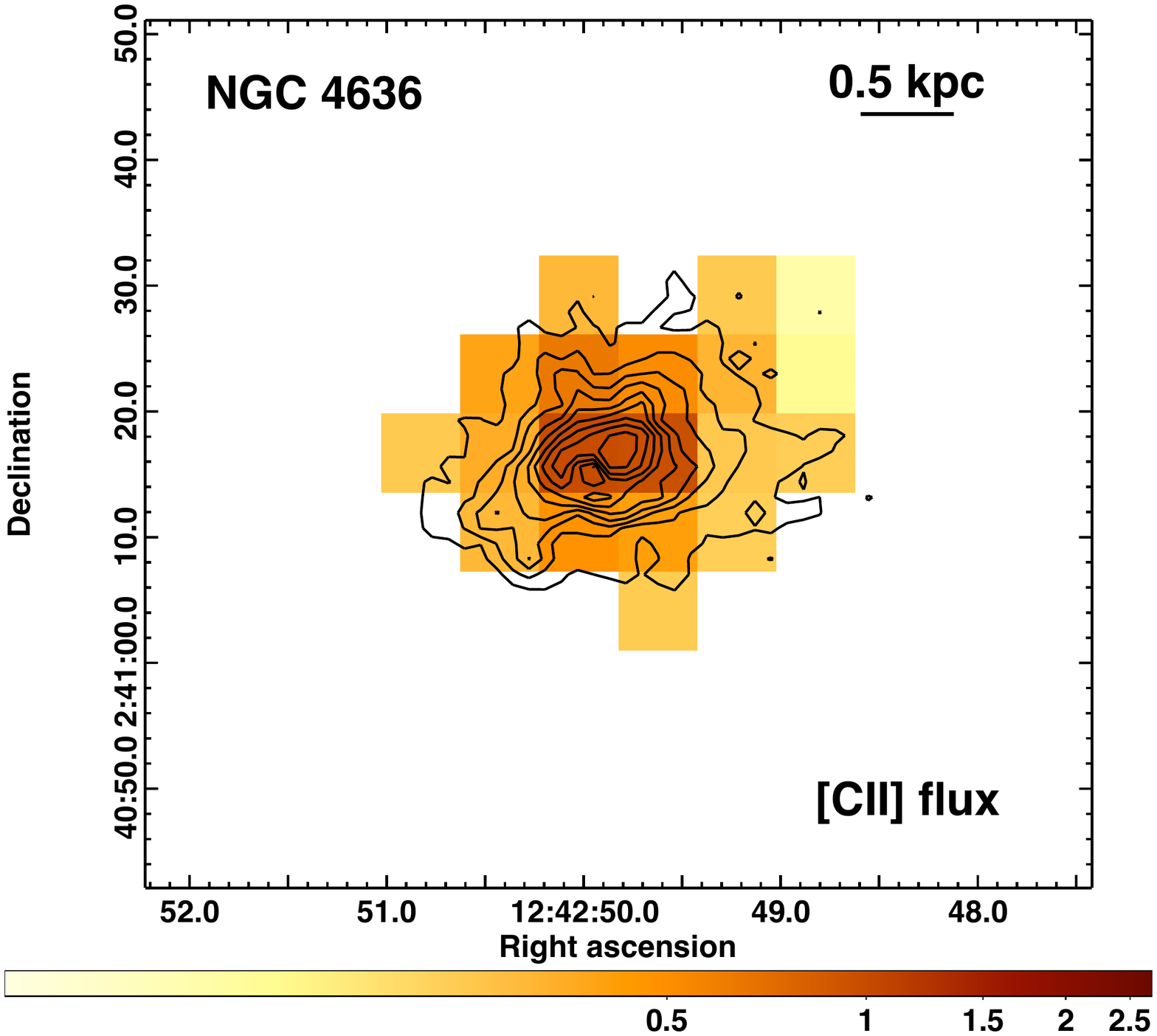}
\end{minipage}
\begin{minipage}{0.32\textwidth}
\vspace{-1.3cm}
 \includegraphics[width=0.95\textwidth,clip=t,angle=0.]{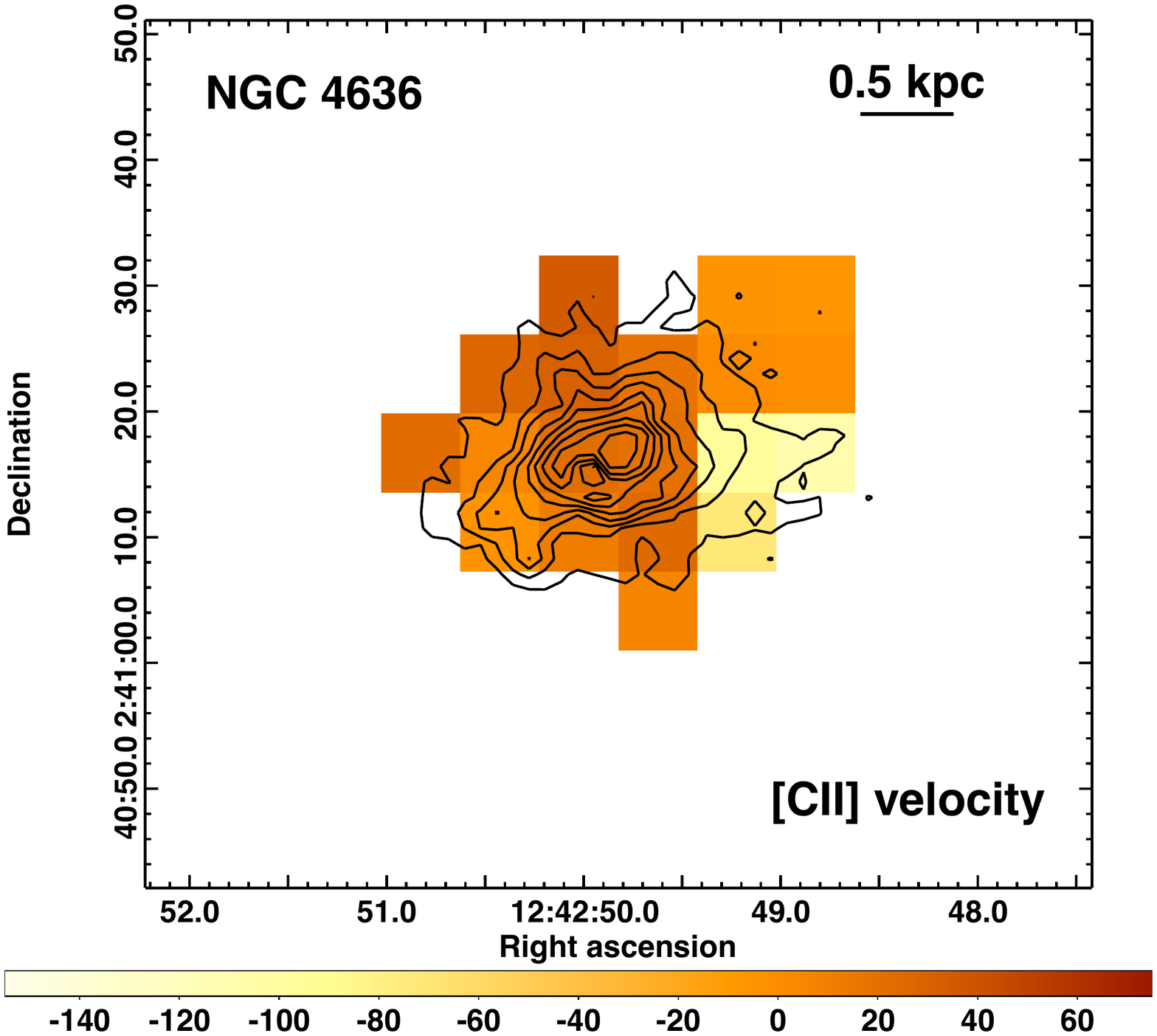}
\end{minipage}
\begin{minipage}{0.32\textwidth}
\vspace{-1.3cm}
\includegraphics[width=0.95\textwidth,clip=t,angle=0.]{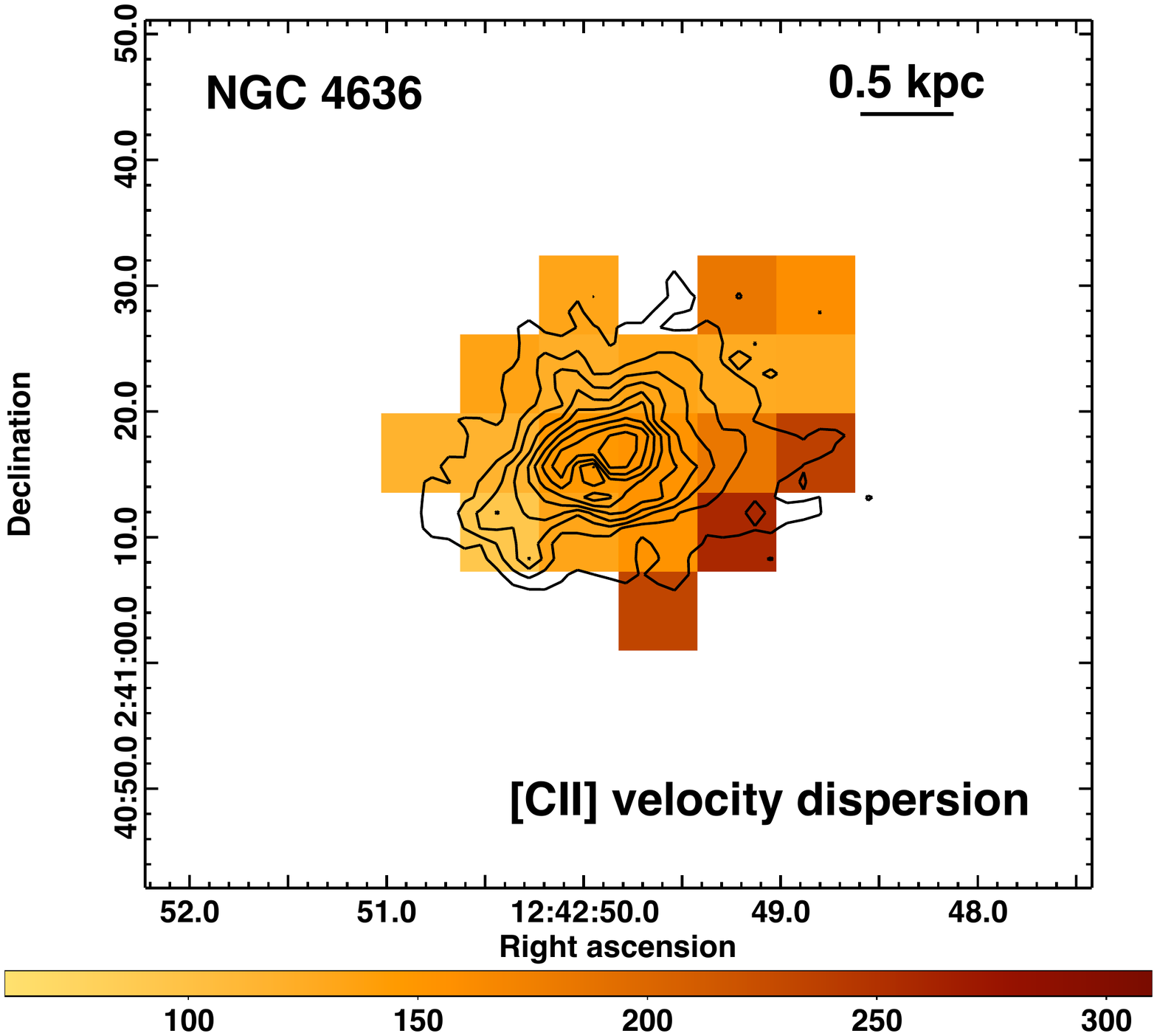}
\end{minipage}
\begin{minipage}{0.32\textwidth}
\vspace{-2.6cm}
\includegraphics[width=1.05\textwidth,clip=t,angle=0.]{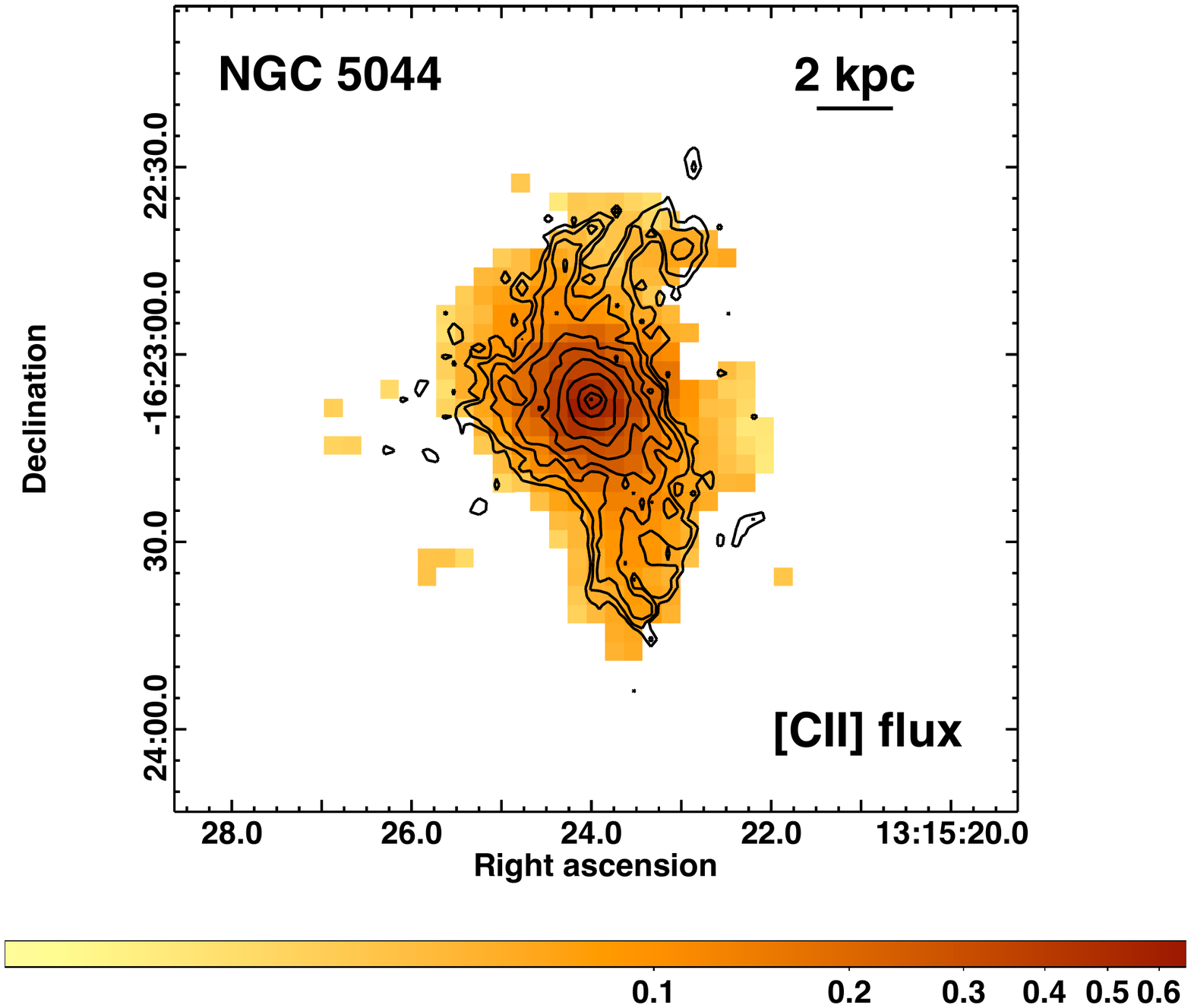}
\end{minipage}
\begin{minipage}{0.32\textwidth}
\vspace{-2.6cm}
 \includegraphics[width=1.05\textwidth,clip=t,angle=0.]{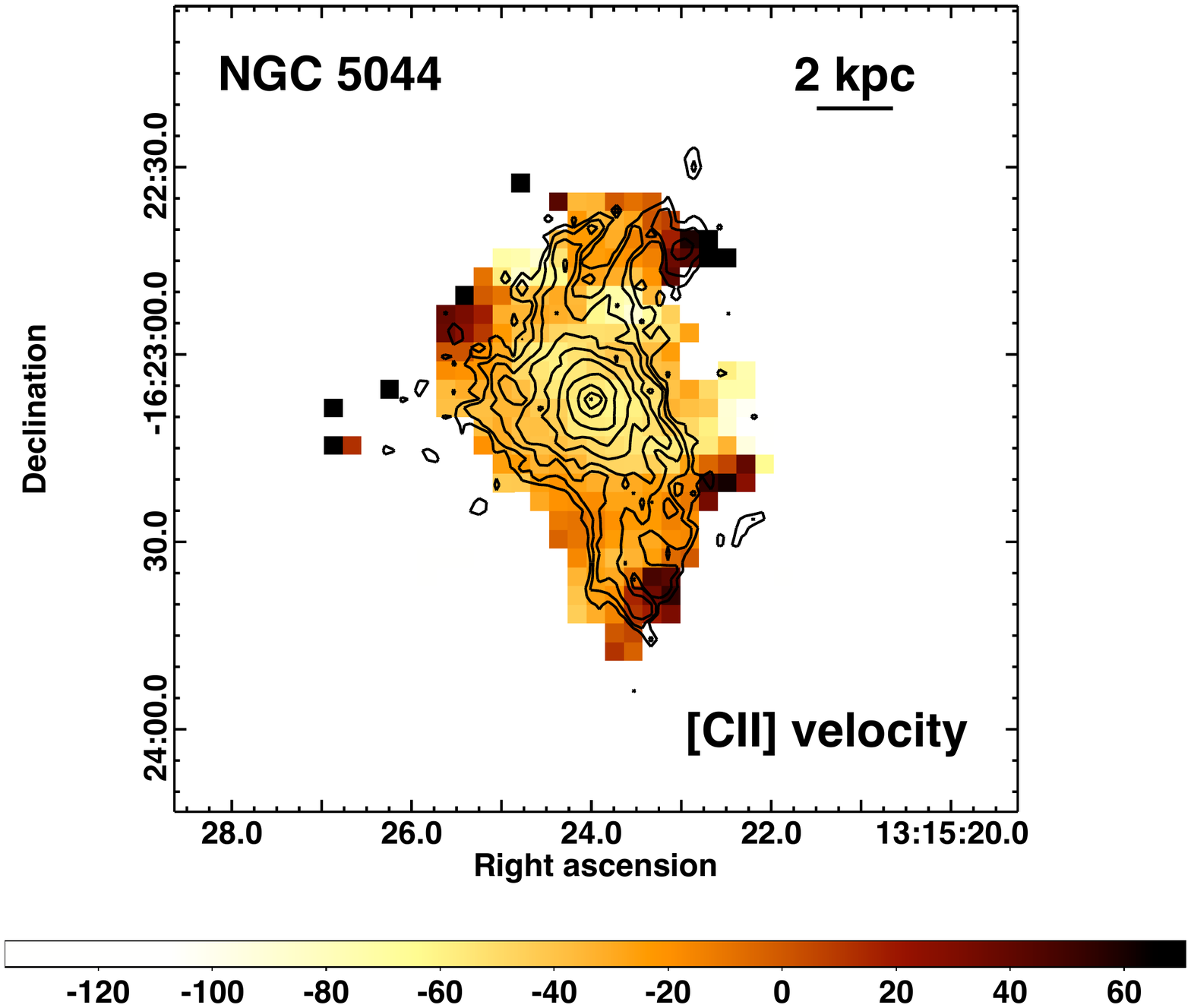}
\end{minipage}
\begin{minipage}{0.32\textwidth}
\vspace{-2.6cm}
\includegraphics[width=1.05\textwidth,clip=t,angle=0.]{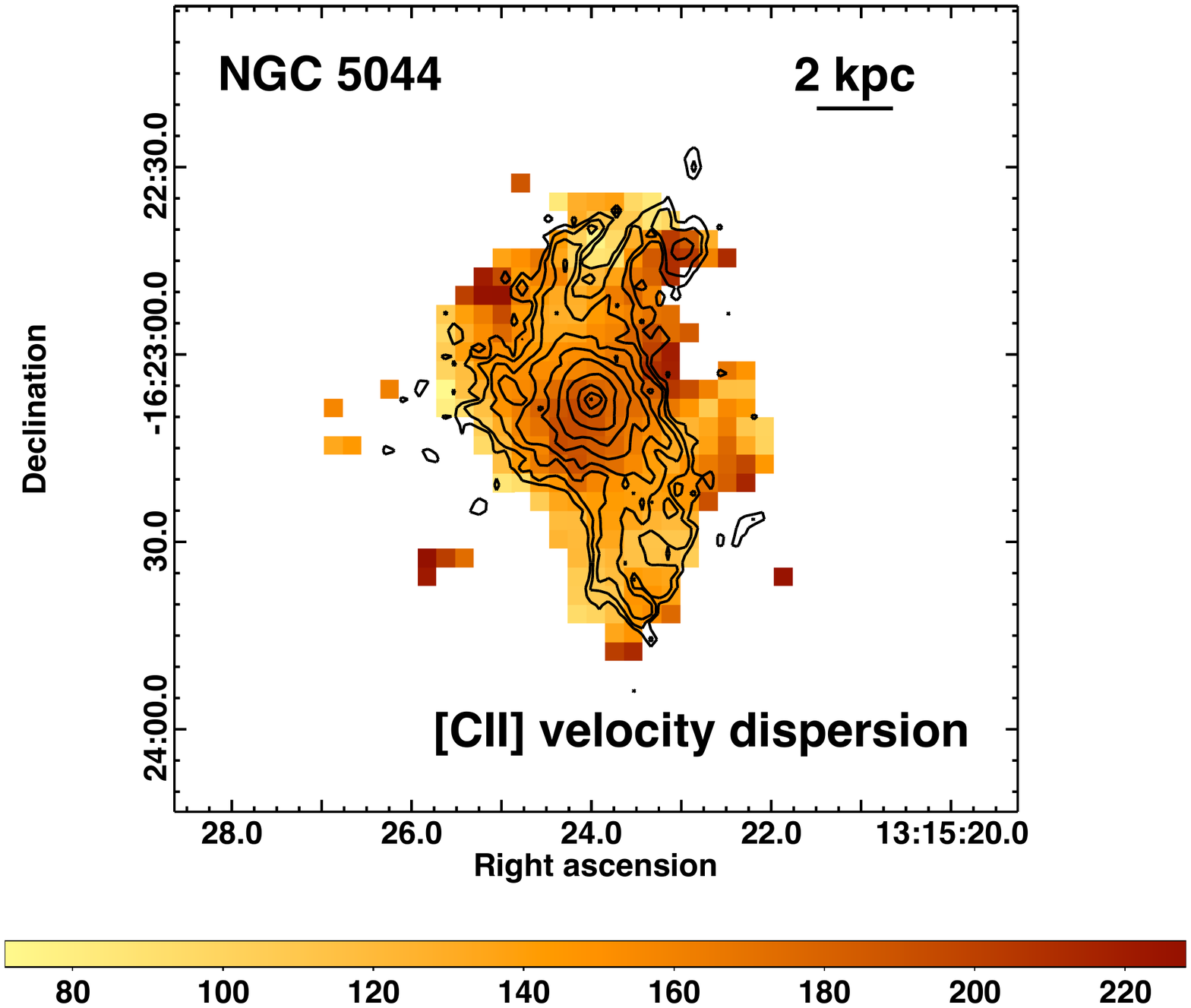}
\end{minipage}
\begin{minipage}{0.32\textwidth}
\vspace{-2.6cm}
\includegraphics[width=1.05\textwidth,clip=t,angle=0.]{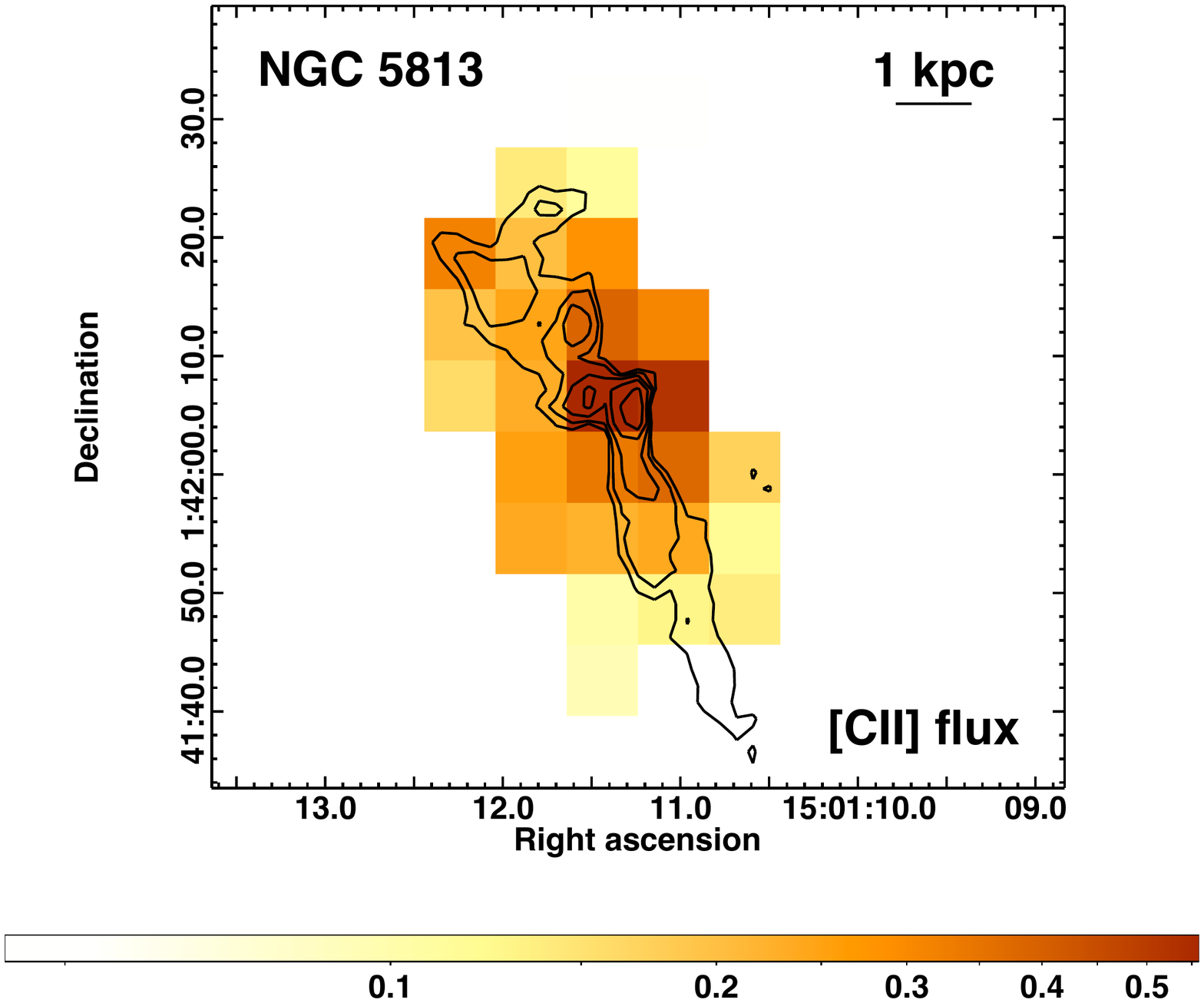}
\end{minipage}
\begin{minipage}{0.32\textwidth}
\vspace{-2.6cm}
\includegraphics[width=1.05\textwidth,clip=t,angle=0.]{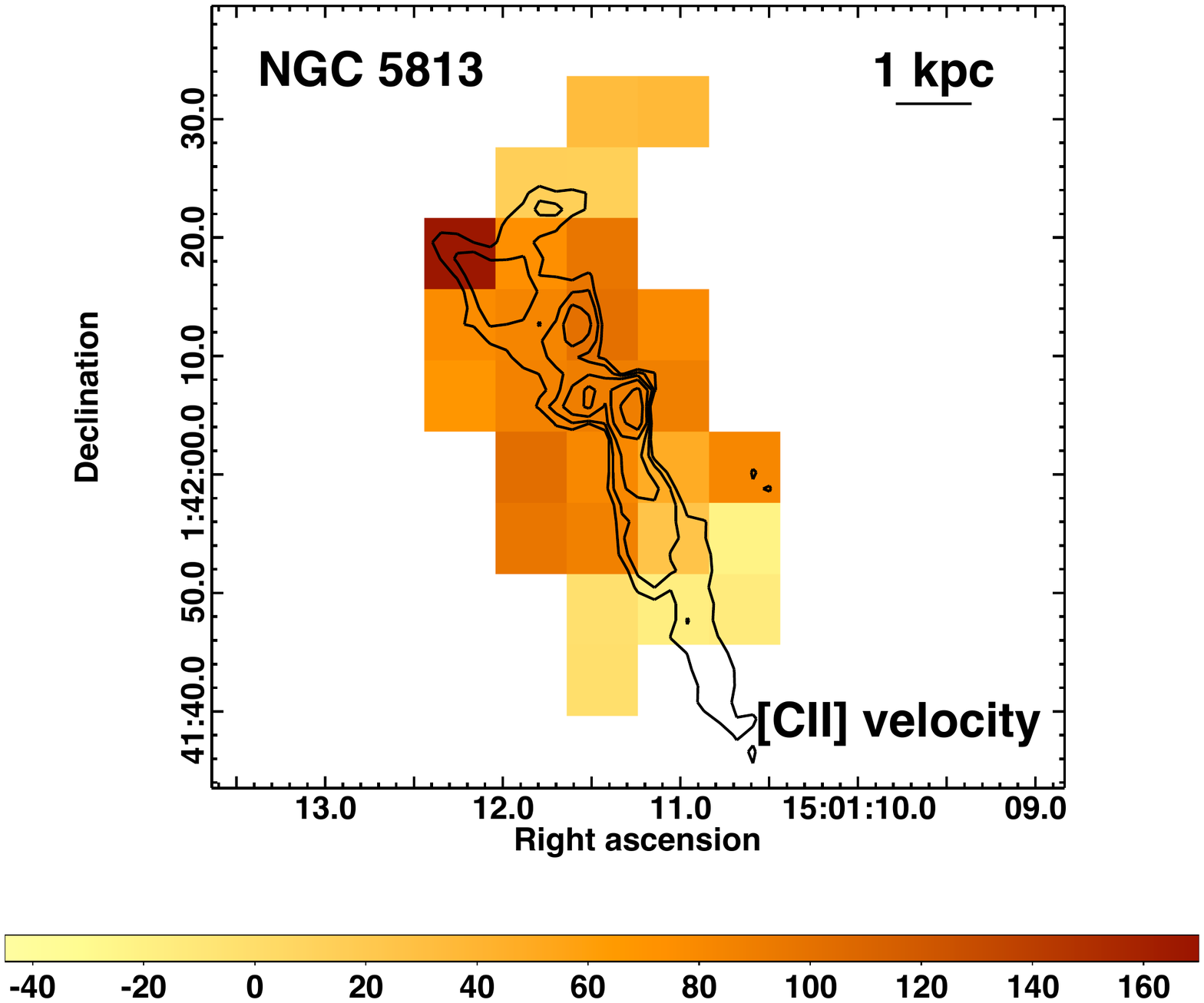}
\end{minipage}
\begin{minipage}{0.32\textwidth}
\vspace{-2.6cm}
\includegraphics[width=1.05\textwidth,clip=t,angle=0.]{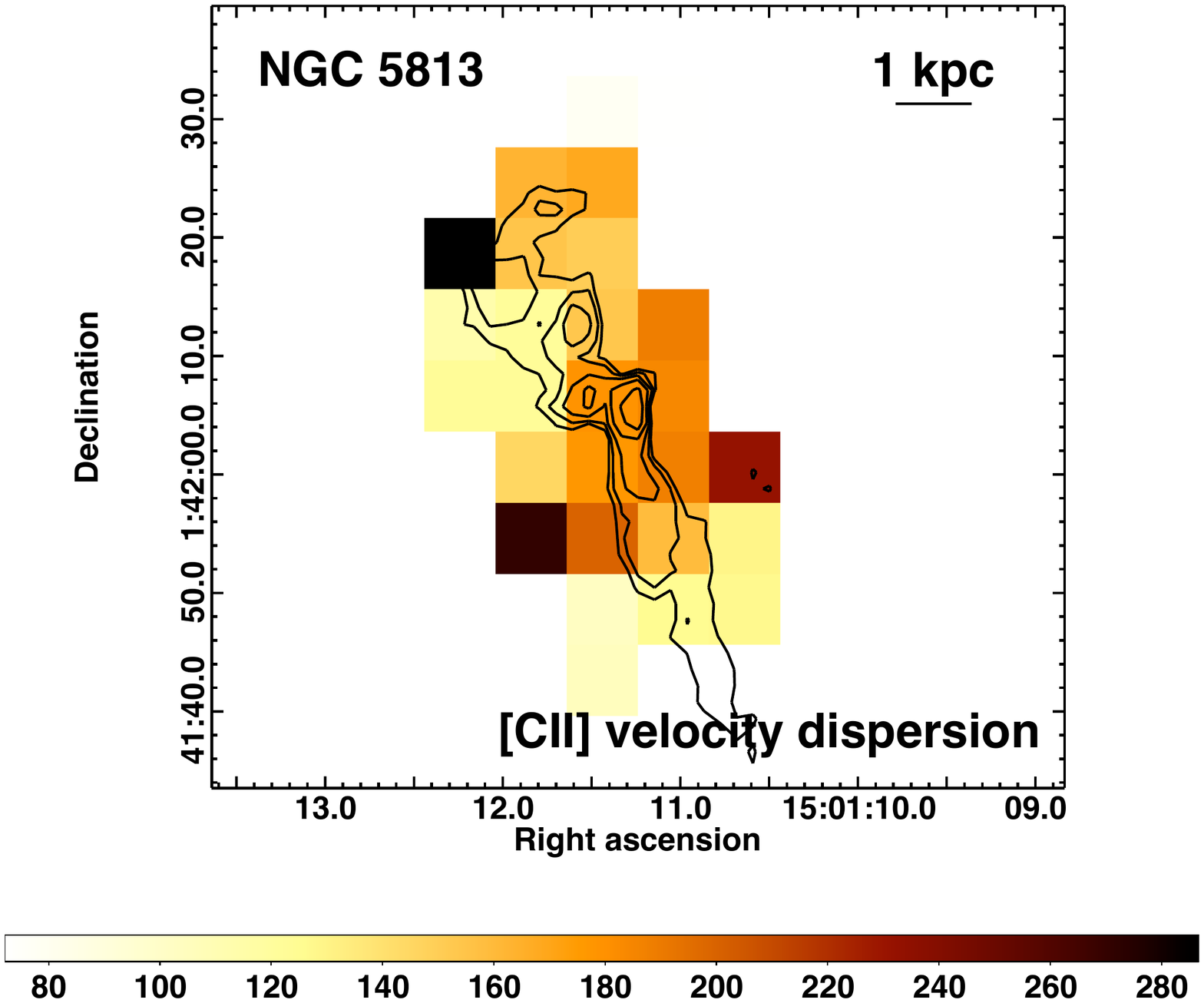}
\end{minipage}
\begin{minipage}{0.32\textwidth}
\vspace{-2.6cm}
\includegraphics[width=1.02\textwidth,clip=t,angle=0.]{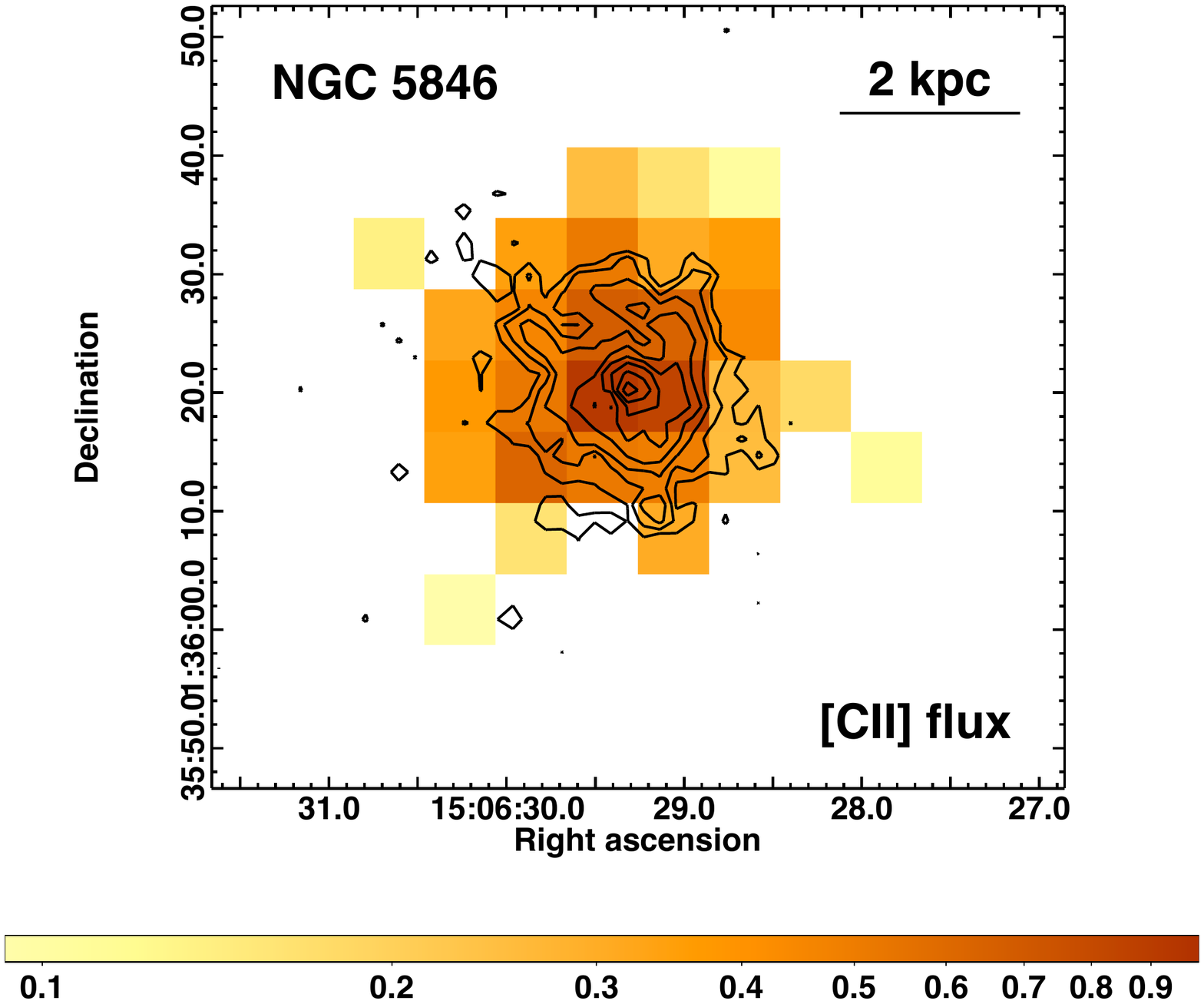}
\end{minipage}
\begin{minipage}{0.32\textwidth}
\vspace{-2.6cm}
 \includegraphics[width=1.02\textwidth,clip=t,angle=0.]{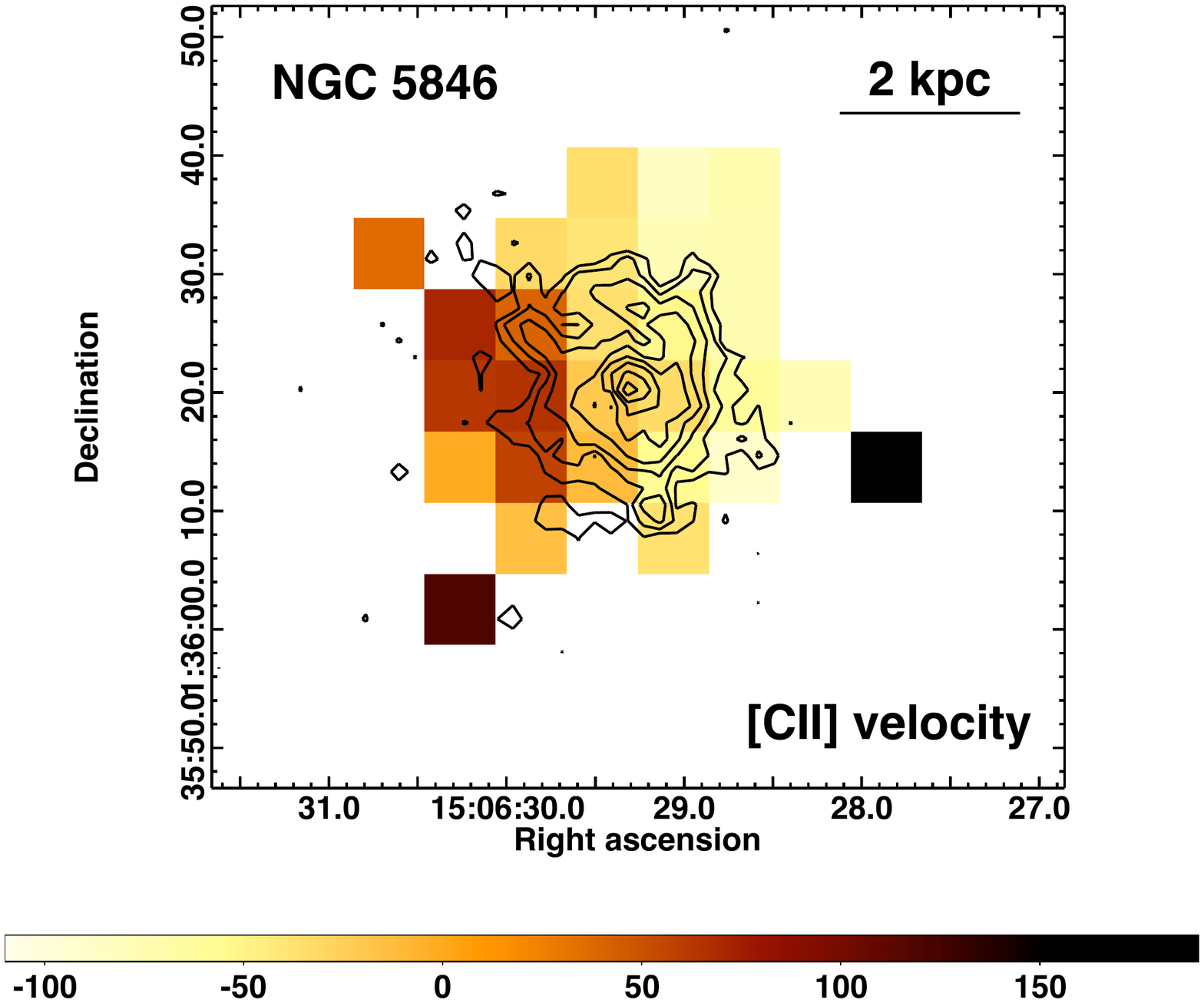}
\end{minipage}
\begin{minipage}{0.32\textwidth}
\vspace{-2.6cm}
\includegraphics[width=1.02\textwidth,clip=t,angle=0.]{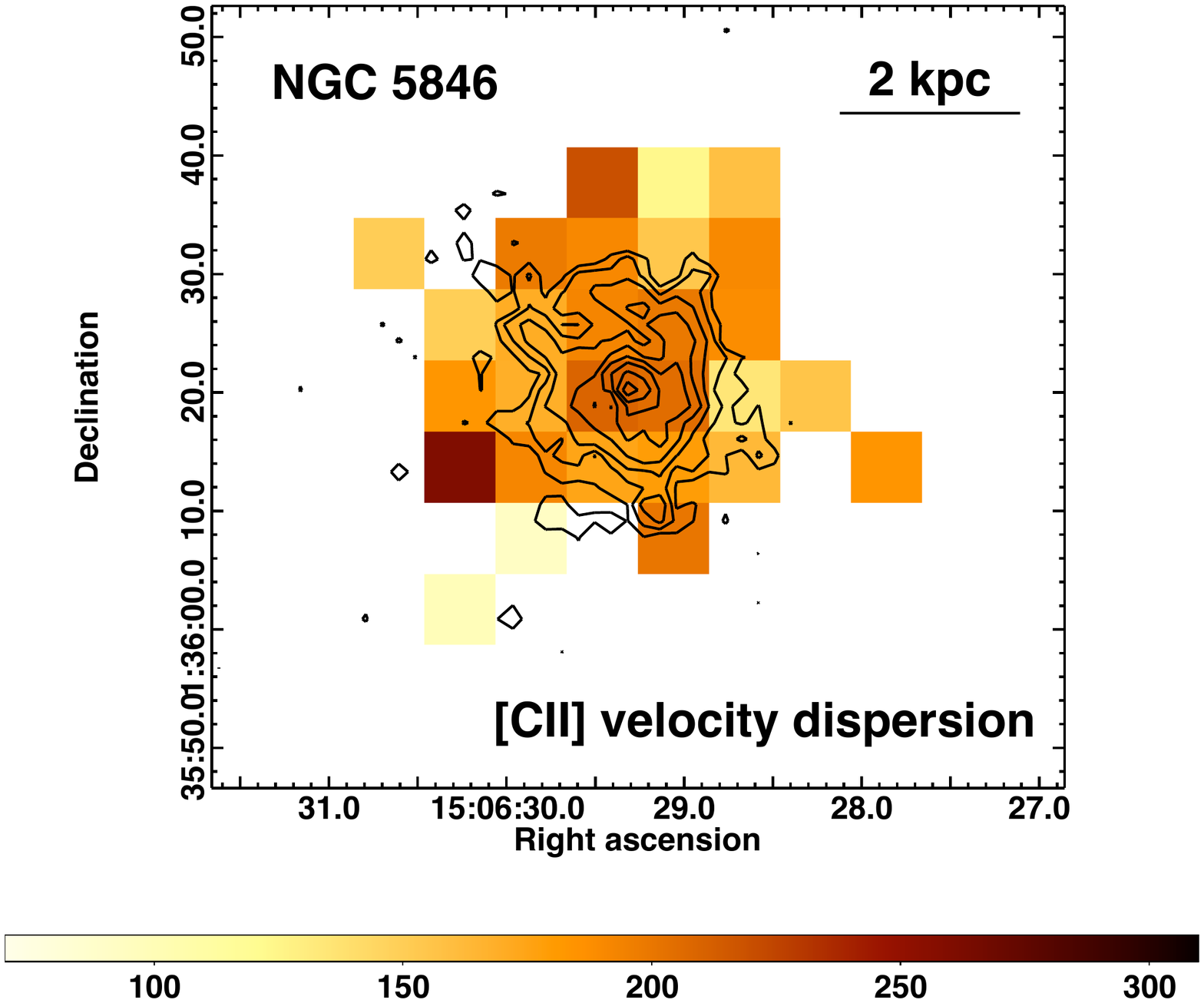}
\end{minipage}
\vspace{-1.2cm}
\caption{Left panels: Maps of the integrated [\ion{C}{ii}] line flux in units of $10^{-14}$~erg~s$^{-1}$~cm$^{-2}$ per $6\arcsec \times 6\arcsec$ spaxel obtained with {\it 
Herschel} PACS for NGC~4636, NGC~5044, NGC~5813, and NGC~5846. Central panels: The velocity distribution of the [\ion{C}{ii}] emitting gas, in units of km~s$^{-1}$, 
relative to the systemic velocity of the host galaxy. Right panels: Map of the velocity dispersion, $\sigma$, of the [\ion{C}{ii}] emitting gas. 
%The images show a $47\times47$~arcsec region of the sky centered on galaxies NGC~4636, NGC~5044, NGC~5813, and NGC~5846. 
Contours of the H$\alpha$+[\ion{N}{ii}] emission are overlaid.} 
\label{fig:C2maps1}
\end{figure*}

\begin{figure*}
\begin{minipage}{0.32\textwidth}
\vspace{-1.3cm}
\includegraphics[width=0.99\textwidth,clip=t,angle=0.]{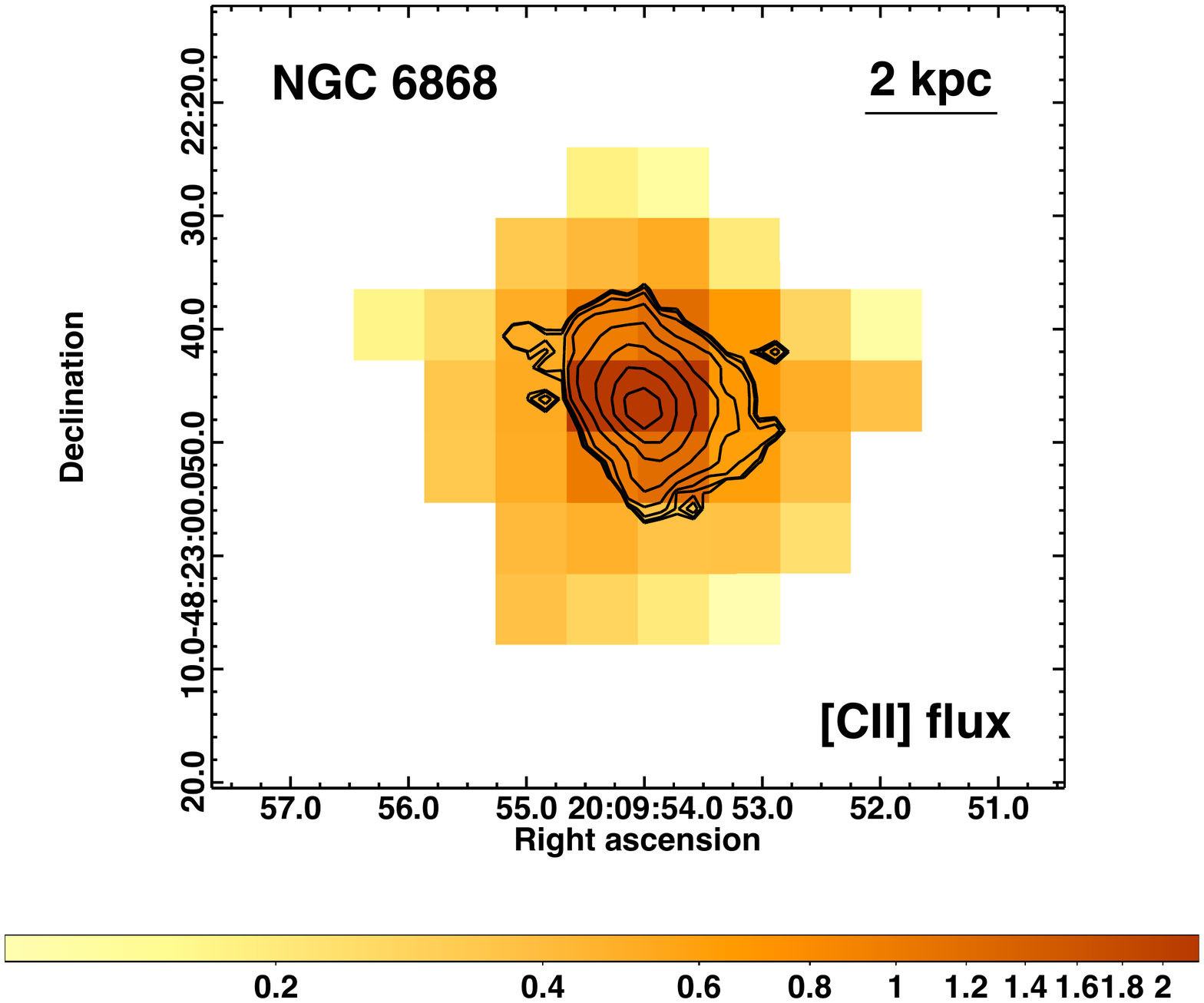}
\end{minipage}
\begin{minipage}{0.32\textwidth}
\vspace{-1.3cm}
 \includegraphics[width=0.99\textwidth,clip=t,angle=0.]{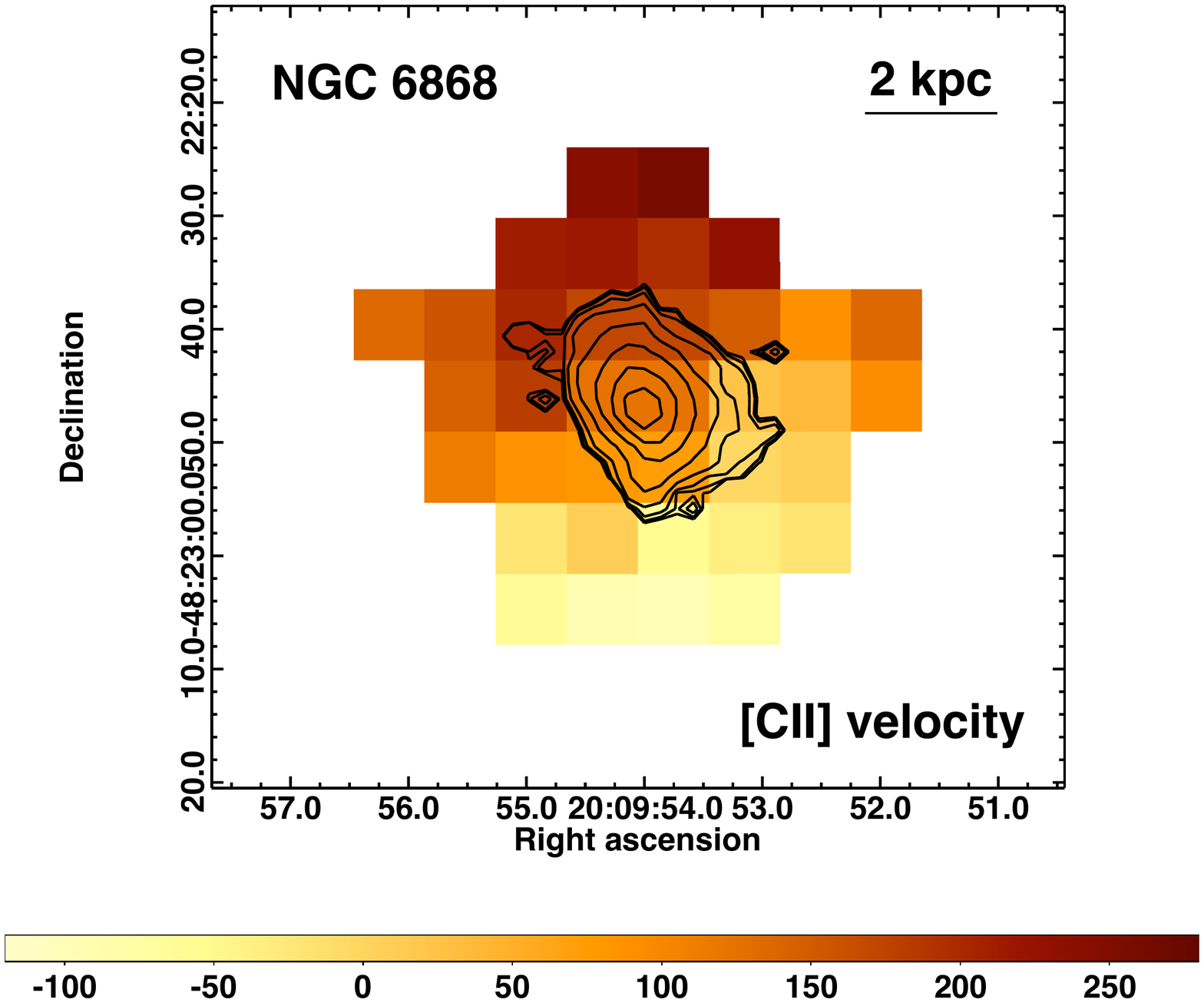}
\end{minipage}
\begin{minipage}{0.32\textwidth}
\vspace{-1.3cm}
\includegraphics[width=0.99\textwidth,clip=t,angle=0.]{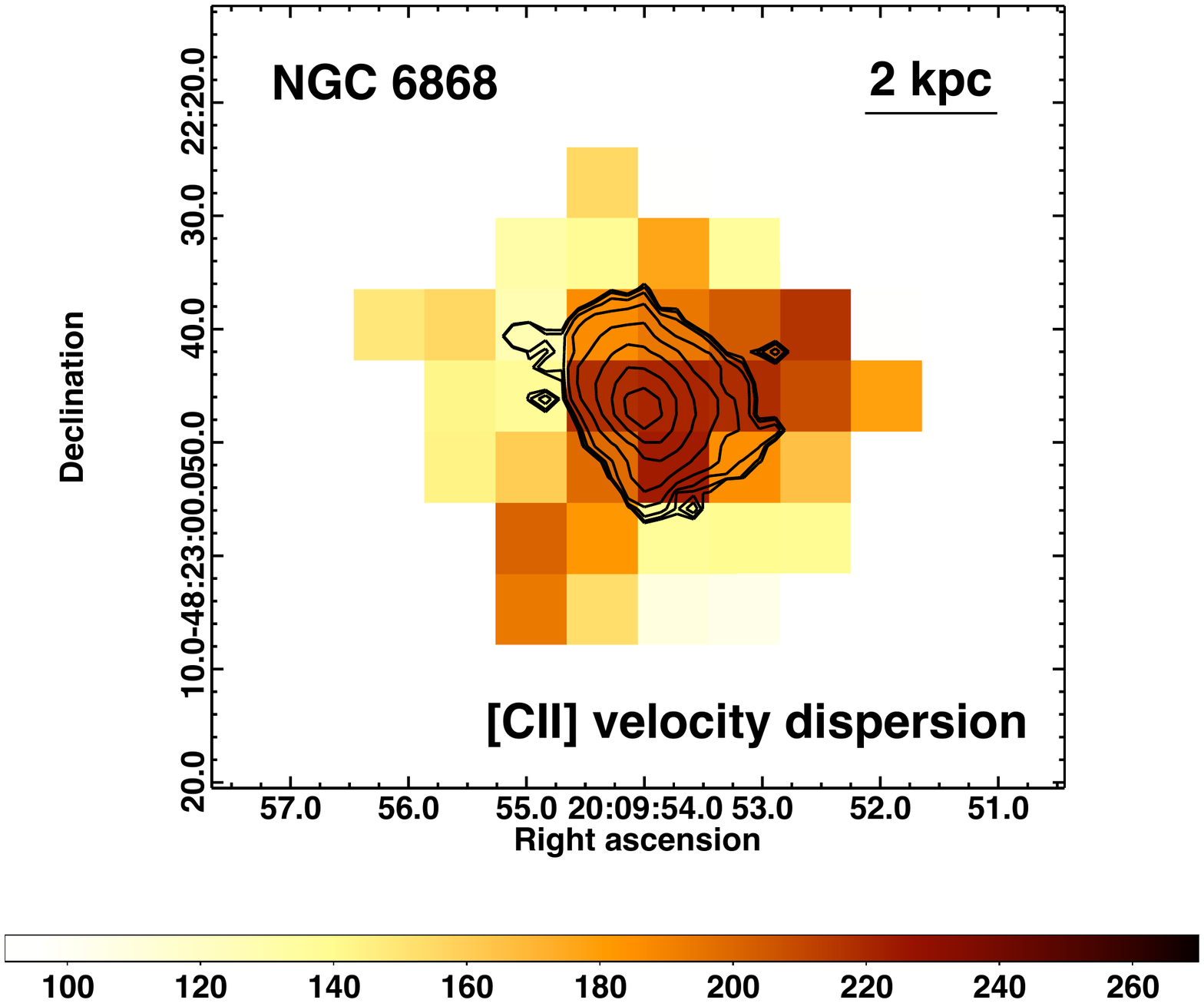}
\end{minipage}
\begin{minipage}{0.32\textwidth}
\vspace{-2.6cm}
\includegraphics[width=1.02\textwidth,clip=t,angle=0.]{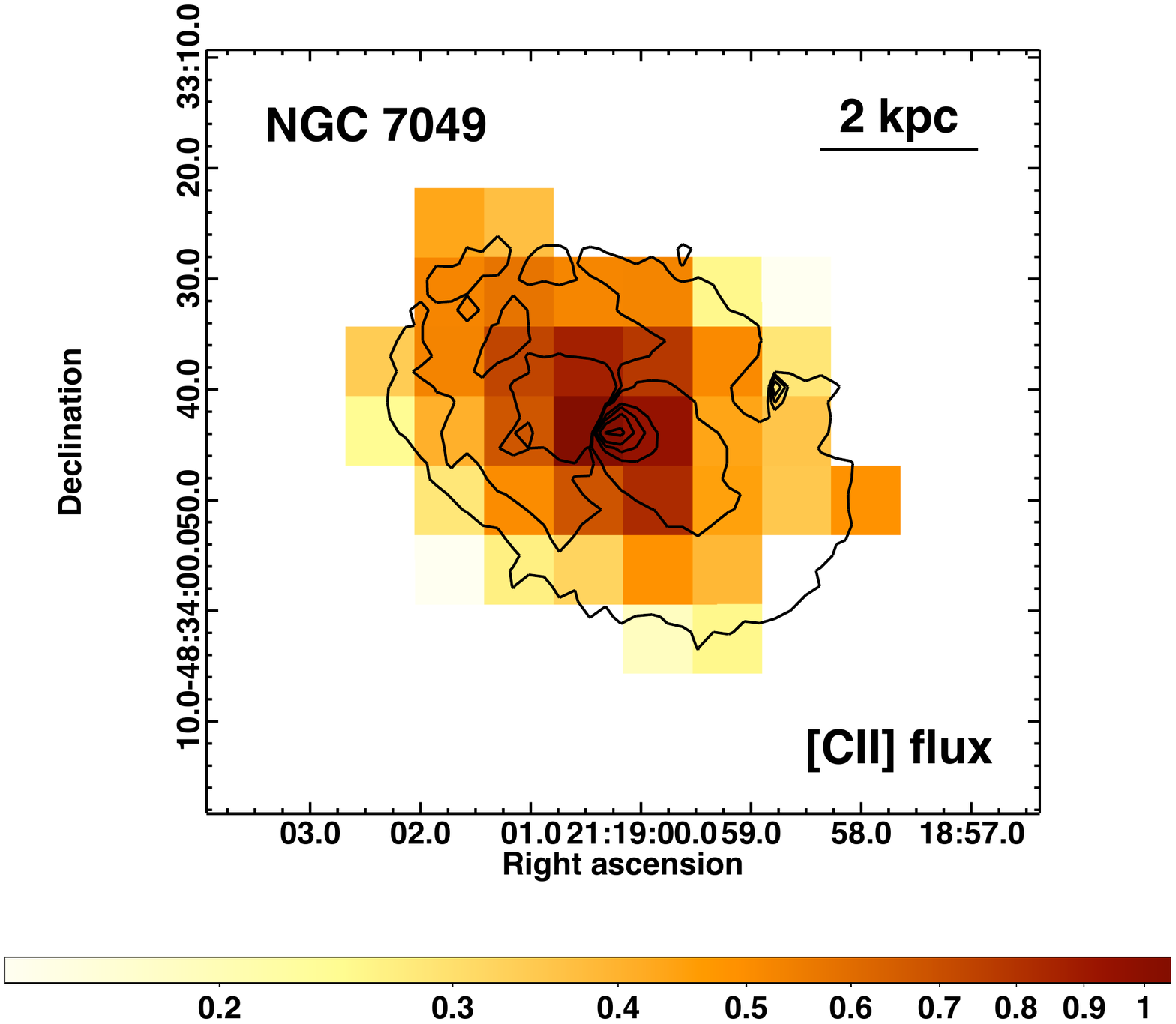}
\end{minipage}
\begin{minipage}{0.32\textwidth}
\vspace{-2.6cm}
 \includegraphics[width=1.02\textwidth,clip=t,angle=0.]{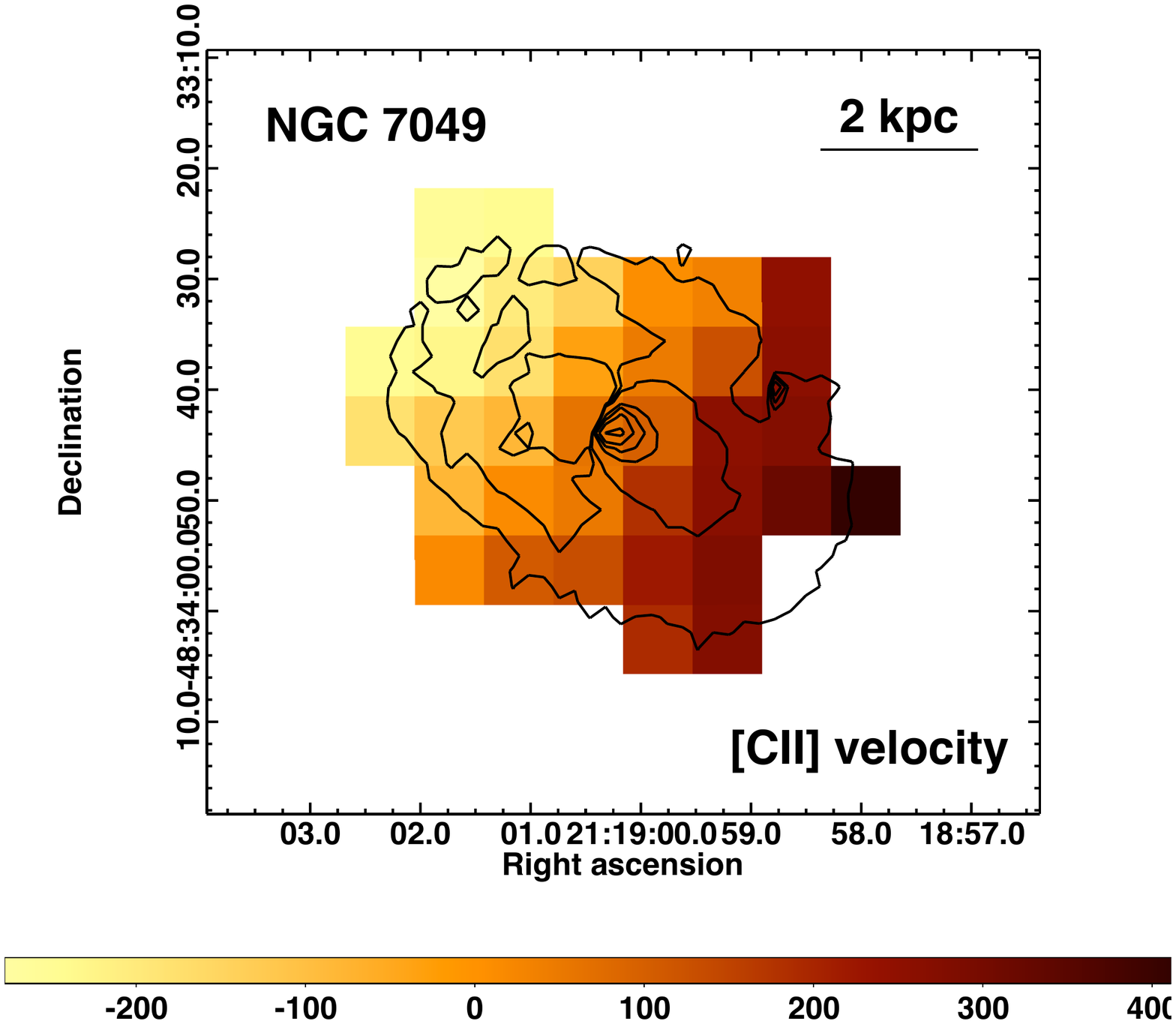}
\end{minipage}
\begin{minipage}{0.32\textwidth}
\vspace{-2.6cm}
\includegraphics[width=1.02\textwidth,clip=t,angle=0.]{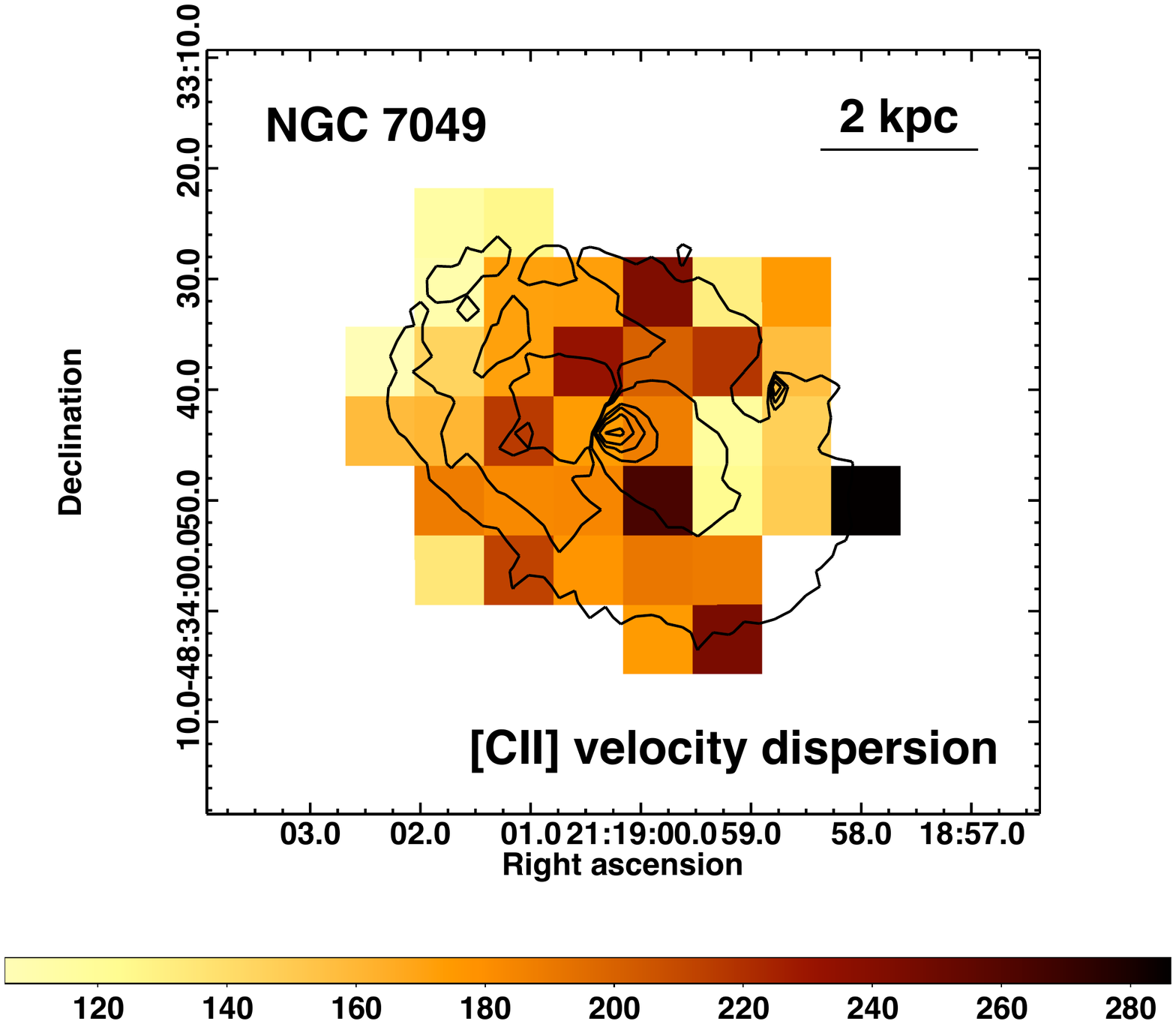}
\end{minipage}
\vspace{-1.2cm}
\caption{Same as Fig.~\ref{fig:C2maps1}, but for NGC~6868 and NGC~7049.} 
\label{fig:C2maps2}
\end{figure*}

\begin{figure*}
\begin{minipage}{0.24\textwidth}
\includegraphics[width=1\textwidth,clip=t,angle=0.,bb=36 109 577 683]{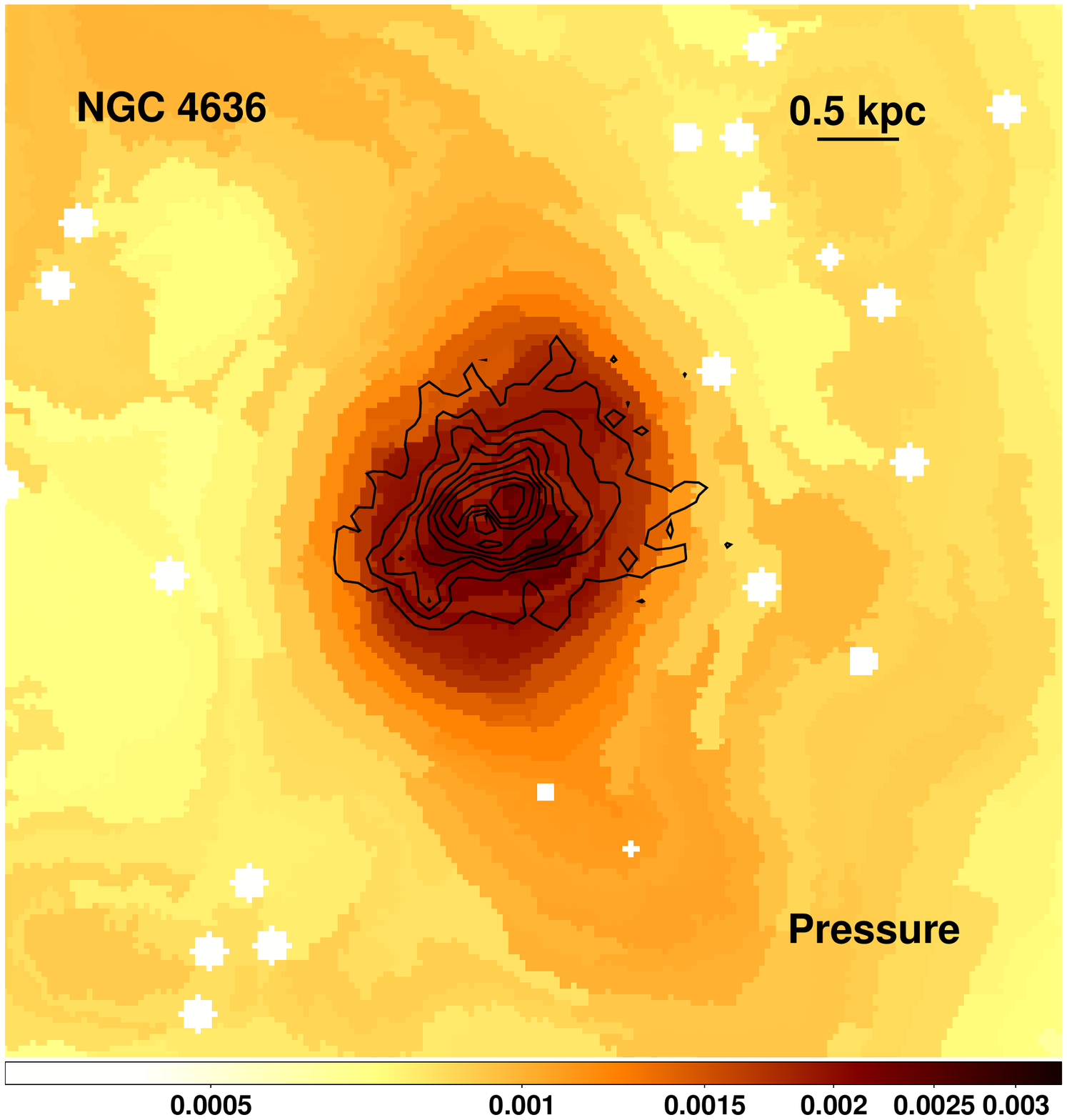}
\end{minipage}
\begin{minipage}{0.24\textwidth}
\includegraphics[width=1\textwidth,clip=t,angle=0.,bb=36 109 577 683]{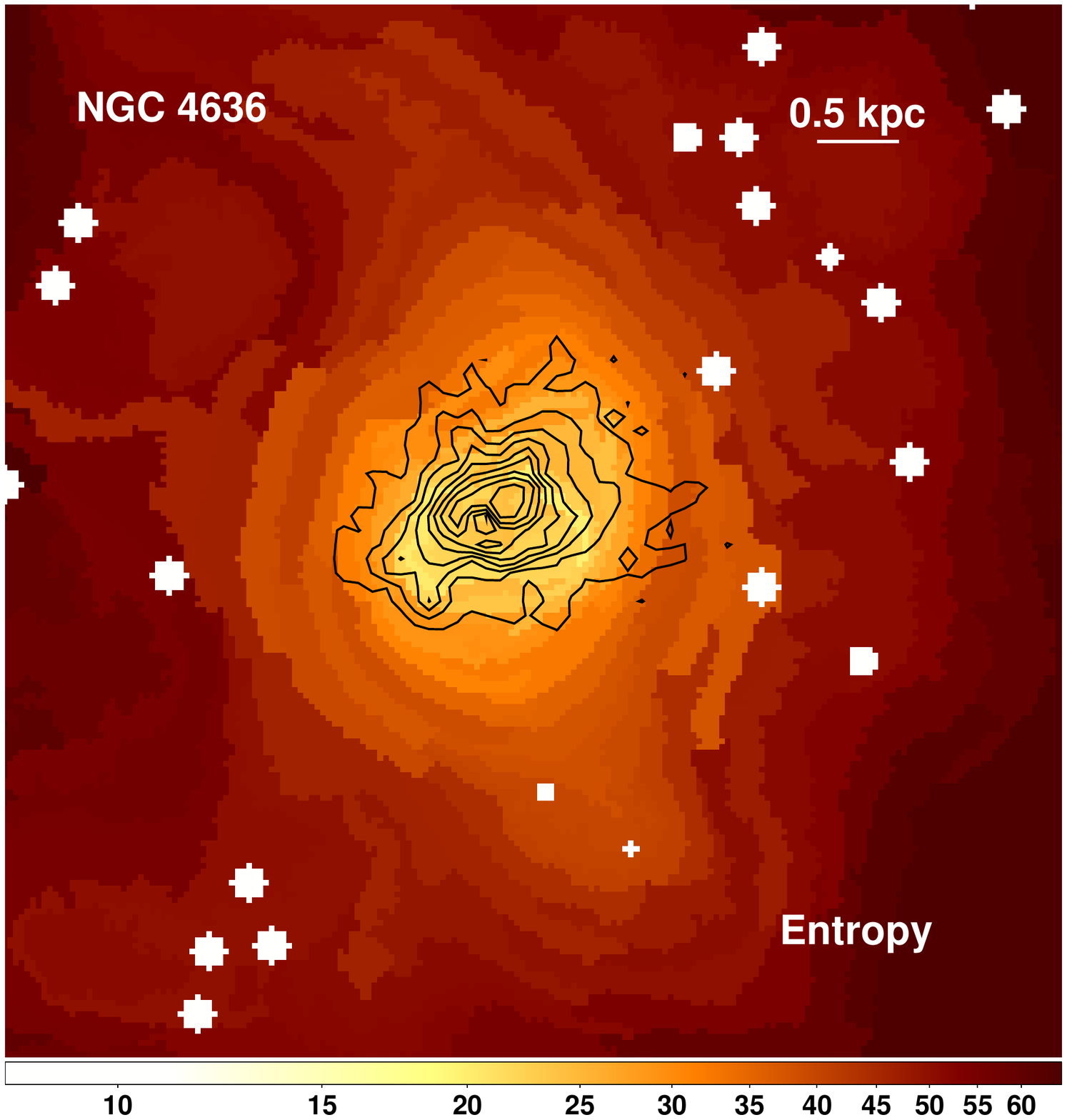}
\end{minipage}
\begin{minipage}{0.24\textwidth}
\includegraphics[width=1\textwidth,clip=t,angle=0.,bb=36 109 577 683]{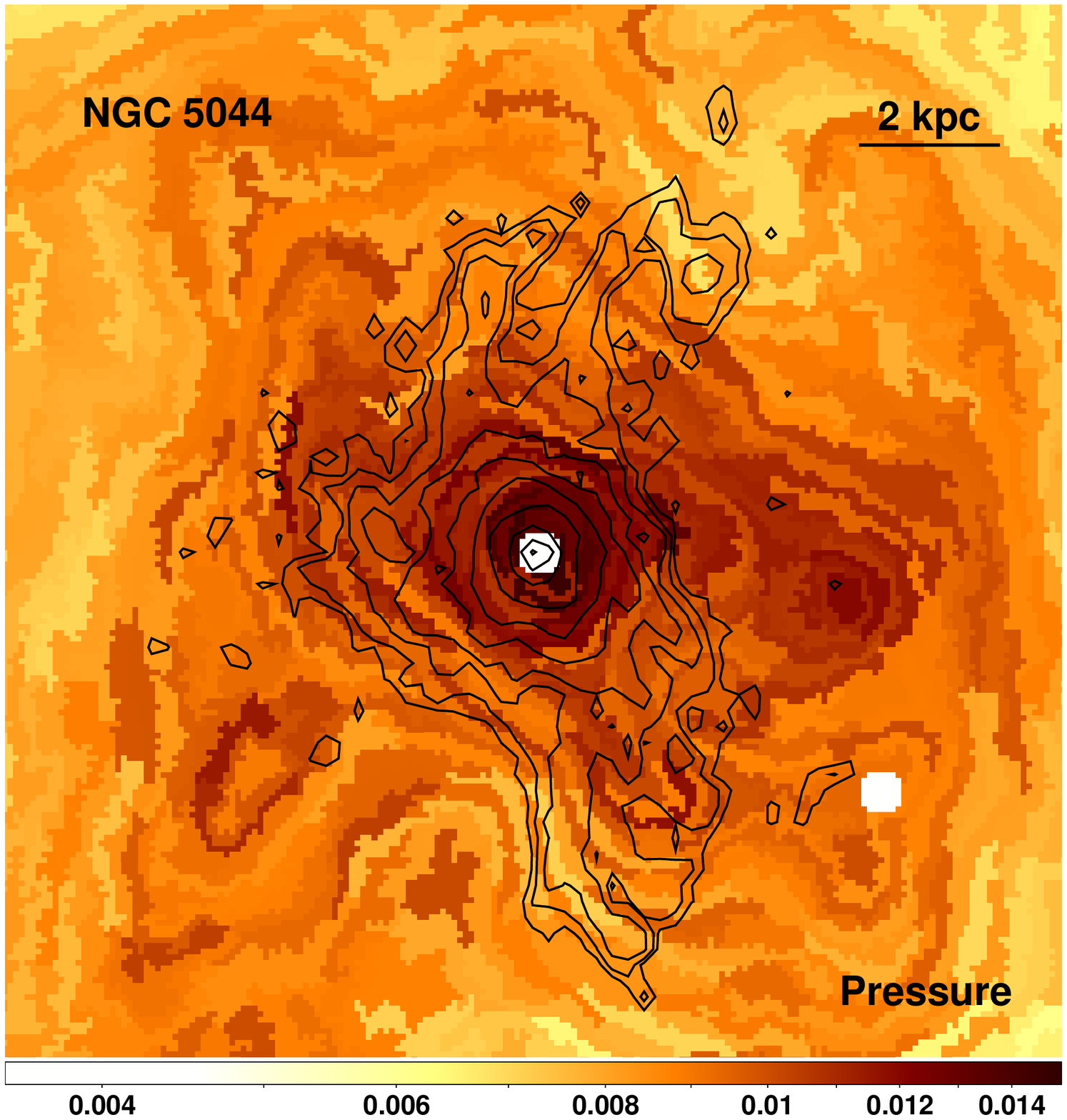}
\end{minipage}
\begin{minipage}{0.24\textwidth}
\includegraphics[width=1\textwidth,clip=t,angle=0.,bb=36 109 577 683]{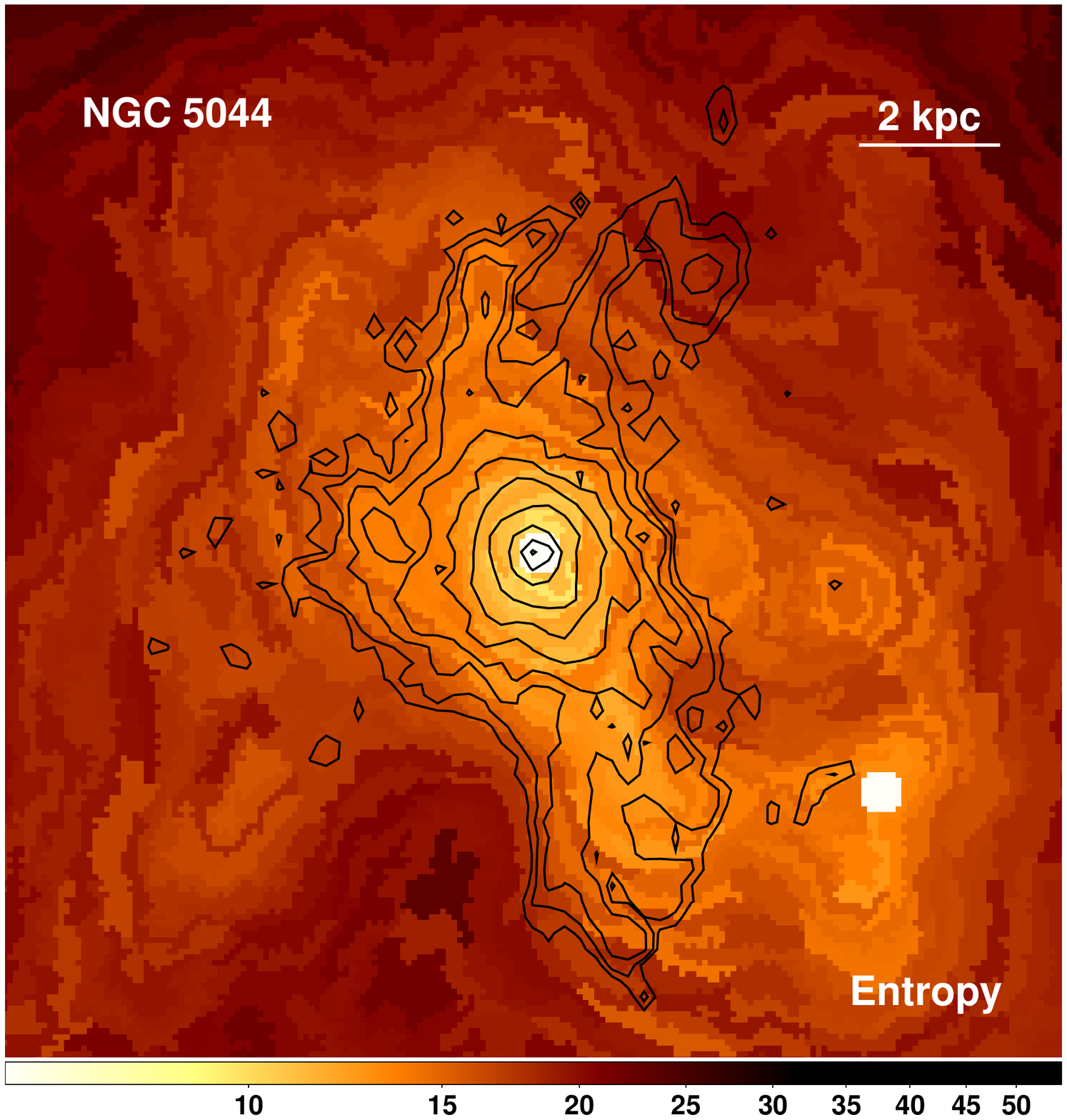}
\end{minipage}
\begin{minipage}{0.24\textwidth}
\includegraphics[width=1\textwidth,clip=t,angle=0.,bb=36 109 577 683]{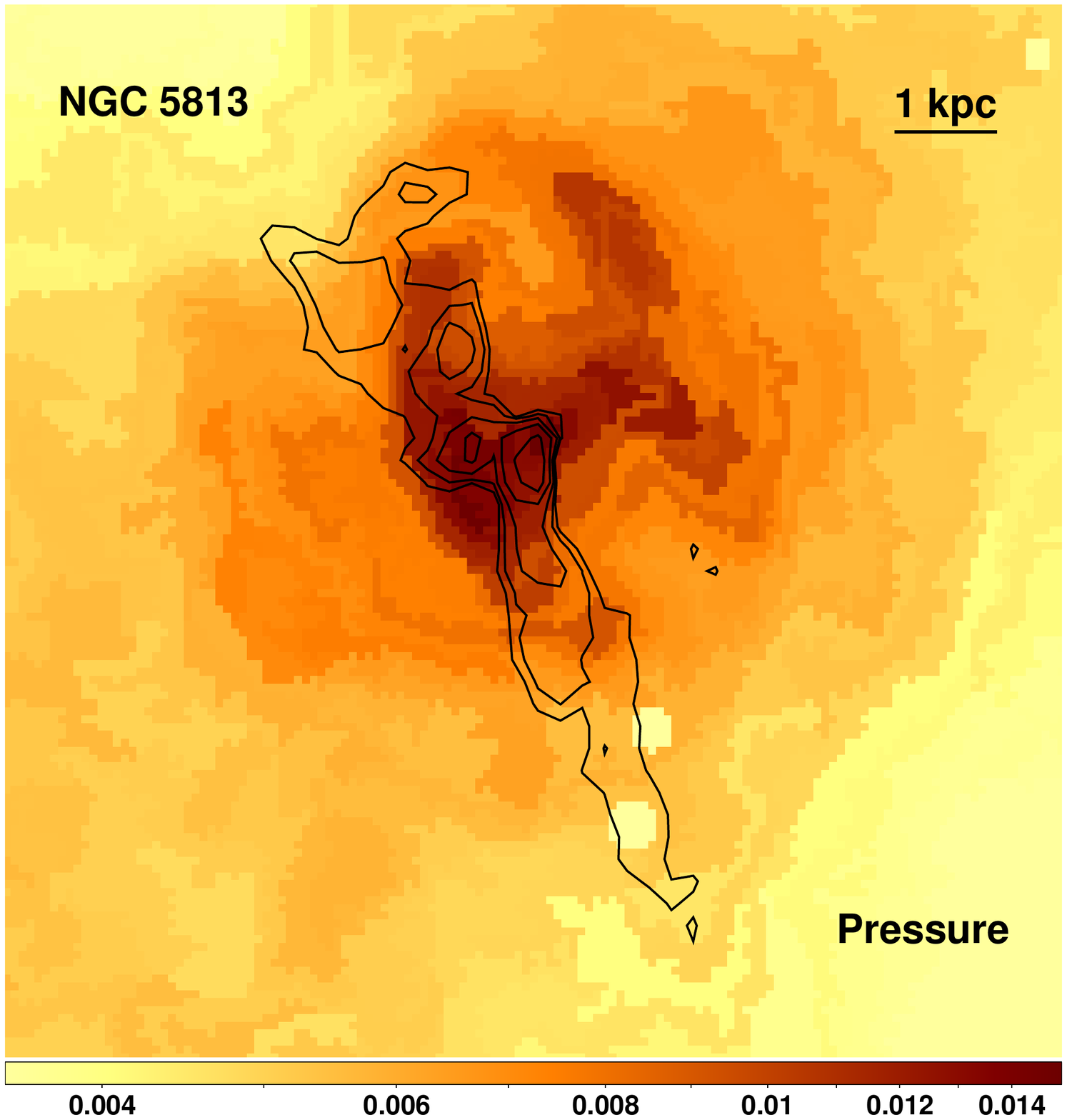}
\end{minipage}
\begin{minipage}{0.24\textwidth}
\includegraphics[width=1\textwidth,clip=t,angle=0.,bb=36 109 577 683]{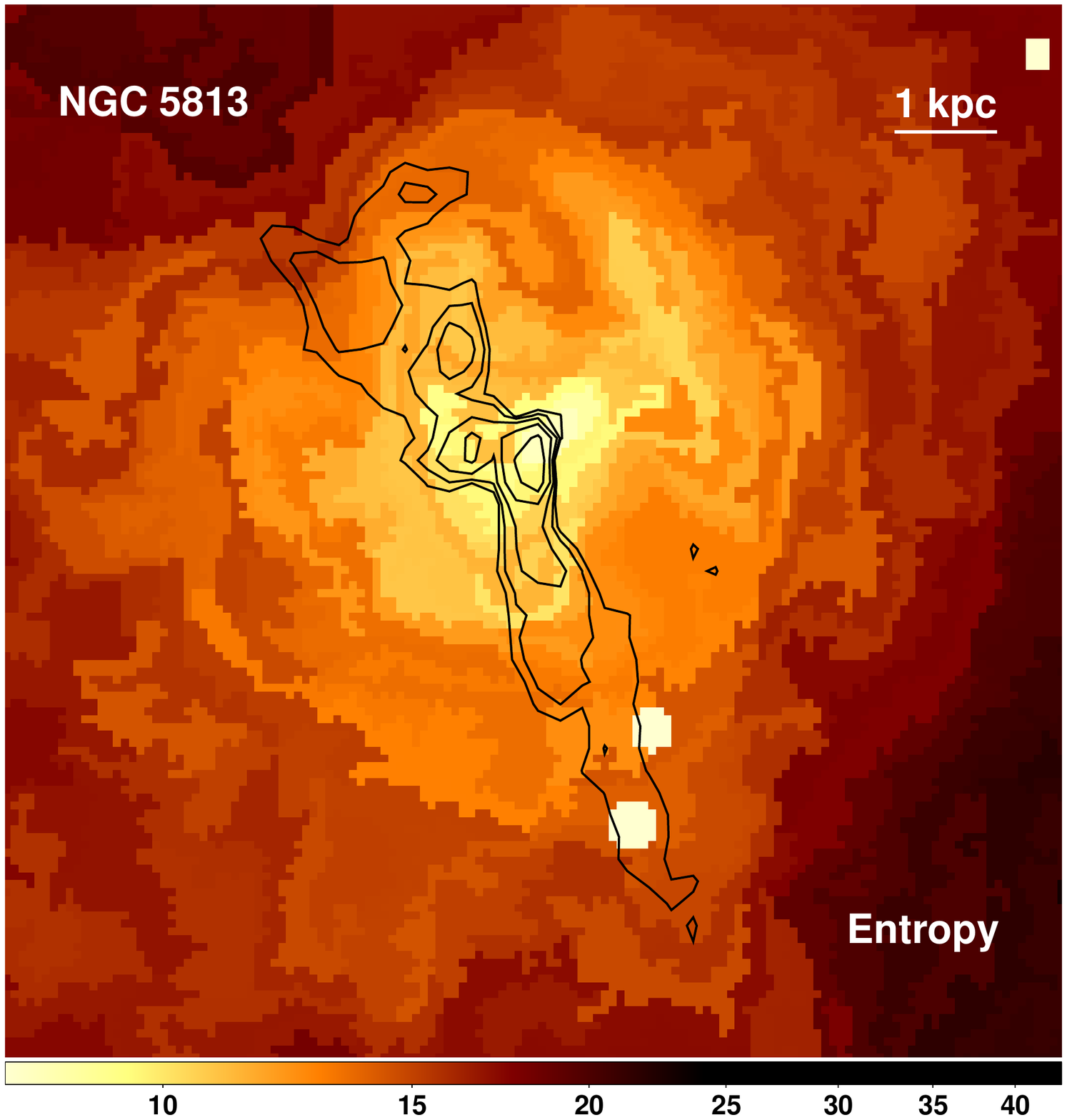}
\end{minipage}
\begin{minipage}{0.24\textwidth}
\includegraphics[width=1\textwidth,clip=t,angle=0.,bb=36 109 577 683]{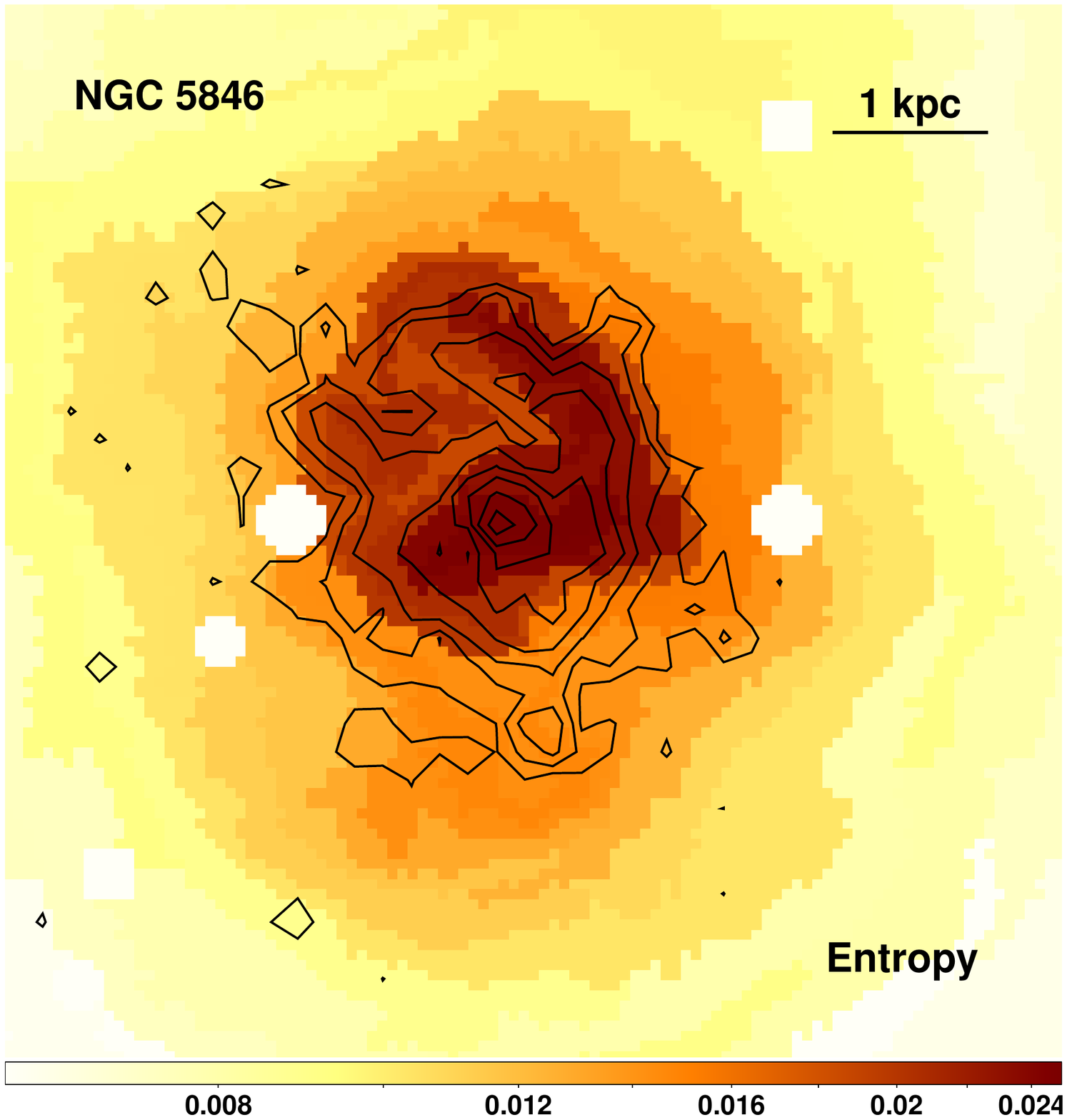}
\end{minipage}
\begin{minipage}{0.24\textwidth}
\includegraphics[width=1\textwidth,clip=t,angle=0.,bb=36 109 577 683]{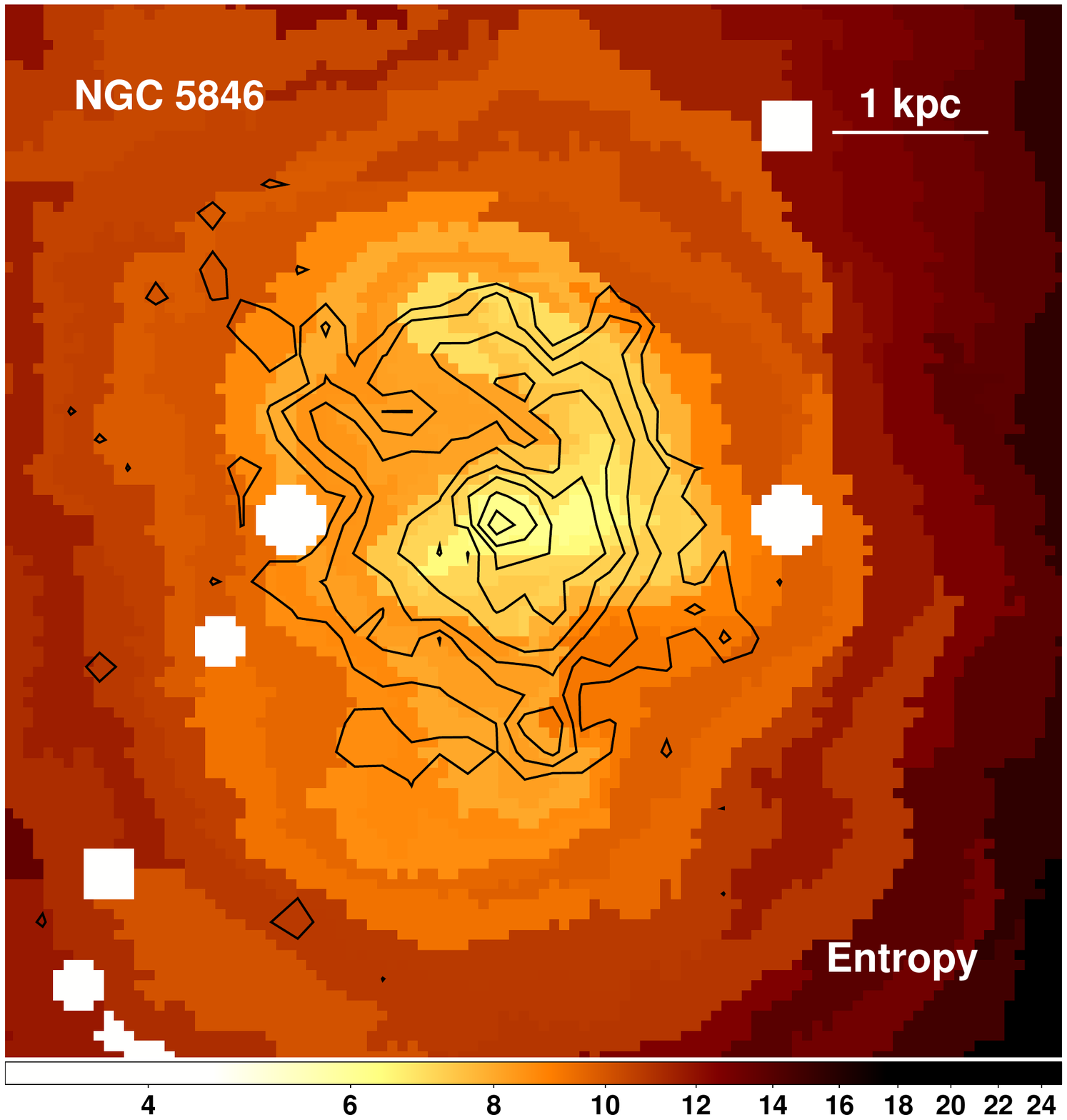}
\end{minipage}
\caption{2D map of the hot gas pressure (in units of keV~cm$^{-3}\times\left(\frac{l}{\mathrm{20kpc}}\right)^{-1/2}$) and entropy (in units of keV~cm$^2\times\left(\frac{l}
{\mathrm{20kpc}}\right)^{1/3}$) with the H$\alpha$+[\ion{N}{ii}] contours overlaid. The maps were obtained by fitting each region independently with a single temperature 
thermal model, yielding 1$\sigma$ fractional uncertainties of $\sim$3--5 per cent in temperature, entropy, and pressure. Point sources were excluded from the spectral 
analysis and therefore appear as white circles on the maps.}
\label{fig:Chandrasample}
\end{figure*}

\begin{figure*}
\begin{minipage}{0.24\textwidth}
\includegraphics[width=1\textwidth,clip=t,angle=0.,bb=36 109 577 683]{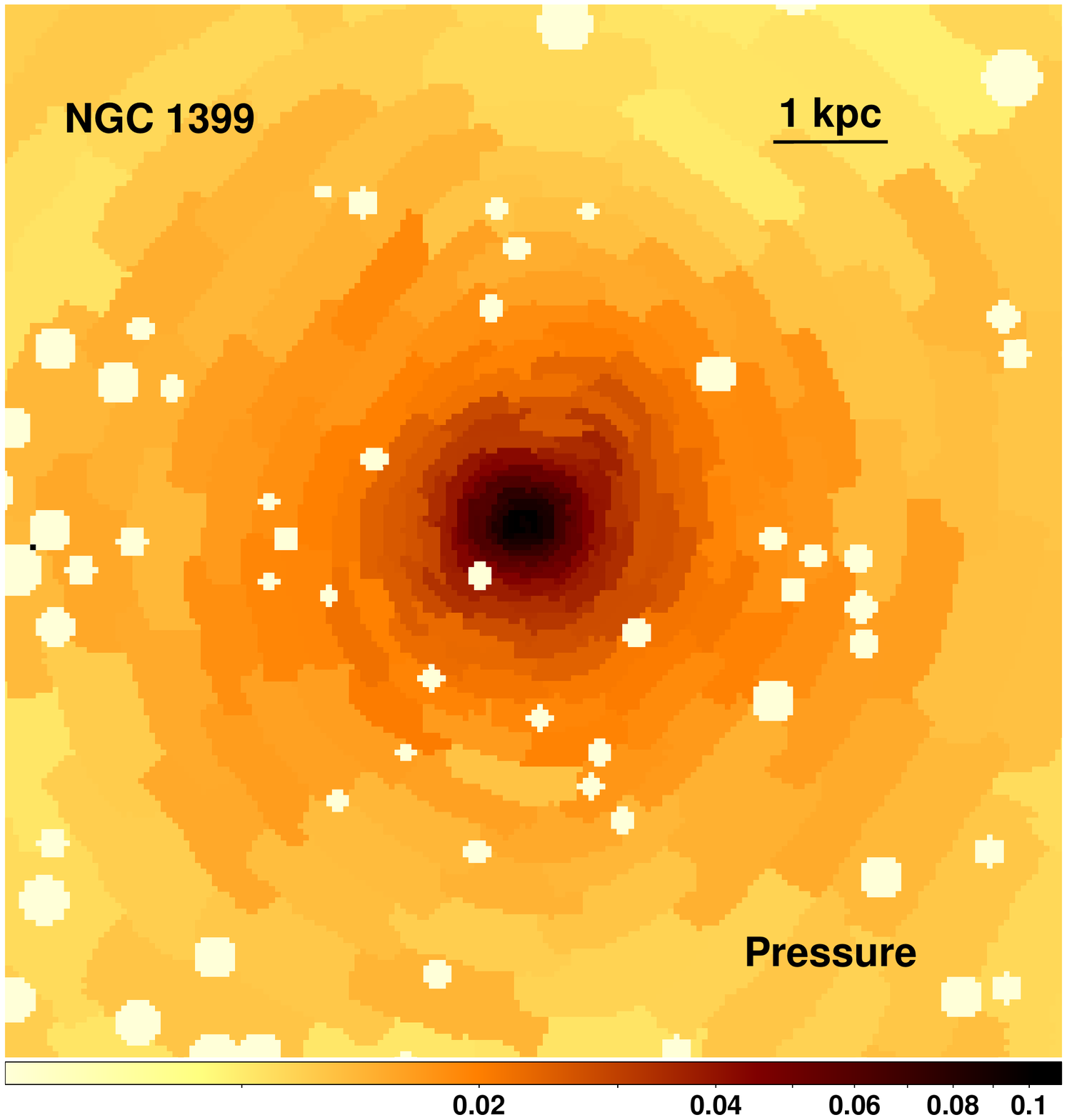}
\end{minipage}
\begin{minipage}{0.24\textwidth}
\includegraphics[width=1\textwidth,clip=t,angle=0.,bb=36 109 577 683]{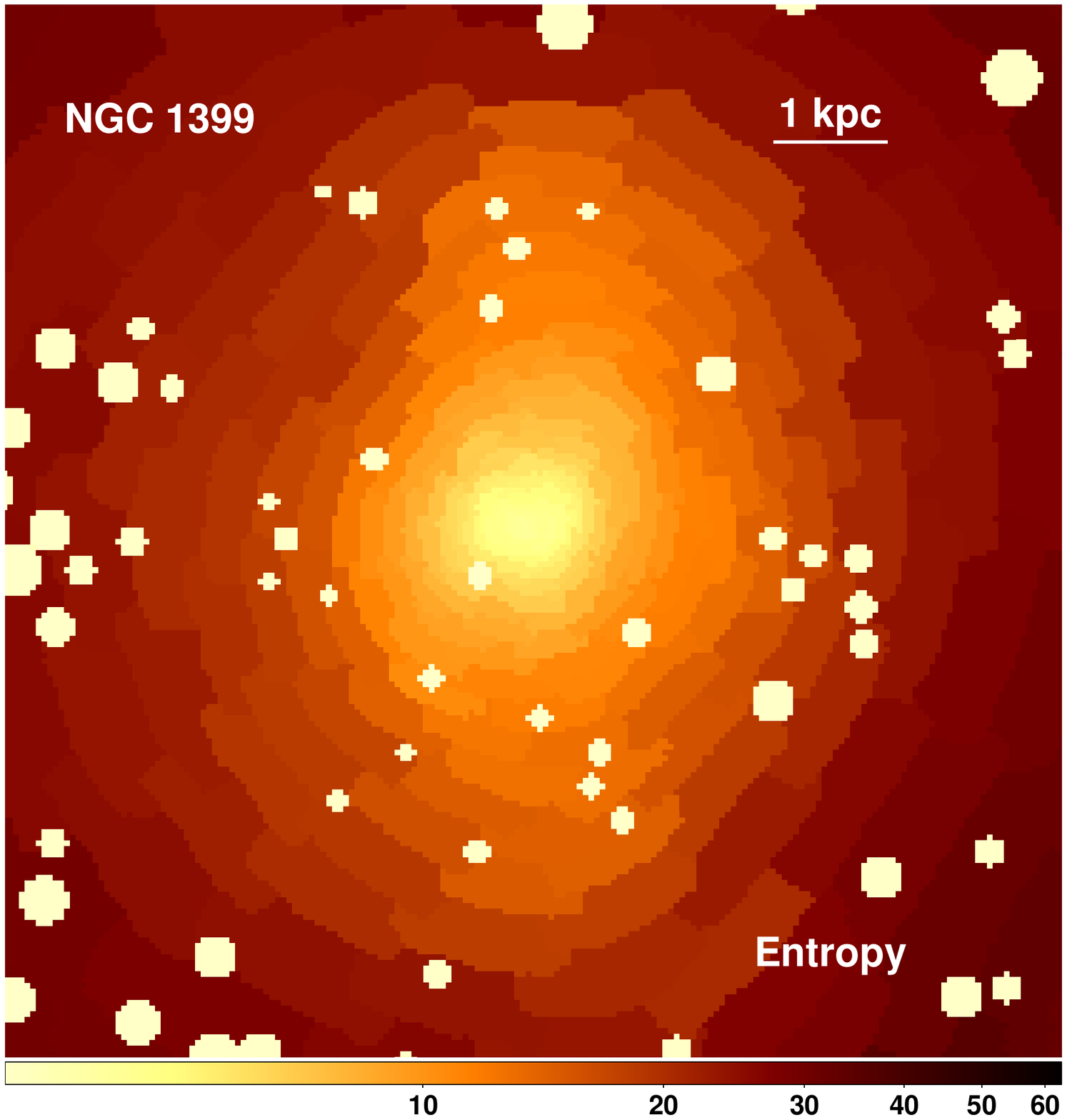}
\end{minipage}
\begin{minipage}{0.24\textwidth}
\includegraphics[width=1\textwidth,clip=t,angle=0.,bb=36 109 577 683]{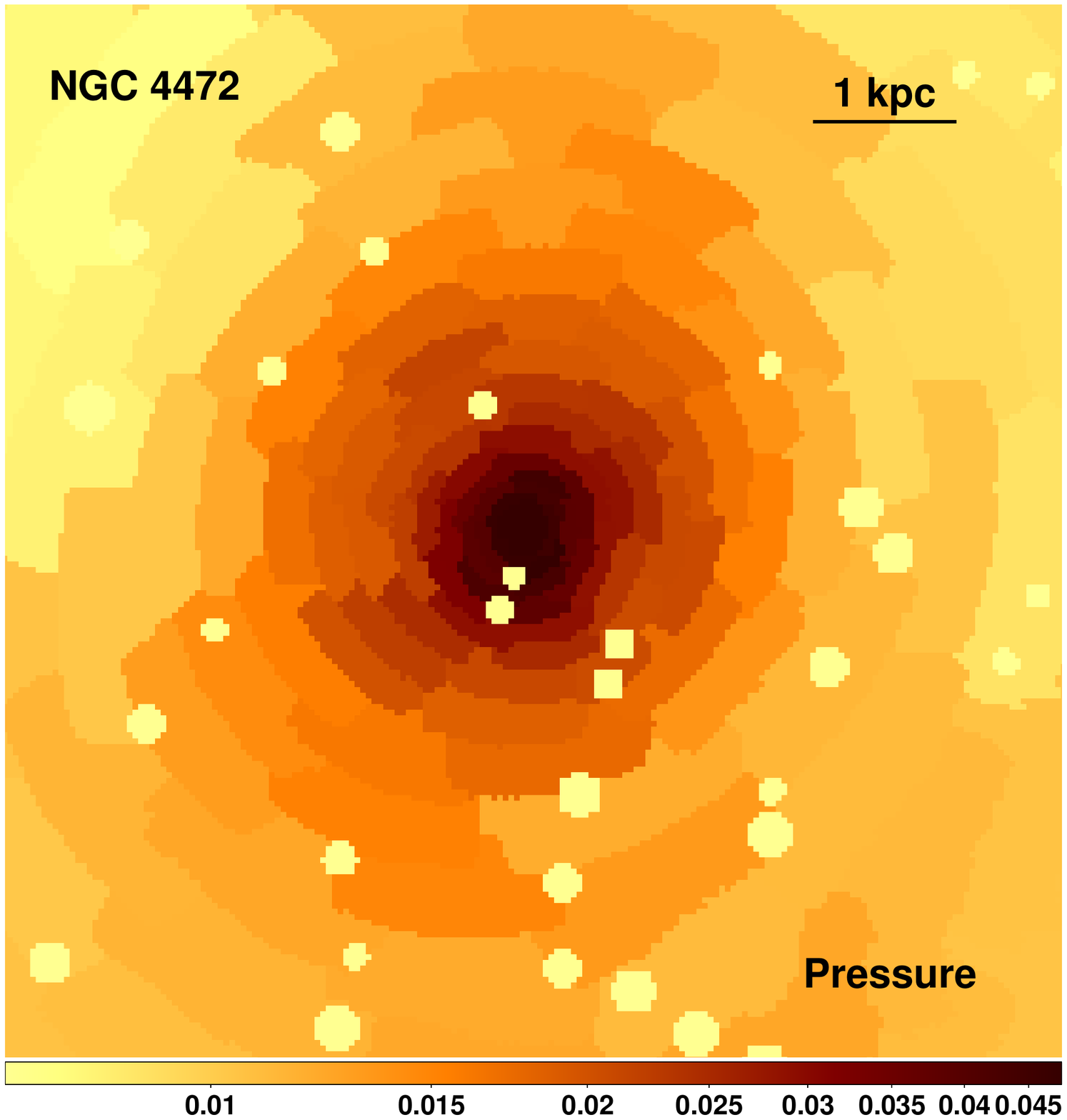}
\end{minipage}
\begin{minipage}{0.24\textwidth}
\includegraphics[width=1\textwidth,clip=t,angle=0.,bb=36 109 577 683]{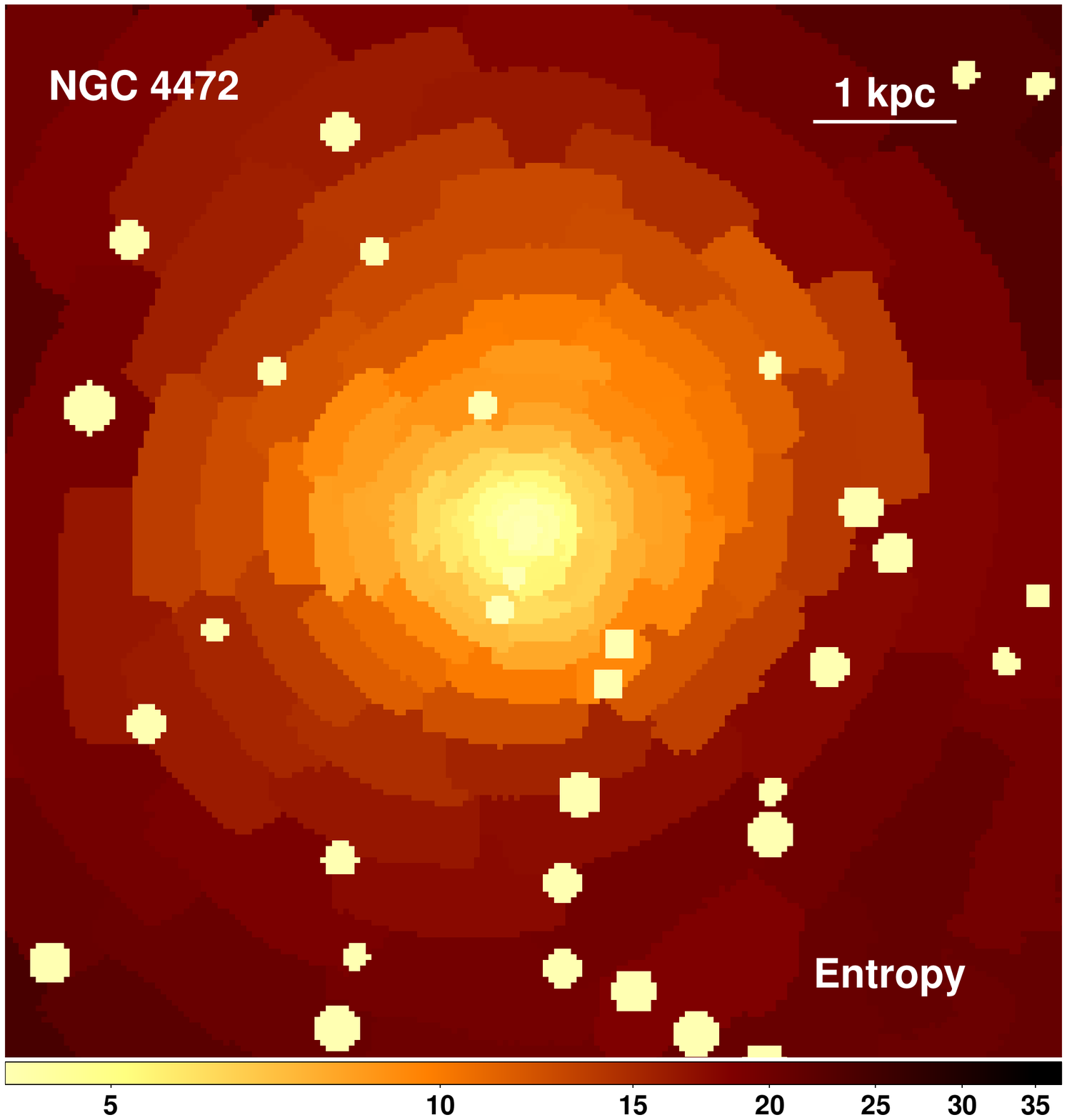}
\end{minipage}
\caption{2D maps of the hot gas pressure and entropy for the two galaxies with little or no [\ion{C}{ii}] emitting cold gas detected. } 
\label{fig:Chandrasample2}
\end{figure*}

Contrary to the results of \citet{macchetto1996}, we detected no H$\alpha$+[\ion{N}{ii}] emission in NGC~1399.  \citet{goudfrooij1994} derived an H$\alpha$+[\ion{N}{ii}] flux of 
$4.2\pm0.2\times10^{-14}$~erg s$^{-1}$~cm$^{-2}$, much smaller than the value measured by \citet{macchetto1996}. Our SOI H$\alpha$+[\ion{N}{ii}] imaging analysis of 
NGC~1399, with either adjacent narrow-band or R band image for the continuum subtraction, results in negative net flux around the center of the galaxy, indicative of H$\alpha
$ absorption. Therefore, we took additional long-slit spectra with the Goodman spectrograph on SOAR. Our Goodman spectra give a 5 $\sigma$ limit on the EW of the 
emission line of 0.1 $\AA$. Assuming that all the H$\alpha$+[\ion{N}{ii}] emission is from the central 12$''$ radius, as shown by the H$\alpha$+[\ion{N}{ii}] image from 
\citet{macchetto1996}, this EW upper limit translates to an H$\alpha$+[\ion{N}{ii}] flux limit of $3.4\times10^{-14}$~erg~s$^{-1}$~cm$^{-2}$. The lack of a spectroscopic 
detection of optical emission lines in NGC~1399 is inconsistent with previous reports of detections through narrow-band filters. We suspect that the previously reported 
detections of morphologically smooth line emission, where the emitting region has the same morphology as the old stellar populations, may be due to color gradients in the 
stellar population of the galaxy. Currently, we do not have SOAR data for NGC~4472.

In Fig.~\ref{Ha_CII}, we plot the H$\alpha$+[\ion{N}{ii}] luminosities (bottom panel), 70/24 $\mu$m infrared continuum ratios \citep[middle panel,][]{temi2007}, and the K-band 
luminosity \citep[top panel, based on the 2MASS survey,][]{jarrett2003} against the spatially integrated [\ion{C}{ii}] luminosity measured by {\it Herschel} PACS. The ratio of the 
[\ion{C}{ii}] over H$\alpha$+[\ion{N}{ii}] luminosity is remarkably similar $\sim0.4$--0.8 for 6/8 galaxies in our sample. For the same 6/8 systems, the [\ion{C}{ii}]$\lambda157\mu
$m line emission is spatially extended and appears to follow the distribution of the H$\alpha$+[\ion{N}{ii}] filaments (see the left panels of Fig. \ref{fig:C2maps1} and 
\ref{fig:C2maps2}). In the extended network of filaments of NGC~5044 at $r\gtrsim1$~kpc, the [\ion{C}{ii}]/(H${\alpha}$+[\ion{N}{ii}]) ratios also remain similar as a function of 
position. In the two FIR faint systems (shown by the red data points) the ionized H$\alpha$ emission is either concentrated in the nucleus of the galaxy \citep[NGC~4472,][]
{macchetto1996} or is not present (NGC~1399). 

The [\ion{C}{ii}] luminosities, which are likely excellent tracers of the cold molecular gas mass \citep{crawford1985}, do not correlate with the K-band luminosities (see Table~
\ref{galaxies}), which trace the stellar masses of the galaxies. There is also no correlation between the [\ion{C}{ii}] luminosities of systems with extended emission line nebulae 
and the 70/24~$\mu$m infrared continuum ratios. In the FIR faint systems indicated by the red data points these ratios are smaller (see the central panel of Fig.~\ref{Ha_CII}). 
The 70/24~$\mu$m infrared continuum ratio is a good tracer of interstellar dust \citep{temi2007}.

Fig.~\ref{Pj_CII} shows that the [\ion{C}{ii}] luminosities in systems with spatially extended emission do not correlate with the jet powers determined from the work required to 
inflate bubbles of relativistic plasma associated with the cavities in the hot X-ray emitting atmospheres of the galaxies listed in Table~\ref{galaxies}. In our relatively small 
sample, the galaxies with little or no cold gas (indicated in red) have higher jet powers than the systems with significant [\ion{C}{ii}] detections. 

The central panels of Fig.~\ref{fig:C2maps1} and \ref{fig:C2maps2} show the velocity distributions of the [\ion{C}{ii}] line emitting gas with respect to the velocities of the host 
galaxies. In NGC~6868 and NGC~7049 the velocity distribution of the [\ion{C}{ii}] emitting gas indicates that the cold gas forms a rotating disk. This is consistent with the 
overall morphology of the optical line emission nebulae. The velocity distribution of the [\ion{C}{ii}] emitting gas in the other four systems with extended emission does not show 
any coherent structure. 

The projected velocity dispersions measured from the [\ion{C}{ii}] lines along our line of sight are generally in the range between 100~km~s$^{-1}$ and 200~km~s$^{-1}$, with 
the highest measured values reaching $\sim280$~km~s$^{-1}$. 

Fig.~\ref{fig:Chandrasample} shows maps of the pressure and entropy in the hot X-ray emitting intra-group/interstellar medium for the galaxies where the cold gas has a 
filamentary morphology. In all of these systems, the filaments are co-spatial with the lowest entropy hot X-ray emitting gas. They also appear to anti-correlate and interact with 
the AGN inflated bubbles of relativistic plasma, which provide significant non-thermal pressure in the X-ray atmospheres of the galaxies and therefore appear as regions of low 
projected X-ray pressure on our maps. This is most clearly seen for the southwestern filament in NGC~5813, which appears to be pushed aside by the large, older, outer 
bubble. The inner portion of the filament also appears on the side and partly on the top of the small, younger, inner southwestern bubble that is currently being inflated in the 
core \citep[for detailed discussion on the bubbles see][]{randall2011}. Similar interaction, with the filaments being pushed around by the bubbles, is seen in NGC~5846 and 
NGC~4636, both of which show `cavities' in the H$\alpha$+[\ion{N}{ii}] images that coincide with pressure decrements in their X-ray spectral maps. NGC~5044 shows a 
strongly disturbed morphology due to both AGN activity and the intra-group medium sloshing in the gravitational potential of the system 
\citep{gastaldello2009,david2009,david2011,osullivan2013}. 

Fig.~\ref{fig:Chandrasample2} shows the pressure and entropy maps for NGC~4472 and NGC~1399 - systems with little or no cold and warm gas. Compared to the systems 
with filamentary H$\alpha$ nebulae, their thermodynamic maps have relatively regular morphologies. These systems are among the nearby giant ellipticals with the most 
relaxed X-ray morphologies at small radii \citep{werner2012}. 

\begin{figure}
\includegraphics[width=0.22\textwidth,clip=t,angle=0.]{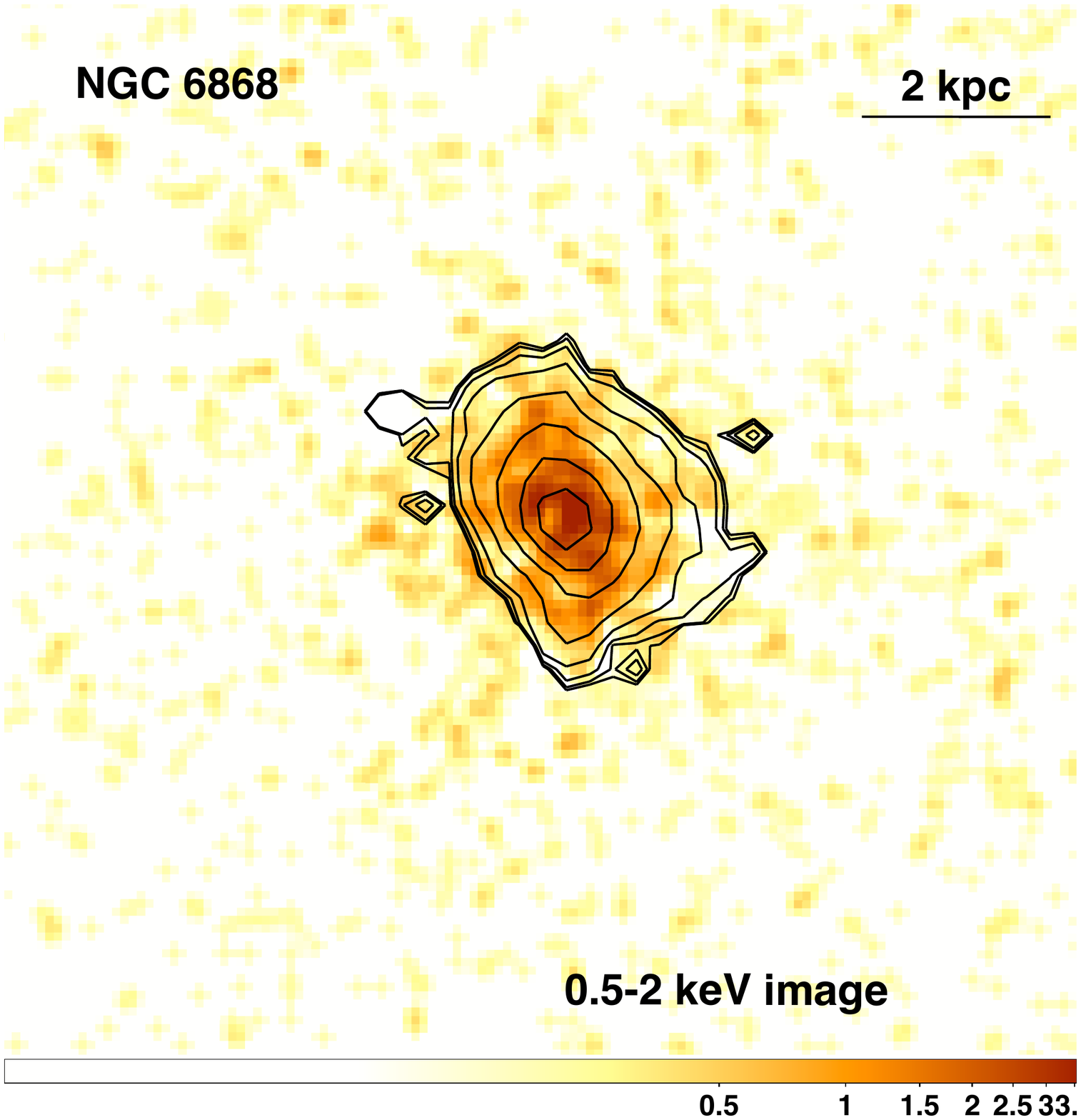}
\includegraphics[width=0.22\textwidth,clip=t,angle=0.]{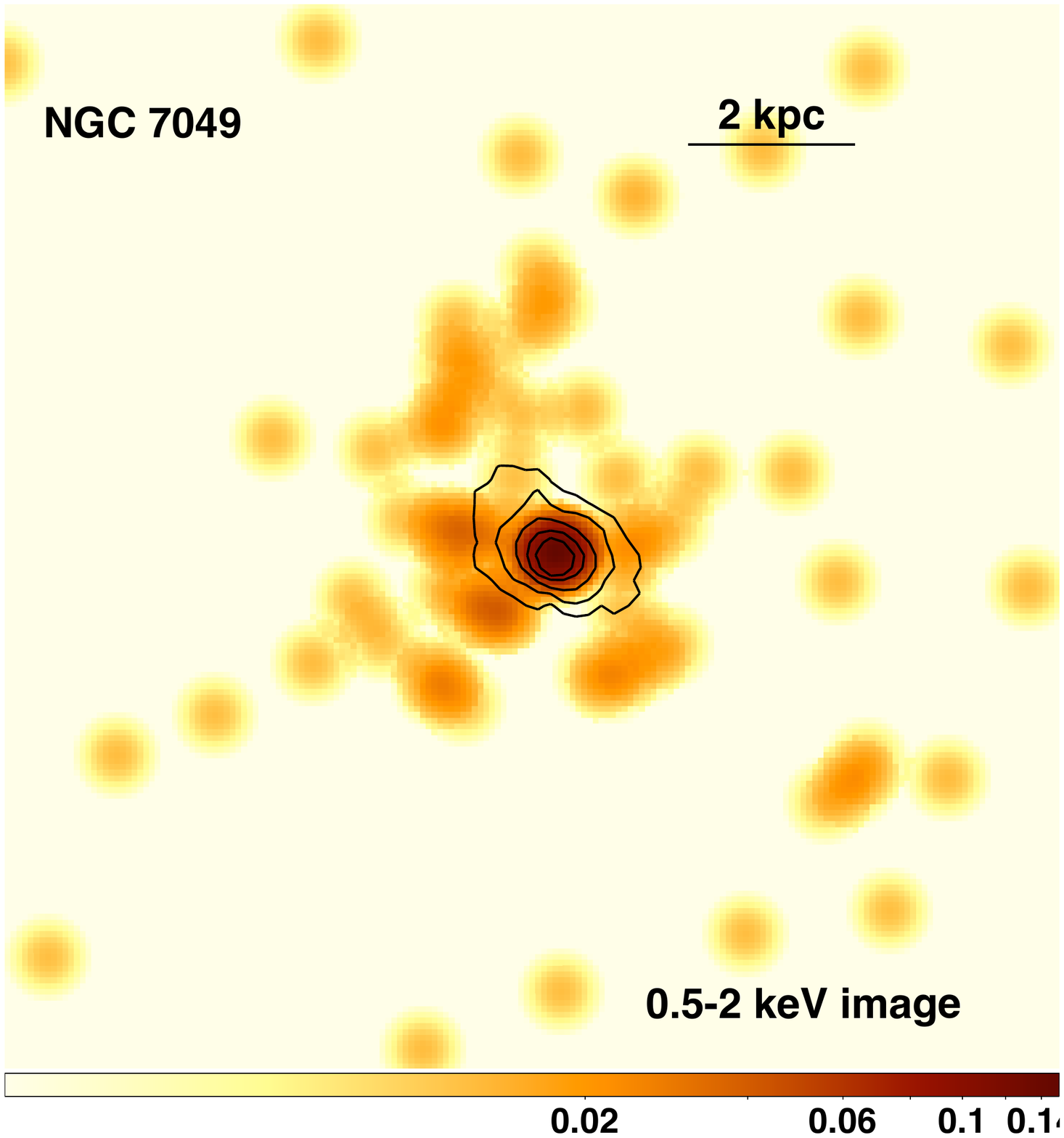}
\caption{0.5--2 keV {\it Chandra} images of the X-ray faintest galaxies in our sample, NGC~6868 and NGC~7049, with the H$\alpha$+[\ion{N}{ii}] contours overlaid. The 
statistical quality of the X-ray data does not allow us to produce 2D maps of thermodynamic properties for these two systems.}
\label{xrayimages}
\end{figure}

NGC~6868 and NGC~7049 do not have X-ray data with the quality that would allow us to extract detailed maps of thermodynamic properties and in Fig.~\ref{xrayimages} we 
only show their X-ray images. The images show extended X-ray haloes, co-spatial with the H$\alpha$+[\ion{N}{ii}] emission, surrounding the galaxies. The X-ray image of 
NGC~6868 shows a relatively complex disturbed morphology \citep[see also][]{machacek2010}. The X-ray data of NGC~7049 are very shallow and only allow us to detect the 
extended X-ray emission. Four galaxies - NGC~5044, NGC~6868, NGC~7049, and NGC~5813 - have an X-ray point source associated with the central AGN. 

\begin{figure}
\includegraphics[width=1\columnwidth,clip=t,angle=0.]{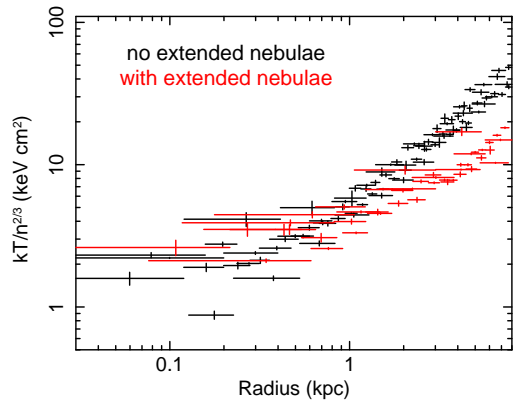}
\caption{Entropy profiles for a sample of morphologically relaxed giant ellipticals \citep[NGC~4472, NGC~1399, NGC~4649, NGC~1407, NGC~4261;][]{werner2012}, with no 
extended optical emission line nebulae (in black), and for five galaxies from our sample with [\ion{C}{ii}] emission indicating the presence of reservoirs of cold gas (NGC~5044, 
NGC~5813, NGC~5846, NGC~4636, NGC~6868 - in red). The entropy in galaxies containing cold gas is systematically lower at $r\gtrsim1$~kpc. The only exception is 
NGC~6868 the entropy profile of which follows that of the cold-gas-poor galaxies.  } 
\label{entropies}
\end{figure}

The deprojected central densities and pressures measured at $r=0.5$~kpc in the morphologically relaxed NGC~1399 and NGC~4472 appear significantly higher than in the 
other, more disturbed, [\ion{C}{ii}] bright systems (see Table~\ref{deprojected}). All [\ion{C}{ii}] bright galaxies appear to have similar central densities and pressures. The 
entropies at $r=0.5$~kpc span a small range of 3.0--4.5~keV~cm$^{2}$ and the central cooling times are 4--$8\times10^7$~yr. At radii $r\gtrsim1$~kpc the entropies of the 
galaxies containing cold gas (except NGC~6868) are systematically lower than those of NGC~1399 and NGC~4472 (see Fig.~\ref{entropies}). The measured relatively low 
central densities and pressures of the morphologically disturbed systems may partly be due to the additional non-thermal pressure support in the central regions of the these 
systems. However, we also suspect a bias due to departures from spherical symmetry. Due to the non-uniform surface brightness within the circular annuli used in the 
deprojection analysis, densities in the outer shells may be somewhat overestimated, leading to underestimated central densities. For the morphologically disturbed systems, 
we repeated the deprojection analysis using wedges and found, that despite the systematic uncertainties, at radii $r\gtrsim1$~kpc the difference in the shapes of the profiles shown in Fig.~\ref{entropies} remains robust.

\section{Discussion}
\label{discussion}

\subsection{Properties of the cold ISM}
The observed high [\ion{C}{ii}] luminosities indicate the presence of large reservoirs of cold gas.
Cold gas is evidently prevalent in massive giant elliptical galaxies with spatially extended H$\alpha$+[\ion{N}{ii}] nebulae, even though their stellar populations are old and 
appear red and dead \citep{annibali2007}.
NGC~5044 has previously also been detected in the CO(2--1) line with the 30~m IRAM telescope with a total flux of $6.6\times10^{-17}$ erg~s$^{-1}$~cm$^{-2}$ (Jeremy Lim 
\& Francoise Combes, private communication). Assuming a CO(2--1) to CO(1--0) ratio of 0.8, we obtain a [\ion{C}{ii}]$\lambda158$$\mu$m over CO(1--0) flux ratio of 3817, 
which is within the range of the ratios (1500--6300) observed in normal and star forming galaxies and in Galactic molecular clouds \citep{crawford1985,stacey1991}.

Assuming the cold $\sim$100~K [\ion{C}{ii}] emitting gas is in thermal pressure equilibrium with the surrounding hot X-ray emitting plasma, it will presumably form filaments 
with densities $\sim10^4$~cm$^{-3}$ and small volume filling fractions. As discussed in \citet{fabian2003b,fabian2008,fabian2011} and \citet{werner2013}, the emission line 
filaments are likely to consist of many strands that have small volume filling fractions but large area covering factors. The soft X-ray emission that is always associated with 
filaments in cool core clusters  \citep[][]{sanders2007,sanders2008,sanders2009,sanders2009b,deplaa2010,werner2011,werner2010,werner2013} indicates the presence of 
cooling hot plasma distributed between the threads. The hot plasma may be cooling both radiatively and by mixing with the cold gas in the filaments. Part of the soft X-ray 
emission from this cooling plasma gets absorbed by the cold gas in the strands \citep{werner2013}. 

The velocities inferred from the [\ion{C}{ii}] line emission are consistent with those measured from the H$\alpha$+[\ion{N}{ii}] lines by \citet{caon2000}, indicating that the 
different observed gas phases form multiphase filaments where all the phases move together. 
The velocity structure measured in NGC~6868 and NGC~7049 indicates that the cold gas in these systems is distributed in large extended rotating disks. 
Given the expected densities of the [\ion{C}{ii}] emitting gas, its volume filling fraction must be low and the disks are also likely to be formed from many small clouds, sheets, 
and filaments.

The observed velocity dispersions of the [\ion{C}{ii}] emitting gas are similar to the range of values measured from CO line widths and using optical and near-infrared integral 
field spectroscopy of 2000--10,000~K gas in the cooling cores of clusters \citep[e.g.][]{edge2001,ogrean2010,oonk2010,farage2012,canning2013}. Observations indicate 
that, due to their small volume filling fraction and large area covering factors, filaments may be blown about by the ambient hot gas, potentially serving as good tracers of its 
motions \citep{fabian2003b}. The measured velocity dispersions of the [\ion{C}{ii}] emitting gas, which reflect the range of velocities of the many small gas clouds/filaments 
moving through the galaxy, may therefore be indicative of the motions in the hot ISM. They are in general consistent with the limits and measurements obtained through X-ray 
line broadening and resonant line scattering observations with the Reflection Grating Spectrometers on {\it XMM-Newton} \citep{werner2009,deplaa2012,sanders2013}.

The mid-infrared {\it Spitzer} spectra of all of our [\ion{C}{ii}] bright galaxies overlapping with the sample of \citet{panuzzo2011} - NGC~4636, NGC~5044, NGC~5813, 
NGC~5846, NGC~6868 - show the presence of dust, warm H$_2$ molecular gas, and for NGC~5044, NGC~6868, and NGC~4636 also PAH emission. \citet{temi2007} show 
that, for NGC~4472 and NGC~1399, galaxies with little or no [\ion{C}{ii}] detected, the infrared spectral energy distributions (SEDs) can be fully explained by circumstellar dust 
in the context of a steady state model of dust production in normal stellar mass loss followed by sputtering by the hot X-ray emitting gas. However, for all of the galaxies with 
extended [\ion{C}{ii}] emission, the infrared SEDs observed with {\it IRAS} and {\it Spitzer} indicate the presence of true interstellar dust, well in excess of the predictions of the 
steady state model \citep{knapp1989,temi2007,temi2007b}. Although there are early reports of detections of \ion{H}{i} emission in NGC~5846 and NGC~4636 
\citep{bottinelli1977,bottinelli1979,knapp1978}, they were not confirmed by later re-observations \citep{krishna1983,lake1984,knapp1985}. 
%No \ion{H}{i} emission has been detected in any of these systems. 

\citet{werner2013} showed that the [\ion{S}{ii}]$\lambda6716/6731$ line ratios of the filaments in M~87 indicate very low densities in the H$\alpha$+[\ion{N}{ii}] emitting 
10,000~K phase. They concluded that, assuming subsonic turbulence in the filaments, the presence of significant magnetic pressure is required to keep this warm ionized gas in pressure equilibrium with the surrounding intra-cluster medium, indicating that the filaments are supported by magnetic fields of $B = 28-73\mu$G. The 
assumption of subsonic turbulence is motivated by the lack of [\ion{O}{iii}] line emission, which would be produced by strong shocks. 
For 5/8 galaxies overlapping with the sample of \citet{annibali2010},  the [\ion{S}{ii}]$\lambda6717/6731$ line ratios point to very low densities, $n_{\rm e}<26$~cm$^{-3}$, in 
the ionized gas, indicating that it is supported by magnetic fields. 
Such low densities appear to be common in the ionized, extended emission line nebulae in the cores of galaxy clusters.
Significant magnetic fields threading the emission line nebulae were also inferred using arguments based on the integrity of the filaments in the Perseus Cluster 
\citep{fabian2008} and based on radio observations of the Faraday rotation measure in cooling core clusters \citep{taylor2001,taylor2007,allen2001,feretti1999}. This magnetic support may be slowing or preventing the gravitational collapse of any molecular gas clouds traced by the [\ion{C}{ii}] line emission that exceed the Jeans 
mass, preventing them from forming stars \citep[e.g. see the discussion for NGC~1275 in ][]{ho2009}.

\subsection{Heating and ionization of the cold ISM}
In the six systems with significant cold gas mass reservoirs, the [\ion{C}{ii}] line emission appears to be co-spatial with the H$\alpha$+[\ion{N}{ii}] emission, and with the lowest 
entropy X-ray emitting plasma. In the same galaxies, the ratios of [\ion{C}{ii}]/(H$\alpha$+[\ion{N}{ii}]) emission appear to be similar (0.4--0.8), indicating that in these 
systems the [\ion{C}{ii}] and H$\alpha$+[\ion{N}{ii}] emission are powered by the same energy source. 

Photoionization by young hot stars or by the central AGN is negligible as a source of excitation for the nebulae in these galaxies. 
\citet{ferland2009} showed that the broad band emission-line spectra of the filamentary nebulae around central galaxies of cooling core clusters most likely originate in gas 
exposed to ionizing particles, either relativistic cosmic rays or hot X-ray emitting plasma penetrating into the cold gas. If the magnetized, ionized H$\alpha$ emitting phase 
forms a thin skin on the underlaying cold neutral gas, then the hot plasma particles  must somehow overcome the obstacle presented by the magnetic fields. Based on the 
observations of the emission line filaments in M~87, \citet{werner2010} and \citet{werner2013} propose that shocks propagating within the hot X-ray emitting plasma as well as 
the movement of the filaments through this ambient hot gas (both as they are being uplifted and as they fall back) induce shearing around the filaments, thereby promoting 
mixing of the cold gas with the ambient hot medium via instabilities \citep[e.g.][]{friedman2012}. \citet{fabian2011} propose that the ionizing hot plasma in the core of the 
Perseus Cluster penetrates the cold filaments through magnetic reconnection diffusion \citep{lazarian2010,lazarian2011}, which may be induced by shearing instabilities and 
turbulence. The penetration of the cold gas by the hot plasma particles implies that the X-ray emitting gas cools through mixing and thus the filaments of cold gas grow 
continuously in mass. If the mixing of the cold and hot gas phases turns more violent and the energy input rate due to penetrating hot plasma becomes larger than the cooling 
rate, this process may also lead to the destruction of the filaments. 

On the other hand, according to the scenario proposed by \citet{churazov2013}, the filaments may be powered by the reconnection of the stretched magnetic fields in the 
wakes of AGN inflated buoyantly rising bubbles, and do not necessarily grow in mass. The fact that the ratios of the [\ion{C}{ii}]/(H$\alpha$+[\ion{N}{ii}]) emission appear similar 
in both filamentary nebulae and rotating disks indicates that, despite the different morphologies, the gas is heated and ionized by the same energy source in both types of 
systems. But because the cold gas in the disky nebulae is not distributed in the wakes of AGN inflated bubbles, the scenario proposed by \citet{churazov2013} could not 
explain their energy source. However, in the reconnection model, the powering of the cold gas could also be driven by the orbiting motion of the magnetized blobs/filaments 
through the hot plasma. This may lead to the sliding of the cold gas along the opposite-polarity magnetic fields and to the subsequent reconnection.

Despite its relatively low X-ray luminosity, the central pressure of the hot X-ray emitting plasma in the disky NGC~6868 is similar to the central pressures in the filamentary 
systems (see Table~\ref{deprojected}) indicating that in the absence of magnetic fields the heat flux from the hot into the cold phase would also be similar in both types of 
systems. It therefore appears that if the cold and hot phase come into direct contact - either due to turbulence, shearing motions, or the magnetic fields tied to the ionized phase 
becoming unstable - hot gas particles penetrating into the cold gas provide a viable mechanism for powering the filaments in all of the systems in our sample. The penetrating particles will heat and increase the degree of ionization of the molecular gas clouds present within the filaments, increasing their Jeans mass and slowing 
down the collapse of the gas clouds in the presence of magnetic fields.

This model, calculated by \citet{ferland2009} considering emission from an optically thin cell of gas, however, predicts [\ion{O}{i}]/[\ion{C}{ii}]$\sim21$, significantly higher than 
our observed range of 0.5--0.7. The ratios previously observed in the Perseus and Centaurus clusters, and in Abell~1068 and Abell~2597 are also low, in the range of 0.3--1 
\citep{edge2010,mittal2011,mittal2012}. \citet{mittal2012} conclude that these line ratios suggest that the lines are optically thick, implying a large reservoir of cold gas, which 
was not accounted for in previous inventories of the filament masses.  Our measured ratios are consistent with the values determined previously in normal and starburst 
galaxies \citep{malhotra2001}.

\subsection{Origin of the cold ISM}

\begin{figure}
\includegraphics[width=1\columnwidth,clip=t,angle=0.]{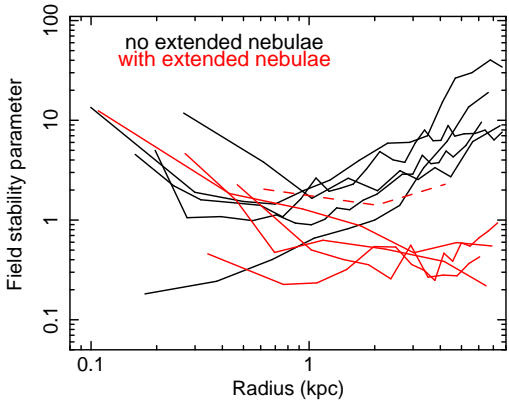}
\caption{The Field criterion,  $\field = {\kappa T/ (\nelec \nh \Lambda(T) r^2)}$, as a function of radius for the galaxies that display extended emission line nebulae and [\ion{C}
{ii}] line emission (shown in red) and for a sample of arguably cold-gas-poor systems (shown in black). The clear dichotomy, with the cold gas-rich systems being thermally 
unstable out to relatively large radii, indicates that the cold gas originates from the cooling hot phase. The galaxy NGC~6868, which shows evidence for a rotating disk of gas, 
is shown with a dashed line. } 
\label{stability}
\end{figure}

A hotly debated question in the literature is whether the cold gas seen in the central giant elliptical galaxies in clusters and groups, and in early type galaxies with X-ray 
haloes, originated from the radiative cooling of the hot plasma, from stellar mass loss, or whether it was accreted in mergers with gas rich galaxies. 

The difference between the entropy profiles of the systems with cool
gas and those without it in Fig.~\ref{entropies} supports
the argument that the cold gas originates by cooling from the hot
phase. In more massive clusters, H$\alpha$ filaments are always co-spatial with soft X-ray emitting $\sim$0.5~keV gas 
\citep{sanders2007,sanders2008,sanders2009,sanders2009b,deplaa2010,werner2011,werner2010,werner2013}, indicating that the cold gas must be related to the hot 
phase.
Furthermore, significant star formation and line
emitting gas around the central galaxy are only found in systems where
the central cooling time is short, or, equivalently, the central
entropy is low \citep{rmn08, cdv08}.  

The parameter
\begin{equation} \label{eqn:field_crit}
\field = {\kappa T \over \nelec \nh \Lambda(T) r^2},
\end{equation}
where $\kappa$ is the thermal conductivity, $T$ is the temperature,
$\Lambda$ is the cooling function, and $\nelec$ and $\nh$ are the
electron and hydrogen number densities, respectively, is a measure of
the ratio of the conductive heating rate to the radiative cooling rate
on scales comparable to $r$, i.e. the ``Field criterion'' for thermal
instability \citep{field1965}. Fig.~\ref{stability} shows the Field criterion, $\field$, as a
function of the radius, where the systems with substantial quantities
of cold gas are plotted in red. There is a clear dichotomy with the cold-gas-rich 
systems remaining unstable out to relatively large radii.  This result strongly indicates that the cold gas is produced
chiefly by thermally unstable cooling from the hot phase. 
A similar result was found for a sample of central galaxies in more massive clusters by
\citet{vcd08}, who found minimum values of $\field \lesssim 5$ in all
systems with star formation. \citet{vcd08} argued that the effective conductivity
is suppressed by a factor of about 5 by magnetic fields, in which case all of their
systems with young stars are thermally unstable by the Field criterion.

The Field criterion alone is insufficient to ensure that the gas is
thermally unstable.  In a spherical atmosphere, the gas is supported by
buoyancy, so that an overdense blob tends to fall inwards on the
free-fall timescale towards its equilibrium position where the
surrounding gas has the same specific entropy and requires the same
specific heating rate. This suppresses the growth of thermal
instability unless the cooling time is comparable to the free-fall
time or shorter \citep{cfn80, bs89}. Using numerical simulations, \citet{smq12}
found that thermal instability is only significant when the cooling
time is less than $\sim 10$ free fall times \citep[see also][]{mccourt2012,kunz2012,gaspari2012,gaspari2013}.  
However, this condition can be circumvented if an overdense gas blob is prevented from falling
to its equilibrium position by the magnetic fields which pin the cloud to the ambient ICM \citep{nulsen1986}, or by rotational support in addition to
buoyancy.  

If a cooling cloud has sufficient angular
momentum that is retained while it cools, it can cool unstably onto a
non-radial orbit.  This provides an alternative condition for thermal
instability, that the viscous diffusion length in a cooling time,
$\sqrt{\nu \tcool}$, where $\nu$ is the kinematic viscosity and
$\tcool$ is the cooling time, must be smaller than the size of a
cloud.  Analogous to the Field criterion, we define the stability
parameter
\begin{equation} \label{eqn:visc_crit}
\viscint = {\nu \tcool \over r^2},
\end{equation}
which needs to be smaller than about unity for thermal instability to
develop on scales comparable to $r$.  Viscous stresses in a plasma
\citep{b65} are much less sensitive to the structure of the magnetic
field than conduction \citep[e.g.,][]{nm13} and, although the stresses
have a different tensor character in magnetized and unmagnetized
plasmas, the viscosity itself is the same.  Because the underlying
process is diffusion controlled by Coulomb collisions in both cases
(electrons for $\field$ and ions for $\viscint$), the ratio of
these two parameters is almost constant at $\viscint / \field \simeq
0.0253$.  Numerical simulations are needed to determine a more
accurate stability threshold, but we expect that the viscous diffusion
length needs to be somewhat smaller than $r$ for instability by this
mechanism.  The condition $\field < 5$ translates to $\viscint
\lesssim 0.13$, or the diffusion length in one cooling time being
$\lesssim 0.36 r$.  Although it is unclear which stability threshold
takes precedence, we conclude that it is sufficient to test only one
of these parameters, which we take to be $\field$, for stability.

Our {\it Herschel} data reveal that in two galaxies, NGC~6868 and NGC~7049, the cold gas forms rotating disks with radii $r\sim3$~kpc. 
The disks of cold gas in these systems rotate with orbital velocities of up to $250$~km~s$^{-1}$.
If this gas cooled from the hot phase, its rotation implies that the hot atmospheres of these systems have significant net angular momentum and the hot gas cools onto non-
radial orbits. In the case of NGC~6868, this angular momentum may have been provided by gas sloshing indicated by the {\it Chandra} data \citep{machacek2010}.
Because the atmosphere is aspherical, heating originating from the center of the galaxy cannot balance the cooling rate locally throughout the atmosphere and so should not 
be expected to stop hot gas from cooling onto an extended disk. Because of this, and because rotational support prevents infall, instabilities may develop more easily and
the gas may become thermally unstable at higher $\field$ than in non-rotating systems. This is consistent with our results presented in Fig.~\ref{stability}, where $\field$ for 
NGC~6868, shown with a dashed line, is higher than $\field$ for the other systems containing non-rotating cold gas. The Field criterion still needs to be met, but the behavior of 
the stability parameter in NGC 6868 also indicates that the viscous stability criterion takes precedence.

However, in the context of the radiative cooling model, the presence of dust and PAHs within the cold gas clouds \citep{panuzzo2011} is intriguing. It suggests that at least 
some of the cold material originated from stellar mass loss \citep[see also][for galaxy clusters]{voitdonahue2011,donahue2011}. Our understanding of dust formation is, 
however, limited and it cannot be ruled out that dust may somehow form within the cold gas clouds \citep{fabian1994b}. 
Recent {\it Herschel} observations suggest that in the most massive galaxies, the dust mass is unconnected to the stellar populations, leading to the suggestion that the dust 
and the cold gas in these systems have been acquired externally \citep{smith2012}. 
\citet{rawle2012} show that the dust-to-stellar mass ratios in BCGs without line emitting nebulae are much smaller than in systems with cold gas. In the absence of cold gas 
clouds, the dust produced by stellar mass loss will get destroyed by sputtering and the tenuous gas, lost by stars, may get more easily mixed with and assimilated into the hot 
ISM. Therefore, if a galaxy loses all of its cold gas content (e.g. due to AGN activity), then stellar mass loss will likely become inefficient in rebuilding its cold gas reservoir. 
A galaxy that loses its cold gas content will therefore remain free of gas and dust until it again accretes cold material from the cooling hot phase or in a wet merger. This may 
have happened to NGC~1399 and NGC~4472. On the other hand, if a galaxy already contains clouds of cold gas (it either never lost its cold ISM or it accreted some cold 
material externally) then these clouds and filaments will help preserving the products of stellar mass loss (both the gas and the dust) by assimilating them as they plow through 
the galaxy. 

In the systems with filamentary cold gas, the AGN jets and the buoyant bubbles of relativistic plasma seem to interact with the filaments, dragging them out from the center of 
the galaxy. As these filaments move through the hot ISM they may get penetrated by hot particles, which deposit energy into the cold gas heating and ionizing it. This 
interaction can both destroy the cold gas and contribute to the growth of its mass, depending on the energy input rate from the hot into the cold phase.  In the context of this 
model, the total gas/dust mass will not correlate with the stellar mass of the galaxy, but it will depend on the balance between heating and cooling and on the interaction of the 
cold nebulae with the AGN \citep[see also][]{mathews2013}.

\subsection{Fueling the AGN}

The jet-powers, determined from the work required to inflate bubbles of relativistic plasma associated with the cavities in the hot X-ray emitting atmospheres, do not 
increase with the amount of cold gas in these galaxies. On the contrary, the two galaxies that lack any significant reservoirs of cold gas have the largest jet-powers in our 
sample. These galaxies, NGC~1399 and NGC~4472, also have the largest hot ISM densities in their cores. Accretion of hot gas in galaxies with high core densities can 
naturally result in higher jet powers, as has been demonstrated by the correlation between the Bondi accretion rates of hot gas and the observed jet powers in such ellipticals 
\citep{allen2006}. The pressure and entropy distributions of the hot ISM in the cores of these two galaxies look remarkably regular, with the X-ray cavities/radio-lobes 
appearing at larger radii of $\sim$4~kpc and $\sim$8~kpc for NGC~4472 and NGC~1399, respectively.

Other nearby giant elliptical galaxies that appear morphologically relaxed in their cores, such as NGC~4649, NGC~1407, NGC~4261, or NGC~1404, also do not display 
extended optical emission line nebulae \citep{macchetto1996,baldi2009,tremblay2009,moustakas2010}, indicating that they too are relatively depleted of cold gas. 
Interestingly, some of these morphologically relaxed cold-gas-poor systems, such as NGC~4261, have very large jet powers, with jets puncturing through the X-ray emitting 
galactic atmosphere to form giant lobes in the surrounding intragroup medium \citep{osullivan2011}. For others, such as NGC~1404, NGC~4649, and NGC~1407, however, 
the current jet powers are more modest \citep{shurkin2008,giacintucci2012}.

Galaxies containing large reservoirs of cold gas, on the other hand, in general display strongly disturbed X-ray morphologies, possibly relating to earlier AGN outbursts 
impacting the cores of these systems. In principle, the presence of cold gas may affect both accretion and jet heating in these systems. Accretion of clumpy multiphase gas (hot 
plasma mixed with cooler, filamentary gas) may result in variable accretion rates and power output of the AGN jets, potentially triggering sporadic, larger outbursts than would 
be typical from the steady accretion of hot gas alone. Because of the relatively small volume filling fraction of this cooler gas, these outbursts may be relatively short 
\citep[arguments for sporadic cold/multi-phase accretion in radio mode AGN in galaxy clusters have also been presented by][]{rafferty2006,russell2013}.

As demonstrated by \citet{morganti2013}, radio-mechanical feedback does not only operate on the hot tenuous atmospheres of galaxies but also on the dense cold gas, which 
can be accelerated by jets to high speeds. 
Recent observations using ALMA also show that radio mode AGN activity can drive massive outflows of molecular gas from BCGs \citep{russell2013b,mcnamara2013}.
In systems with large cold gas reservoirs, jets may be more easily slowed by coupling to this material, and may therefore deposit most of their energy at smaller radii. This will 
result in more disturbed central thermodynamic distributions and, in particular may decrease the central hot ISM density, resulting in lower hot gas accretion rates between the 
relatively brief periods of multiphase accretion. Any reduction in the accretion rates from the hot gas would result in smaller jet powers, consistent with the observations 
presented here.

Numerical simulations \citep{wagner2011,wagner2012} have shown that radio jets can be efficient in clearing the central regions of galaxies of their cold gas, provided that the 
cold gas clouds are porous, clumpy, and have small volume filling fractions, with individual clouds being relatively small, which again are characteristics consistent with the 
observations discussed here. Powerful jets may thus eventually destroy or remove most of the cold gas in the cores of the galaxies, allowing the jets to propagate further out 
and deposit their energy at larger radii, increasing the entropy of the hot galactic atmospheres. This in turn will allow the density of the hot gas in the core of the galaxy to 
increase, thus increasing the jet power. The central regions of galaxies cleared of cold gas may acquire relaxed morphologies within the central $r\sim5$~kpc over $
\sim5\times10^7$~yr or a few sound crossing times.
The similarity of the thermodynamic profiles of elliptical galaxies with highly relaxed X-ray morphologies \citep[see black data points in Fig.~\ref{entropies} and][]{werner2012} 
indicates that, after elliptical galaxies are cleared of cold gas, jet heating and cooling of their hot atmospheres comes to an equilibrium in an approximately universal manner. 
The galaxies may maintain these equilibria until they are disturbed by a merger event. 

\section{Conclusions}
\label{conclusions}
Using FIR, optical, and X-ray data, we study the ISM in eight nearby, X-ray and optically bright, giant elliptical galaxies, all central dominant members of relatively low 
mass groups. We find that:
\begin{itemize}
\item All systems with extended H$\alpha$ emission in our sample (6/8 galaxies) display significant [\ion{C}{ii}] line emission that traces $\sim$100~K gas. 
\item This emission is co-spatial with the optical H$\alpha$+[\ion{N}{ii}] emitting nebulae and the lowest entropy soft X-ray emitting plasma.
\item These systems have similar [\ion{C}{ii}]/(H$\alpha$+[\ion{N}{ii}]) ratios of 0.4--0.8, indicating that the [\ion{C}{ii}] and H$\alpha$+[\ion{N}{ii}] emission are powered by the 
same energy source. The likely dominant source of energy is the hot X-ray emitting plasma penetrating into the cold gas. \\
\item The entropy profiles of the hot galactic atmospheres show a clear dichotomy, with the systems displaying extended emission line nebulae having lower entropies beyond 
$r\gtrsim1$~kpc than the cold-gas-poor systems. We show that while the hot atmospheres of the cold-gas-poor galaxies are thermally stable outside of their innermost cores, 
the atmospheres of the cold-gas-rich systems are prone to cooling instabilities. This provides considerable weight to the argument that cold gas in giant ellipticals is produced 
chiefly by cooling from the hot phase. We show that cooling instabilities may develop more easily in rotating systems and discuss an alternative condition for thermal instability 
for this case.
\item The hot atmospheres of cold-gas-rich galaxies display disturbed morphologies indicating that the accretion of clumpy multiphase gas in these systems may result in 
variable power output of the AGN jets, potentially triggering sporadic, larger outbursts. The jets may be more easily slowed by coupling to the dense cold material and may 
therefore deposit most of their energy at smaller radii, resulting in the observed disturbed central thermodynamic distributions. 
\item In the two cold-gas-poor, X-ray morphologically relaxed galaxies of our sample, NGC~1399 and NGC~4472, powerful AGN outbursts may have destroyed or removed 
most of the cold gas from the cores, allowing the jets to propagate and deposit most of their energy further out, increasing the entropy of the hot galactic atmospheres and 
leaving their cores relatively undisturbed. These observations indicate that radio-mechanical AGN feedback is likely to play a crucial role in clearing giant elliptical galaxies of 
their cold gas, keeping them `red and dead'.
\end{itemize}

\section*{Acknowledgments}
This work is based in part on observations made with Herschel, a European Space Agency Cornerstone Mission with significant participation by NASA. Support for this work 
was provided by NASA through award number 1428053 issued by JPL/Caltech. This work is based in part on observations obtained at the Southern Astrophysical Research 
(SOAR) telescope, which is a joint project of the Minist\'{e}rio da Ci\^{e}ncia, Tecnologia, e Inova\c{c}\~{a}o (MCTI) da Rep\'{u}blica Federativa do Brasil, the U.S. National 
Optical Astronomy Observatory (NOAO), the University of North Carolina at Chapel Hill (UNC), and Michigan State University (MSU). SWA acknowledges support from the U.S. 
Department of Energy under contract number DE-AC02-76SF00515. MR acknowledges the NSF grant AST 1008454 and NASA ATP grant (12-ATP12-0017).

\bibliographystyle{mnras}
\bibliography{clusters}

\appendix
\section{Plots of the FIR line spectra\\}

\newpage

\begin{figure*}
\begin{minipage}{0.24\textwidth}
\includegraphics[width=1.2\textwidth,clip=t,angle=180.]{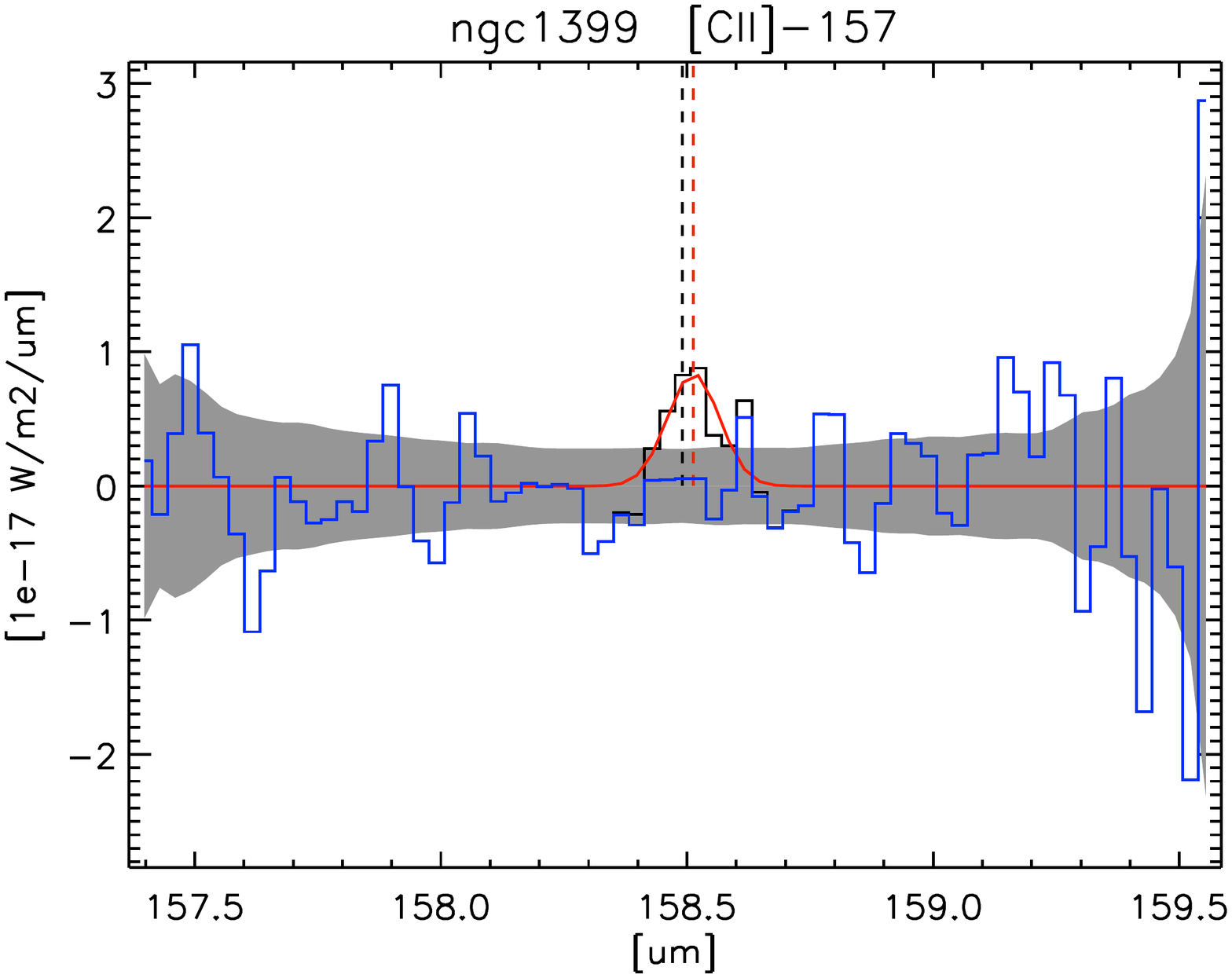}
\end{minipage}
\begin{minipage}{0.24\textwidth}
\includegraphics[width=1.2\textwidth,clip=t,angle=180.]{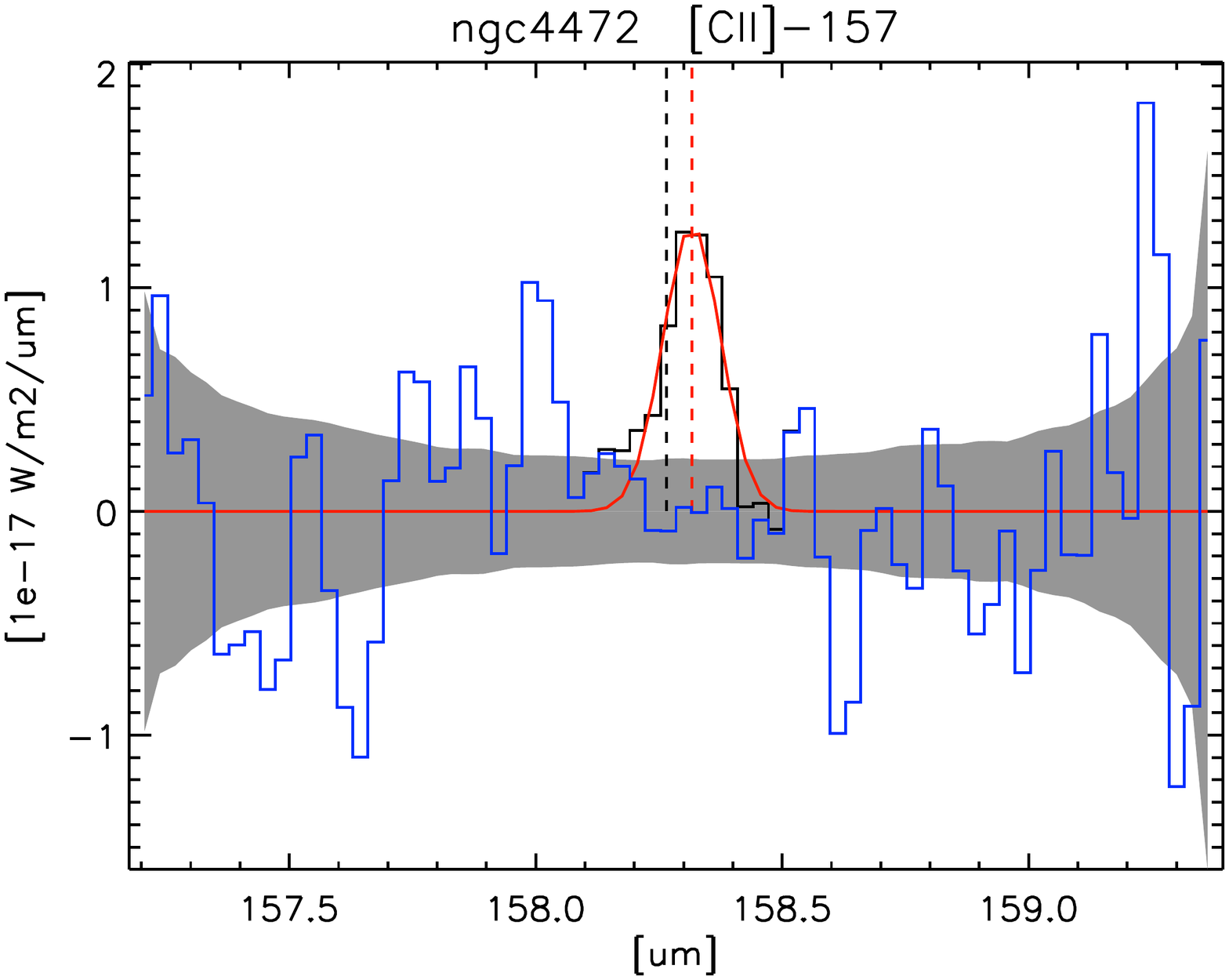}
\end{minipage}
\begin{minipage}{0.24\textwidth}
\includegraphics[width=1.2\textwidth,clip=t,angle=180.]{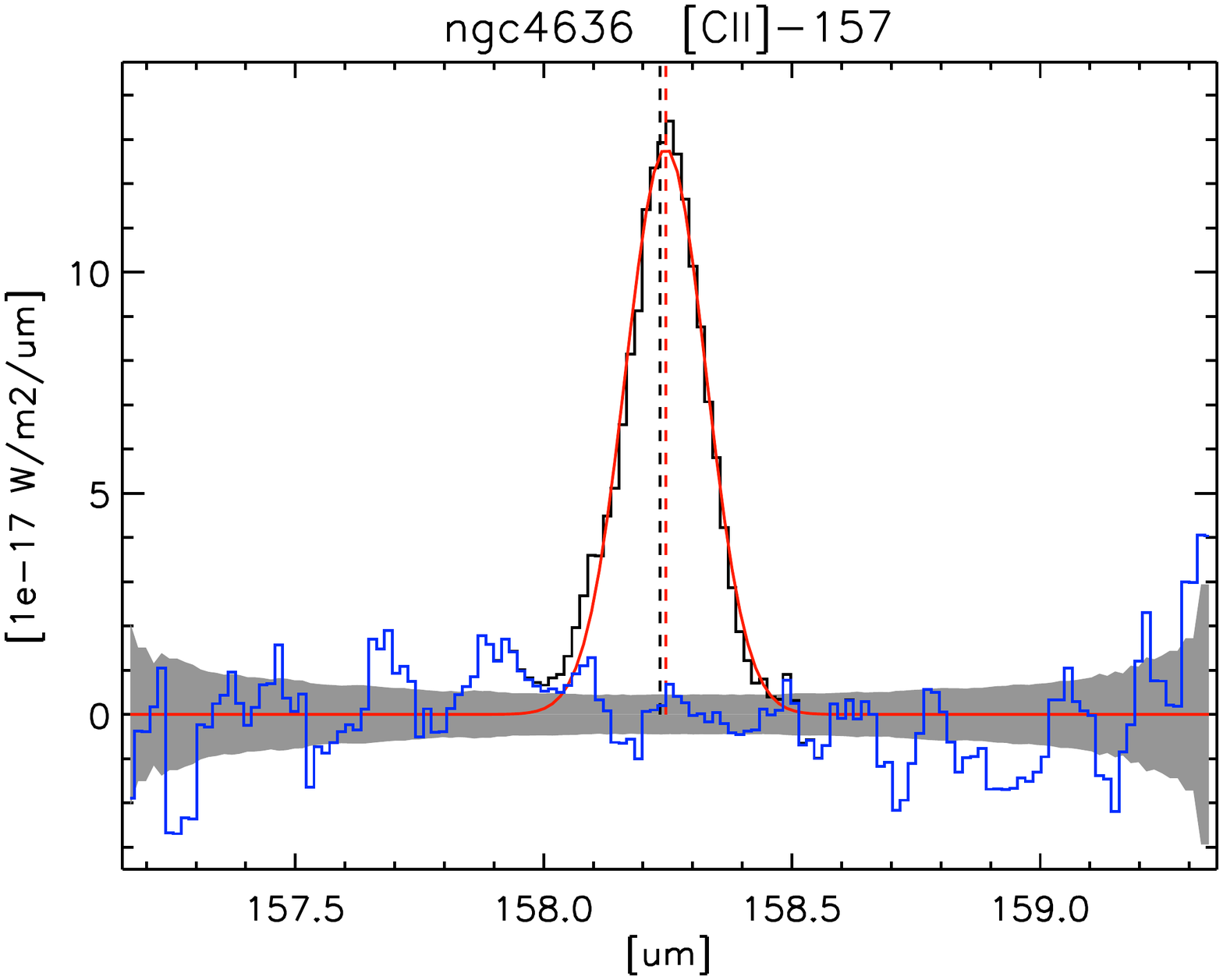}
\end{minipage}
\begin{minipage}{0.24\textwidth}
\begin{overpic}[width=1.2\textwidth,clip=t,angle=180.]{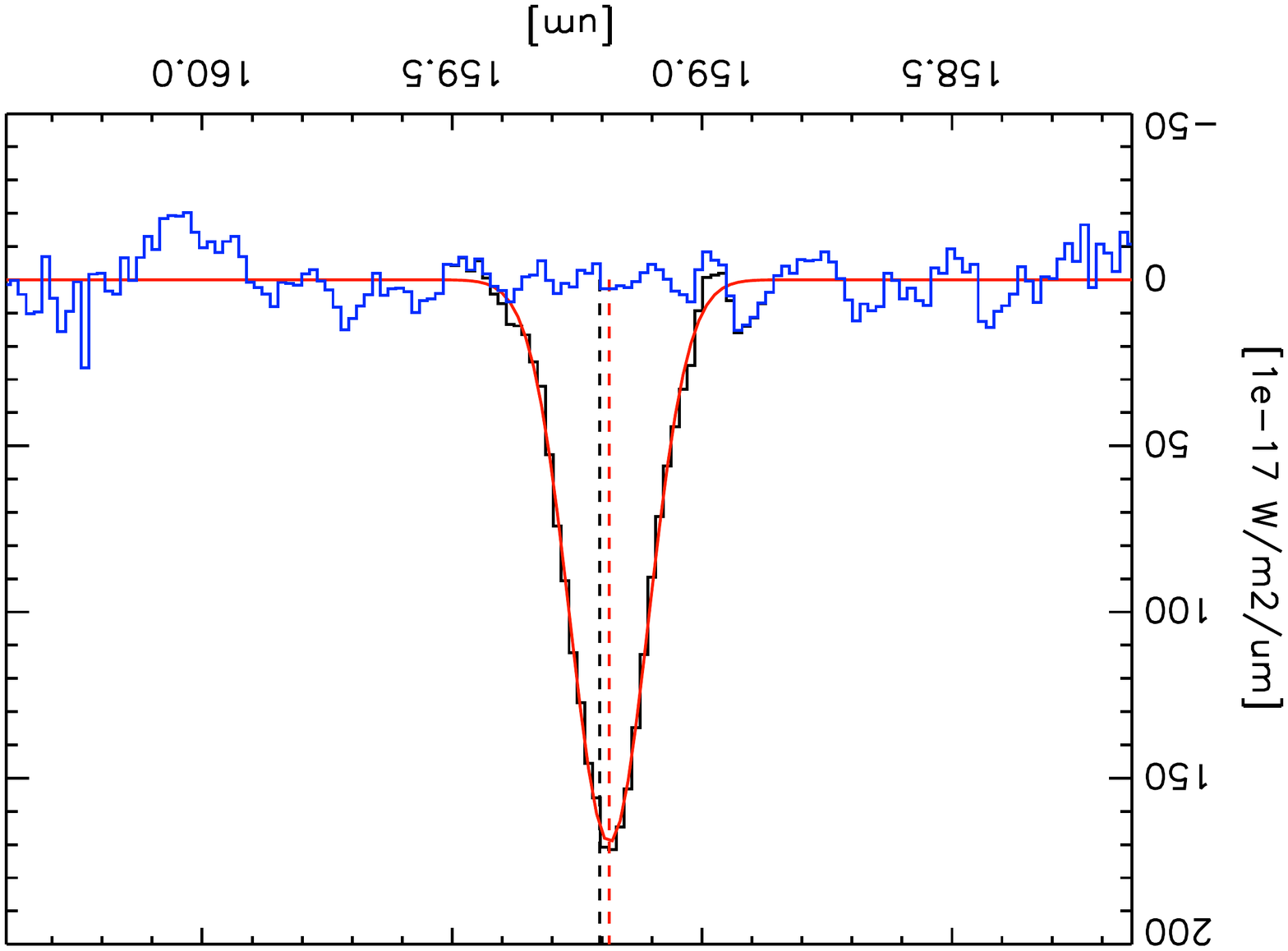}
\put(50,68){[CII]}
\put(65,68){NGC 5044}
\end{overpic}
\end{minipage}
\begin{minipage}{0.24\textwidth}
\includegraphics[width=1.2\textwidth,clip=t,angle=180.]{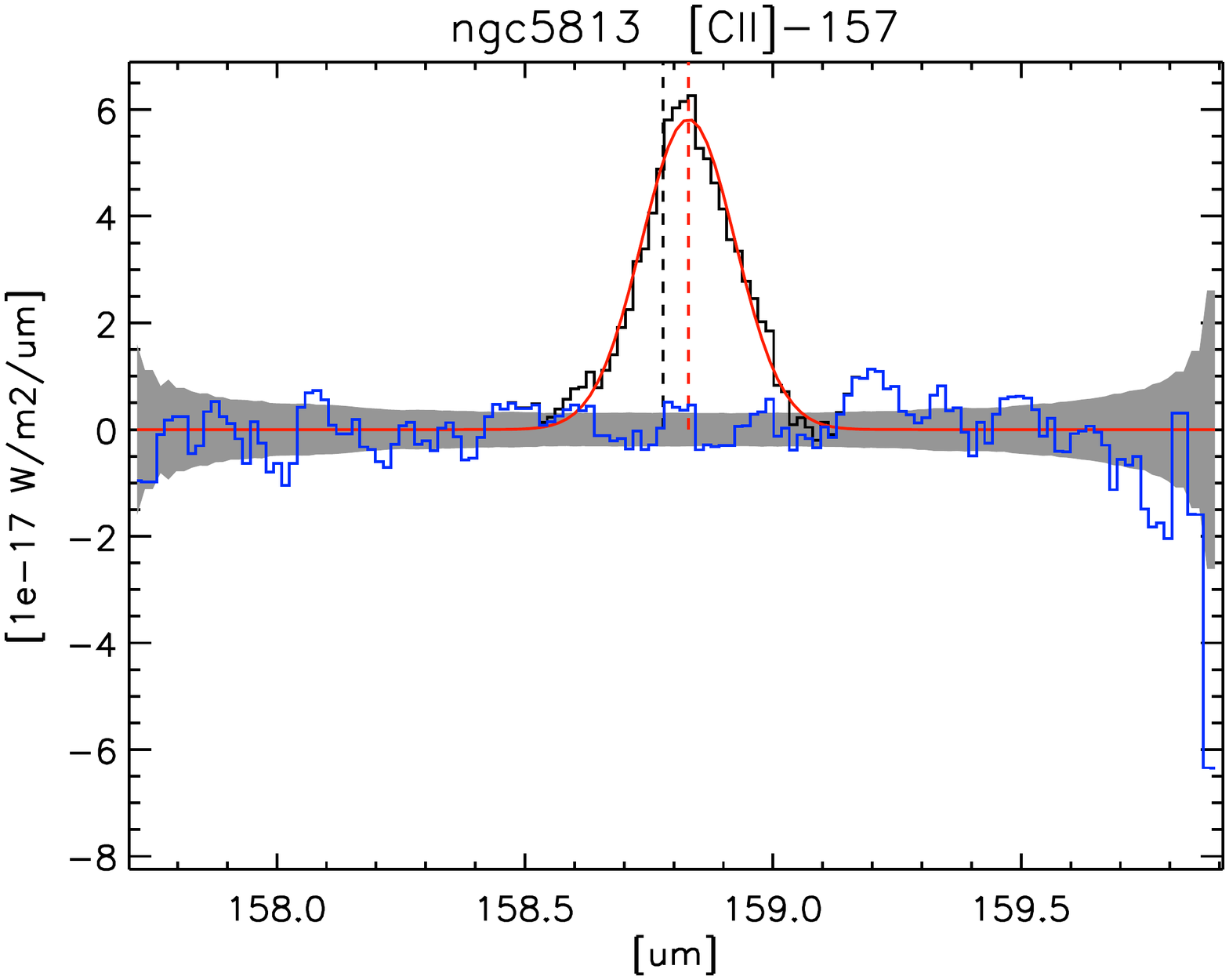}
\end{minipage}
\begin{minipage}{0.24\textwidth}
\includegraphics[width=1.2\textwidth,clip=t,angle=180.]{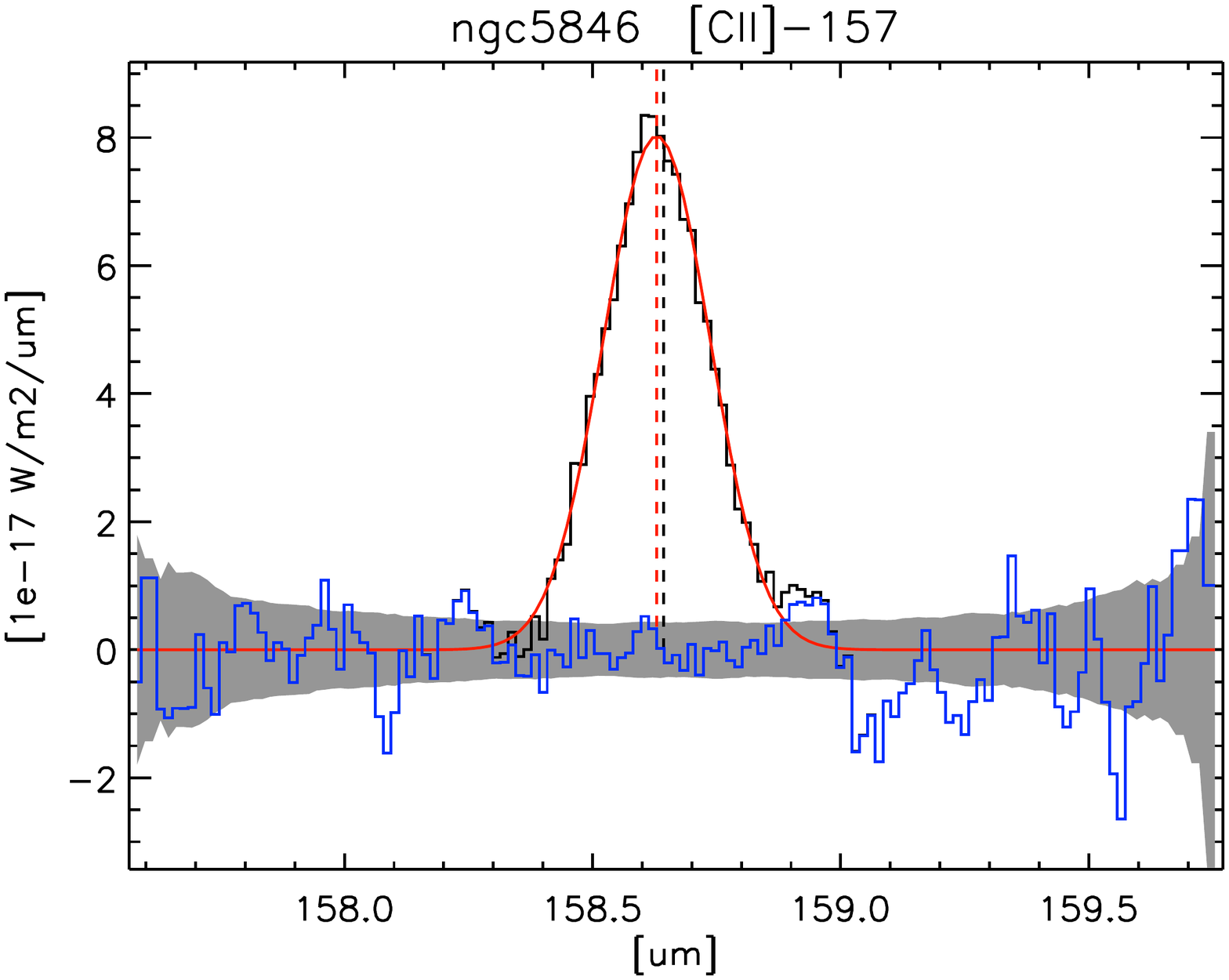}
\end{minipage}
\begin{minipage}{0.24\textwidth}
\includegraphics[width=1.2\textwidth,clip=t,angle=180.]{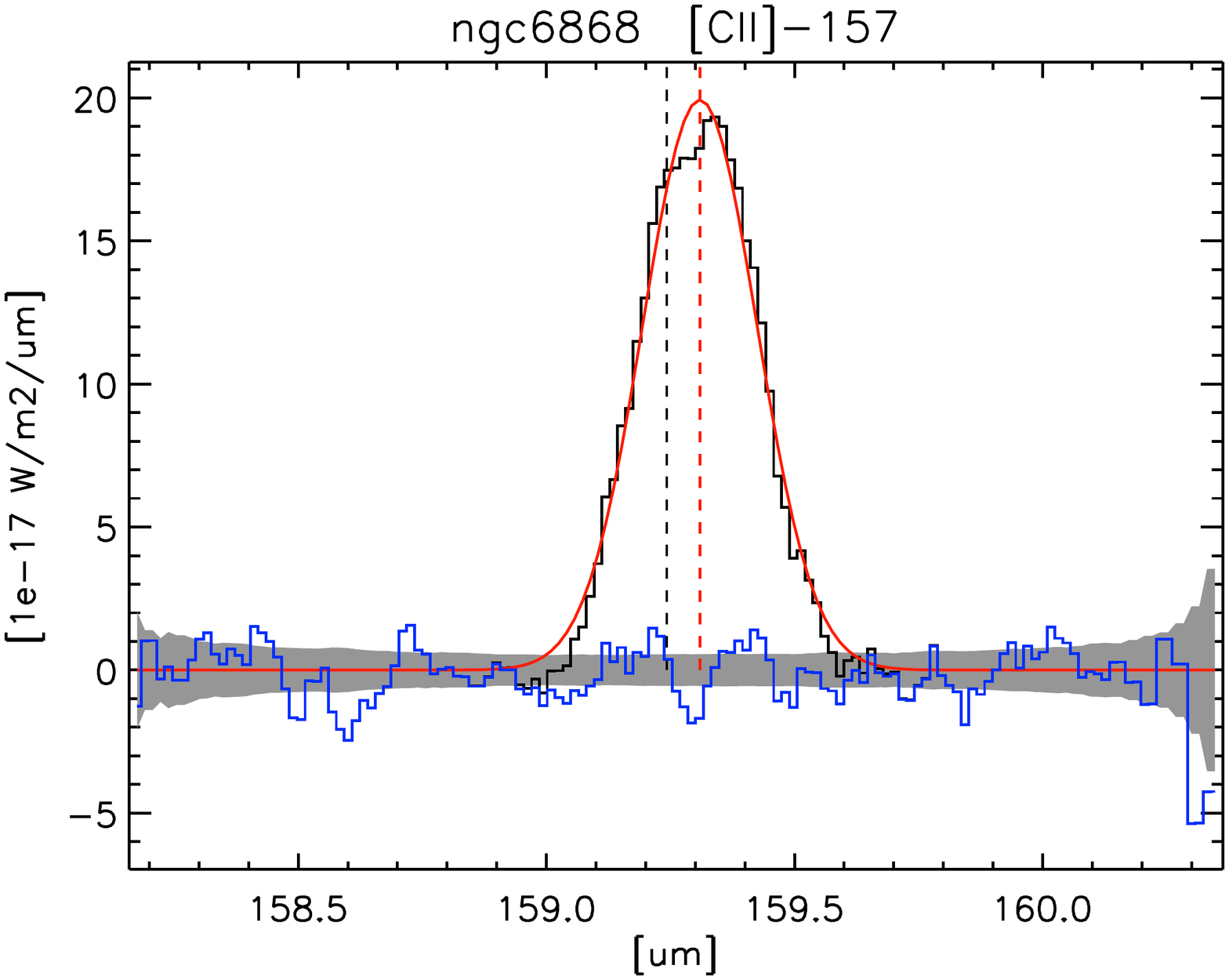}
\end{minipage}
\begin{minipage}{0.24\textwidth}
\includegraphics[width=1.2\textwidth,clip=t,angle=180.]{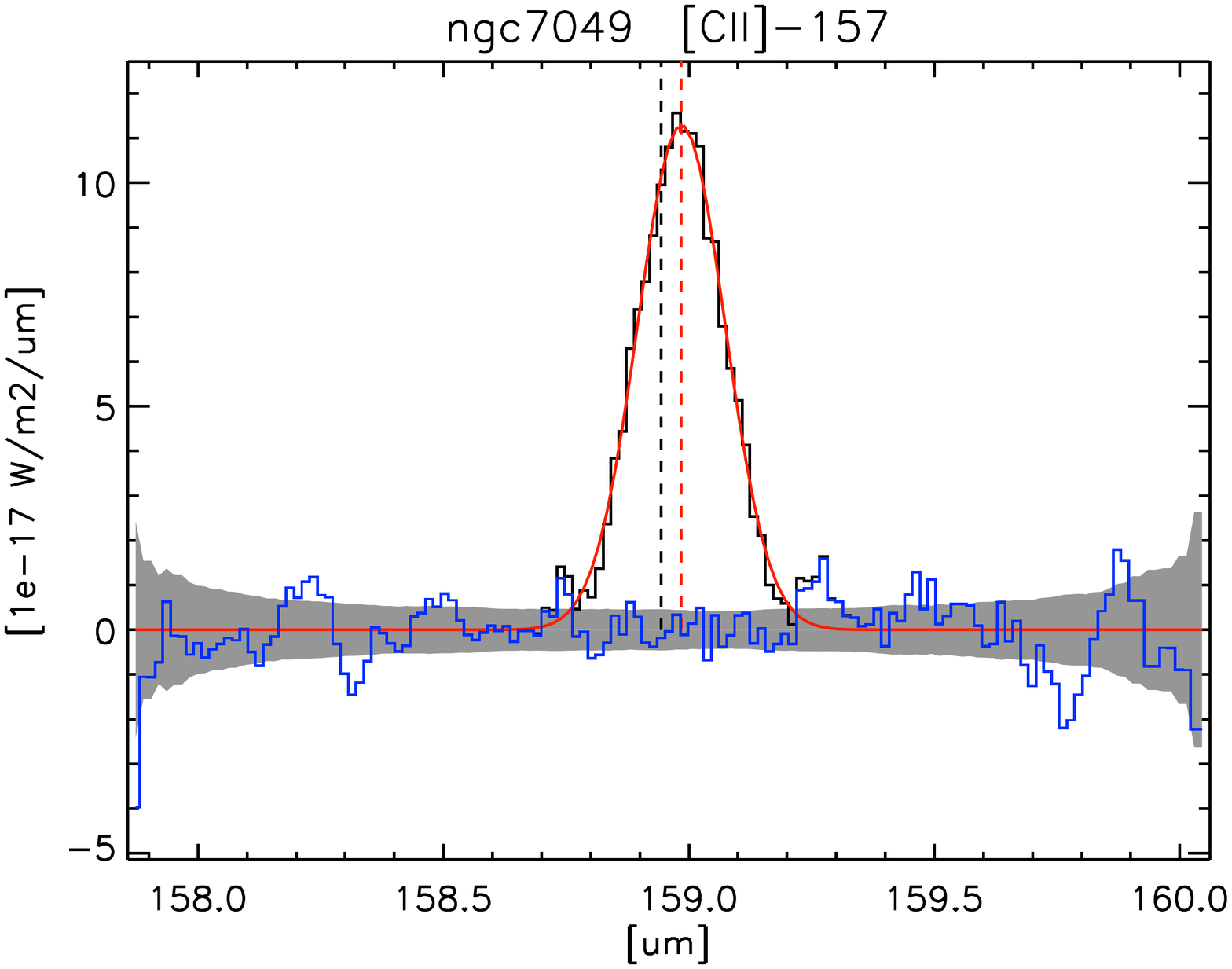}
\end{minipage}
\caption{The FIR [\ion{C}{ii}]$\lambda157\mu$m line obtained from the central spaxel ($9.4\times9.4$~arcsec$^2$) of the {\it Herschel} PACS rebinned data cube, except 
for NGC~5044 where the spectrum was obtained from an $80\times80$ arcsec wide box. The spectra for NGC~1399 and NGC~4472 were re-binned in wavelength.
The blue line indicates the residuals between the data and the model. The grey shaded area indicates the detector noise. No reliable noise product is available for the 
raster scan data of NGC~5044. The red and black dashed lines show the expected (for gas at rest at the redshift of the galaxy) and the observed best fit line centroids, 
respectively.} 
\label{fig:C2sample}
\end{figure*}

\begin{figure*}
\begin{minipage}{0.24\textwidth}
\includegraphics[width=1.2\textwidth,clip=t,angle=180.]{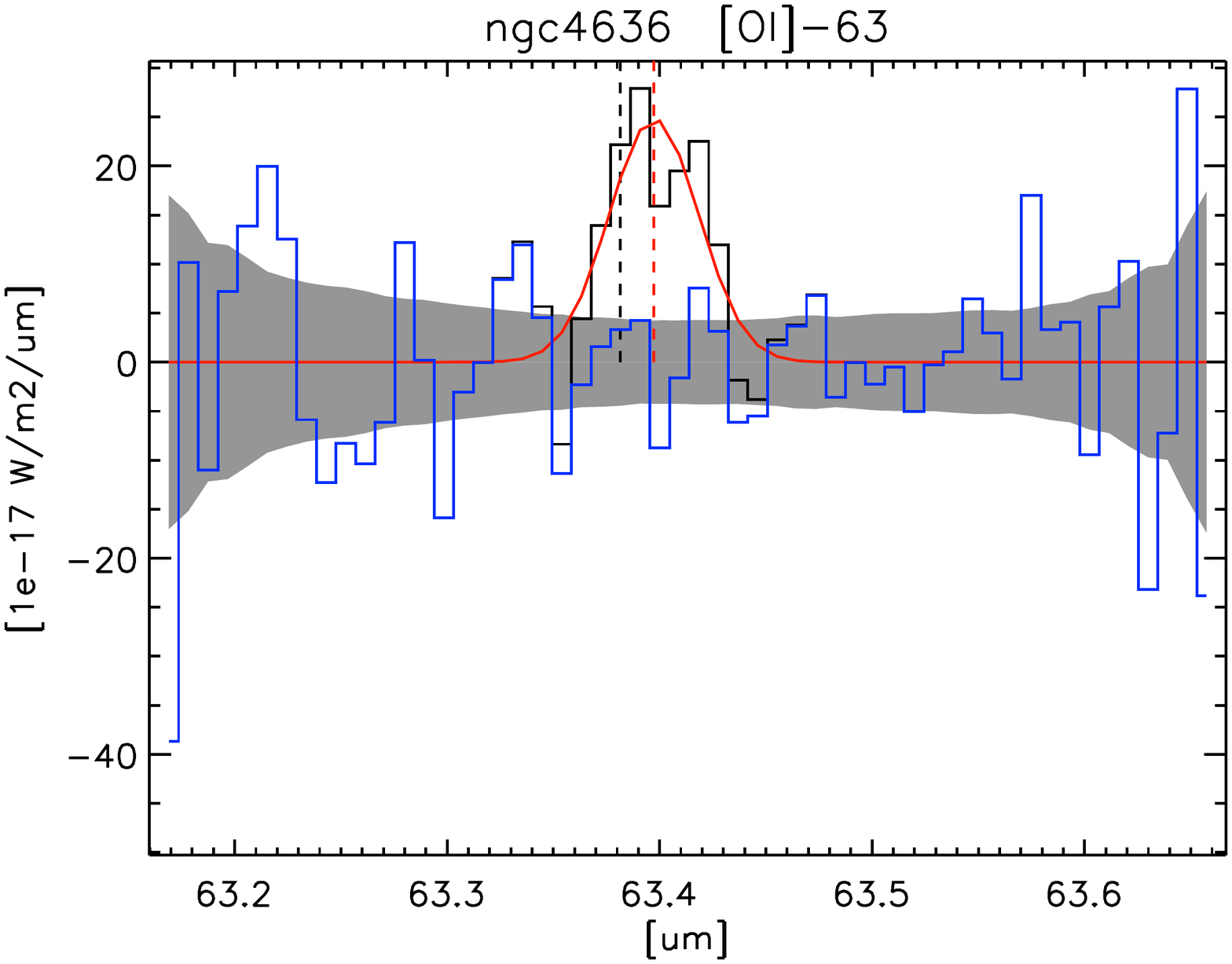}
\end{minipage}
\begin{minipage}{0.24\textwidth}
\includegraphics[width=1.2\textwidth,clip=t,angle=180.]{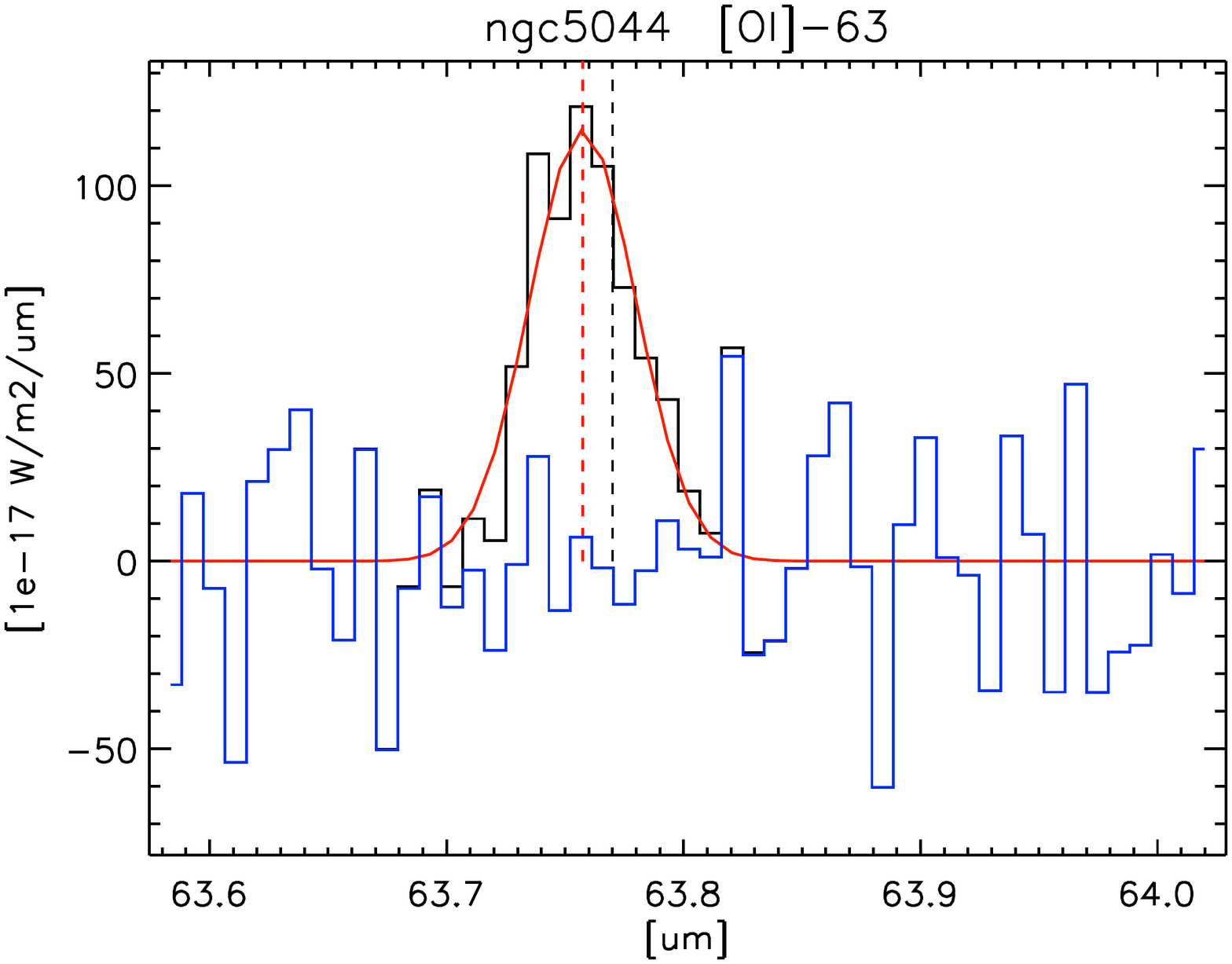}
\end{minipage}
\begin{minipage}{0.24\textwidth}
\includegraphics[width=1.2\textwidth,clip=t,angle=180.]{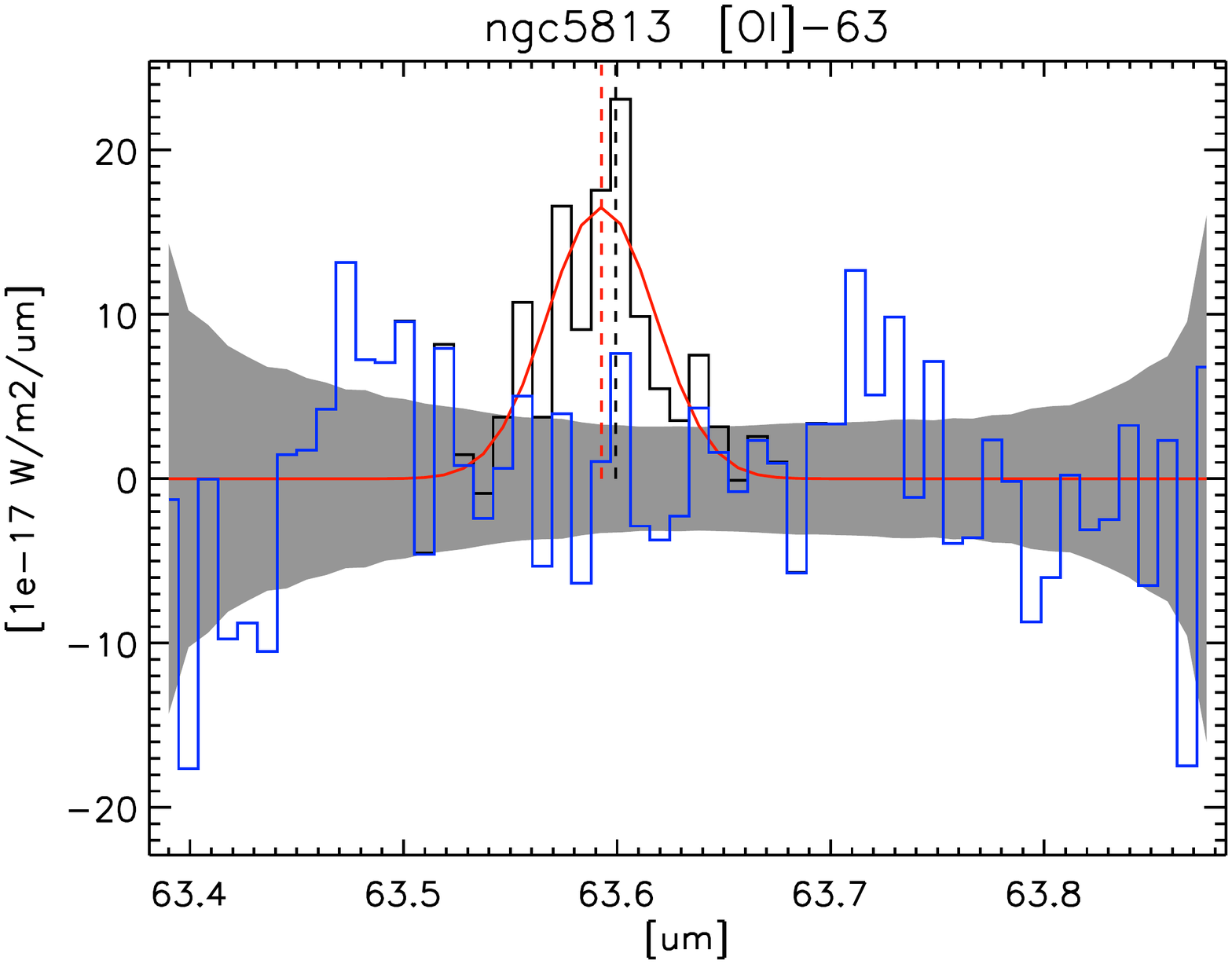}
\end{minipage}
\begin{minipage}{0.24\textwidth}
\includegraphics[width=1.2\textwidth,clip=t,angle=180.]{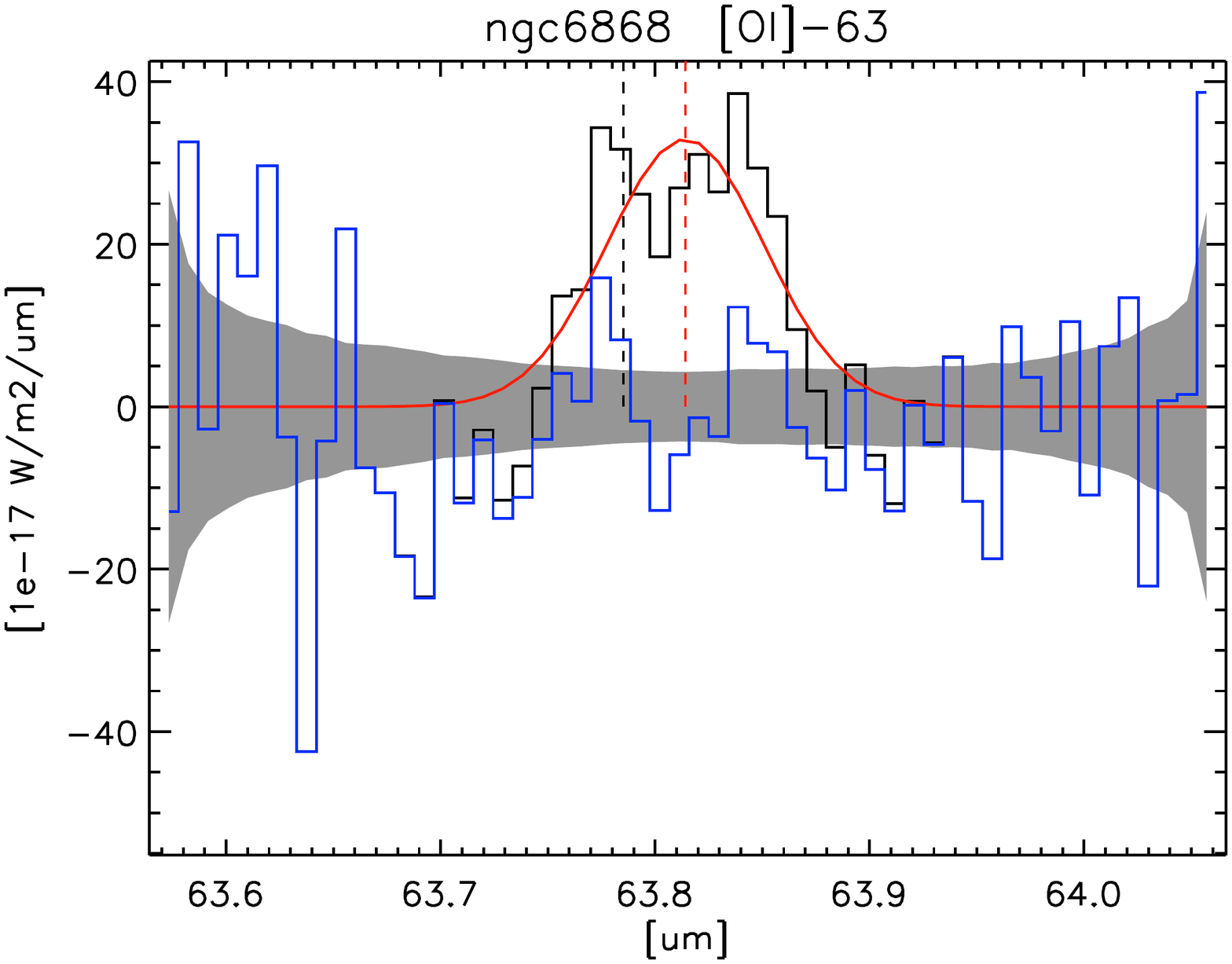}
\end{minipage}
\caption{The FIR [\ion{O}{i}]$\lambda63.2\mu$m line obtained from the central spaxel ($9.4\times9.4$~arcsec$^2$) of the {\it Herschel} PACS rebinned data cube, except 
for NGC~5044 where the spectrum was obtained from a 24 arcsec wide box. The spectra were re-binned in wavelength. The blue line indicates the residuals between the 
data and the model. The grey shaded area indicates the detector noise. No reliable noise product is available for the raster scan data of NGC~5044. Because the 
noise is a strong function of the wavelength, increasing towards the edges of the observed wavelength band, the noise in the adjacent bands may reach an 
amplitude that is comparable to that of the spectral lines, with the lines still remaining statistically significant. The red and black dashed lines show the expected (for gas at rest 
at the redshift of the galaxy) and the observed best fit line centroids, respectively. } 
\label{fig:O1sample}
\end{figure*}

\begin{figure*}
\begin{minipage}{0.32\textwidth}
\includegraphics[width=1.2\textwidth,clip=t,angle=180.]{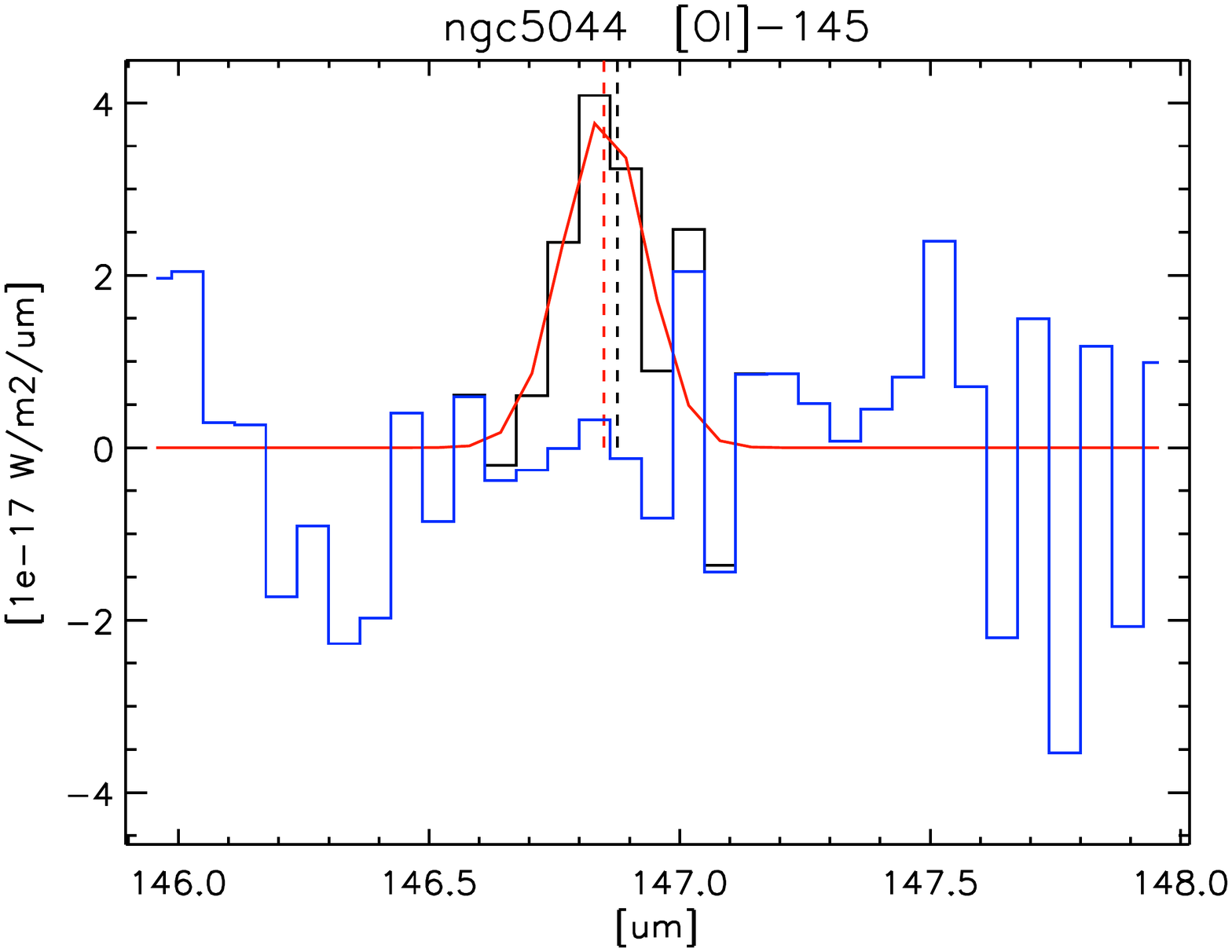}
\end{minipage}
\begin{minipage}{0.32\textwidth}
\includegraphics[width=1.2\textwidth,clip=t,angle=180.]{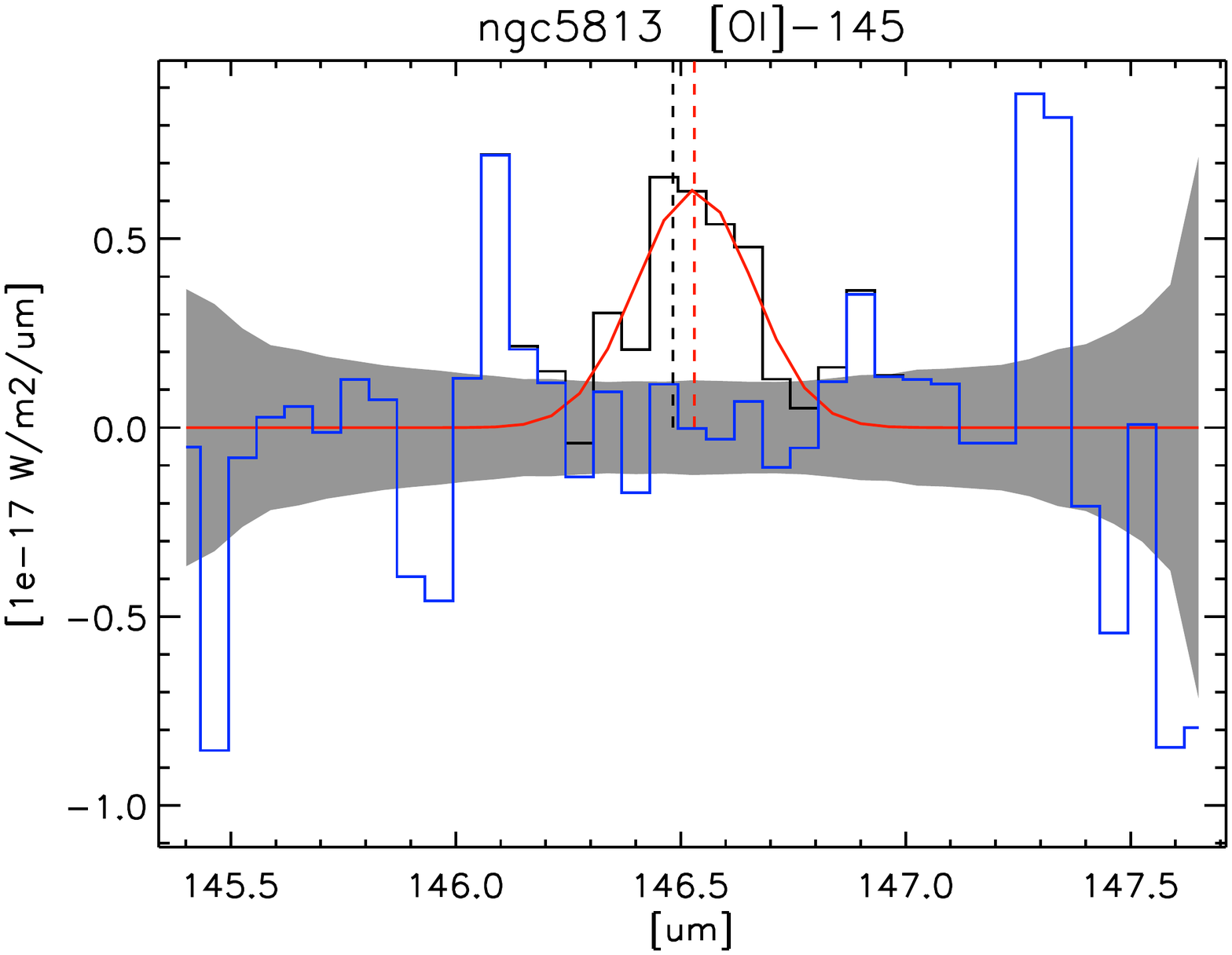}
\end{minipage}
\begin{minipage}{0.32\textwidth}
\includegraphics[width=1.2\textwidth,clip=t,angle=180.]{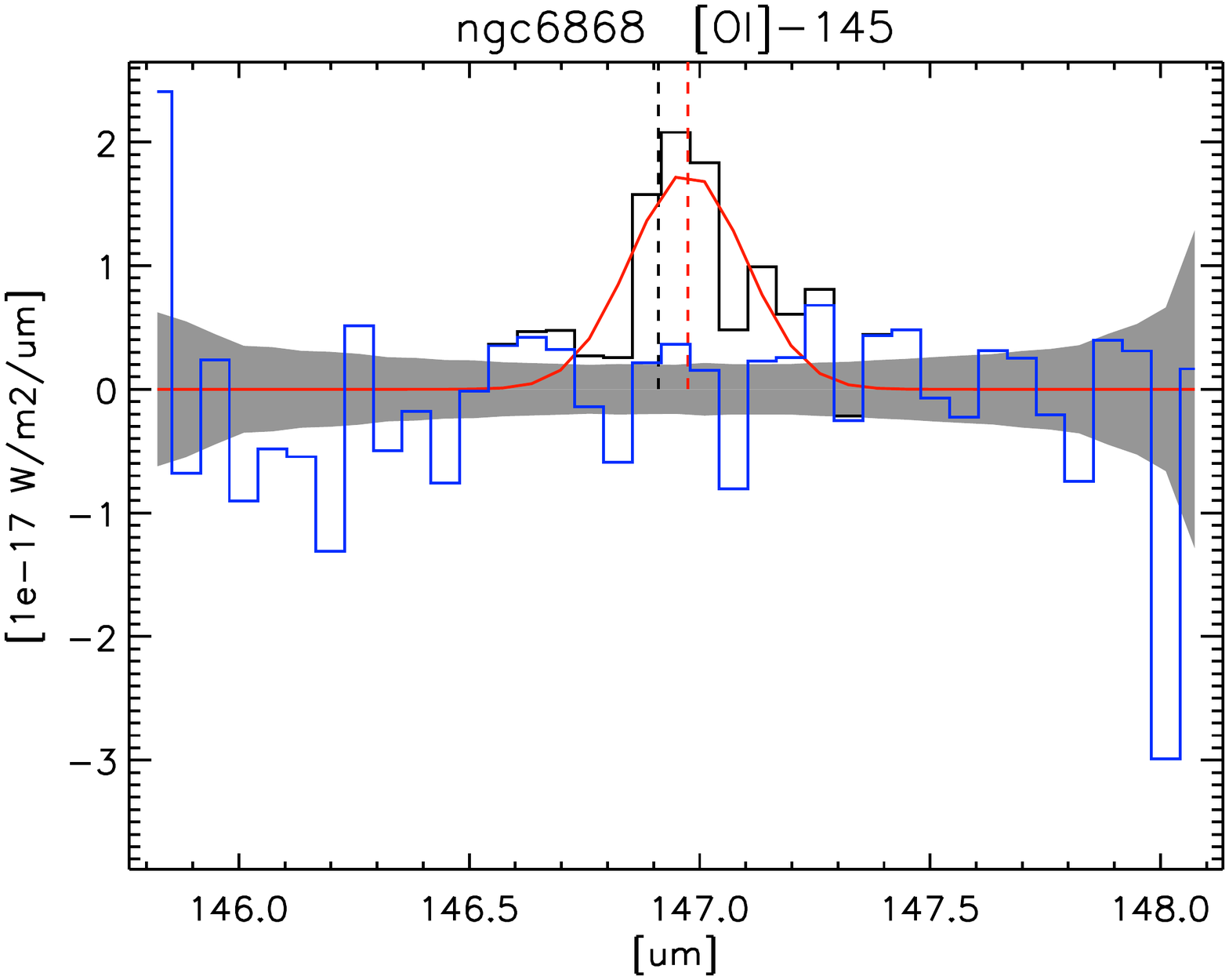}
\end{minipage}
\caption{The FIR [\ion{O}{i}b]$\lambda145.5\mu$m line obtained from the central spaxel ($9.4\times9.4$~arcsec$^2$) of the {\it Herschel} PACS rebinned data cube,
except for NGC~5044 where the spectrum was obtained from a 24 arcsec wide box. The spectra were re-binned in wavelength. The blue line indicates the residuals 
between the data and the model. The grey shaded area indicates the detector noise. No reliable noise product is available for the raster scan data of NGC~5044.
Because the noise is a strong function of the wavelength, increasing towards the edges of the observed wavelength band, the noise in the adjacent bands 
may reach an amplitude that is comparable to that of the spectral lines, with the lines still remaining statistically significant. The red and black dashed lines show the expected 
(for gas at rest at the redshift of the galaxy) and the observed best fit line centroids, respectively.} 
\label{fig:O1bsample}
\end{figure*}

\end{document}